\documentclass[12pt]{article}
\usepackage{xr}
\makeatletter

\newcommand*{\addFileDependency}[1]{
 \typeout{(#1)}
 \@addtofilelist{#1}
 \IfFileExists{#1}{}{\typeout{No file #1.}}
}
\makeatother

\usepackage{amsmath}
\usepackage{graphicx}
\usepackage{natbib}
\usepackage{url} % not crucial - just used below for the URL 
\usepackage[dvips]{epsfig}
\usepackage{graphicx}
\usepackage{amssymb,amsmath,bm}
\usepackage{setspace}
\usepackage{enumitem}
\usepackage{subfigure}
\usepackage{setspace}
\usepackage{booktabs} 
\usepackage{chngpage} 
\usepackage{mdframed}
\usepackage{comment}
\usepackage[noend]{algpseudocode}
\usepackage{algorithmicx,algorithm}
\usepackage[flushleft]{threeparttable}
\usepackage{mathrsfs}

\usepackage{hyperref}

\usepackage[T1]{fontenc}
\usepackage{tabularx,ragged2e,booktabs,caption}
\usepackage{array}
\newcolumntype{L}[1]{>{\raggedright\let\newline\\\arraybackslash\hspace{0pt}}m{#1}}
\newcolumntype{C}[1]{>{\centering\let\newline\\\arraybackslash\hspace{0pt}}m{#1}}
\newcolumntype{R}[1]{>{\raggedleft\let\newline\\\arraybackslash\hspace{0pt}}m{#1}}

\usepackage[english]{babel}
\usepackage{multirow}
\usepackage{theorem}
\usepackage{float}
\usepackage{caption}
\usepackage{listings}
\usepackage{xcolor}
\usepackage{framed}
\usepackage{lipsum}
\usepackage{authblk}
\usepackage{ulem}

\def\00{\mathrm{0}}

\newtheorem{thm}{Theorem}

\newtheorem{lem}{Lemma}

\newcommand{\blind}{1}

\begin{document}

\def\spacingset#1{\renewcommand{\baselinestretch}%
{#1}\small\normalsize} \spacingset{1}

%%%%%%%%%%%%%%%%%%%%%%%%%%%%%%%%%%%%%%%%%%%%%%%%%%%%%%%%%%%%%%%%%%%%%%%%%%%%%%

\if1\blind
{
 \title{\textbf{Fast robust location and scatter estimation: a depth-based method}}
\author[a]{Maoyu Zhang\thanks{Email: zhangmaoyu@ruc.edu.cn.}}
\author[a]{Yan Song\thanks{Email: 2018000743@ruc.edu.cn.}}
\author[a]{Wenlin Dai\thanks{Corresponding author; email: wenlin.dai@ruc.edu.cn.}}
\affil[a]{\small Center for Applied Statistics, Institute of Statistics and Big Data, Renmin University of China, Beijing, China}
 \date{}
  \maketitle
}
 \fi

\if0\blind
{
  \bigskip
  \bigskip
  \bigskip
  \begin{center}
    {\LARGE\bf Fast robust location and scatter estimation: a depth-based method}
\end{center}
  \medskip
} \fi

\bigskip
\begin{abstract}

The minimum covariance determinant (MCD) estimator is ubiquitous in multivariate analysis, the critical step of which is to select a subset of a given size with the lowest sample covariance determinant. The concentration step (C-step) is a common tool for subset-seeking; however, it becomes computationally demanding for high-dimensional data. To alleviate the challenge, we propose a depth-based algorithm, termed as \texttt{FDB}, which replaces the optimal subset with the trimmed region induced by statistical depth. We show that the depth-based region is consistent with the MCD-based subset under a specific class of depth notions, for instance, the projection depth. With the two suggested depths, the \texttt{FDB} estimator is not only computationally more efficient but also reaches the same level of robustness as the MCD estimator. 
Extensive simulation studies are conducted to assess the empirical performance of our estimators. We also validate the computational efficiency and robustness of our estimators under several typical tasks such as principal component analysis, linear discriminant analysis, image denoise and outlier detection on real-life datasets. A R package \textit{FDB} and potential extensions are available in the Supplementary Materials.

\end{abstract}

\noindent%
\vfill

\newpage
\spacingset{1.8} % DON'T change the spacing!

\section{Introduction}
\label{sec:intro}
The Minimum Covariance Determinant (MCD) estimator \citep{rousseeuw1984least} is among the first affine equivariant and highly robust estimators of multivariate location and scatter. Specifically, for a collection of multivariate data, MCD seeks a subset of samples that leads to a sample covariance matrix with the minimum determinant out of all the candidate sets of a specific size. The location and scatter estimators are then defined as the average and a scaled covariance matrix of these samples, respectively.  \cite{butler1993asymptotics} and \cite{cator2012central} established the consistency and asymptotic normality of the MCD estimator.

MCD has been applied in various fields such as quality control, medicine, finance, image analysis, and chemistry \citep{hubert2008high,hubert2018minimum}. Estimating the covariance matrix is the cornerstone of many multivariate statistical methods, so MCD has also been used to develop robust and computationally efficient multivariate techniques, such as principal component analysis \citep{croux2000principal,hubert2005robpca}, factor analysis \citep{pison2003robust}, classification \citep{hubert2004fast}, clustering \citep{hardin2004outlier}, multivariate regression \citep{rousseeuw2004robust}, and others \citep{hubert2008high}. 
To cater to its broad applications, extensive effort has been made to improve the computational efficiency of the approximation algorithm. For example, \cite{rousseeuw1999fast} propose the first computationally efficient algorithm, termed FASTMCD; \citep{hubert2012deterministic} suggest an improved version of FASTMCD, termed DetMCD; \cite{de2020real} accelerates DetMCD by refinement of the calculation steps and parallel computation. Furthermore, \cite{boudt2020minimum} generalizes the MCD to high-dimensional cases as the minimum regularized covariance determinant (MRCD). Other variants include the orthogonalized Gnanadesikan-Kettenring \citep{maronna2002robust}, the minimum (regularized) weighted covariance determinant \citep{roelant2009minimum, kalina2021minimum}, and kernel MRCD for non-elliptical data \citep{Schreurs2021OutlierDI}.

Practically, the MCD-type algorithms are limited by two factors. First, the computational complexity of the concentration step (C-step), which is critical for such algorithms, is $O(np^2+p^3)$, and this severely limits the scalability of the algorithm for massive high-dimensional data. Second, the approximation to the true MCD subset becomes less accurate due to the curse of dimensionality. We note that the asymptotic trimmed region induced by a class of statistical depth shares the same form with the asymptotic MCD subset when the data are elliptically symmetric distributed \citep{butler1993asymptotics,zuo2000structural}. This motivates us to investigate the possibility of finding the MCD subset directly by utilizing statistical depth. 
Statistical depth was first considered for ranking multivariate data from the center outward \citep{mahalanobis1936generalized,tukey1975mathematics,oja1983descriptive,liu1990notion, zuo2000general, vardi2000multivariate}. Usually, a statistical depth is an increasing function of the centrality of observations, taking values in $[0,1]$.

Motivated by the connection mentioned above, we propose a fast depth-based algorithm, denoted as $\texttt{FDB}$, which approximates the MCD subset with a depth-induced trimmed region. Specifically, we investigate $\texttt{FDB}$ based on two representative depth notions, projection depth and $L_2$ depth, and denote the estimators as, $\texttt{FDB}_{\rm pro}$ and $\texttt{FDB}_{\rm L_2}$, respectively. Four main advantages of the proposed algorithm are worth mentioning: 1) Asymptotically, $\texttt{FDB}_{\rm pro}$ leads to a trimmed region equivalent to the MCD subset for elliptically symmetric distributions. 
2) Both $\texttt{FDB}_{\rm pro}$ and $\texttt{FDB}_{\rm L_2}$ achieve the same level of robustness as the MCD estimator. 
3) Empirically, $\texttt{FDB}$ reveals comparable or even better performance than the MCD estimator regarding estimation accuracy. 4) Furthermore, the computational efficiency is dramatically improved by using $\texttt{FDB}$, especially for high-dimensional cases.

The rest of the paper is organized as follows. Section 2 reviews the MCD estimator and some related theoretical properties. 
Section 3 introduces the idea of \texttt{FDB} estimators and demonstrates the theoretical equivalence between the MCD subsets and the depth-trimmed regions. 
Section 4 investigates the invariance, robustness, and computational complexity of the proposed $\texttt{FDB}$ estimators.  
In Section 5, we conduct extensive simulation studies to assess the performance of the $\texttt{FDB}$ algorithm and compare it with the existing ones regarding estimation accuracy and computational efficiency. Section 6 applies the proposed methods to several real applications through typical multivariate analysis tasks, including principal component analysis, linear discriminant analysis, image denoise, and outlier detection. 
We end the paper with some discussion in Section 7. Proofs of the theoretical results and additional simulation results are provided in the Supplementary Material.

\section{MCD Estimators}

In this section, we review the theoretical property of the MCD estimator as well as three widely utilized approximation algorithms. Let $\boldsymbol{x} \in \mathbb{R}^p$ be a random variable from an elliptically symmetric distribution, denoted as ${\rm ES}(f; \boldsymbol{\mu},\boldsymbol{\Sigma})$, whose density is of the form
$$
g(x)=c|\boldsymbol{\Sigma}|^{-1 / 2} f\left((\boldsymbol{x}-\boldsymbol{\mu})^{T} \boldsymbol{\Sigma}^{-1}(\boldsymbol{x}-\boldsymbol{\mu})\right),
$$
where $\boldsymbol{\Sigma}$ is a symmetric positive definite matrix, and the function $f:\mathbb{R}_{+} \rightarrow \mathbb{R}_{+}$ is assumed to be non-increasing so that $g(x)$ is unimodal.

Considering random samples $\boldsymbol{x}_1,\dots,\boldsymbol{x}_n$ independently generated from the above distribution, MCD aims to solve the following optimization problem
\begin{equation}\label{mcd}
\hat{H}_{\alpha_n,{\rm MCD}}=\underset{H \in \mathcal{H}_{h}}{\operatorname{argmin}}\left(\operatorname{det}\mathbf{\hat{\Sigma}}(\boldsymbol{x}_H)\right),
\end{equation}
where $H$ is an index set of $h$ observations (with $\lfloor(n+p+1) / 2\rfloor \leqslant h \leqslant n$, where $\lfloor a \rfloor$ means the largest integer smaller than or equal to $a$), $\alpha_n=h/n$ and $\mathcal{H}_{h}$ is the collection of all such sets. Observations with corresponding  indices $\hat{H}_{\alpha_n,{\rm MCD}}$ constitute the final MCD subset, termed $\hat{E}_{\alpha_n,{\rm MCD}}$,. Define $\Delta(A,B)=A\cup B -A\cap B$ as the difference between sets $A$ and $B$. The convergence property of ${\hat E}_{\alpha_n {\rm MCD}}$ is revisited in Lemma \ref{thm1}, with the proof provided in Section S1.1 of the Supplementary Material.

\vskip 10pt

\begin{lem}\label{thm1}
Assume that the random samples $\boldsymbol{x}_i \sim ES(f; \boldsymbol{\mu},\boldsymbol{\Sigma}),i=1,\dots ,n$. Then for $\alpha>0$, we have 
$$\mathbb{P}\left(\Delta\left(\hat{E}_{\alpha_n, \text {\rm MCD}}, E_{\alpha}\right)\right) \rightarrow 0$$ 
for any sequence $\alpha_n\to \alpha$ as $n\to \infty$, 
where $E_{\alpha}=\left\{\boldsymbol{x} \in \mathbb{R}^{p} \mid(\boldsymbol{x}-\boldsymbol{\mu})^{T} \boldsymbol{\Sigma}^{-1}(\boldsymbol{x}-\boldsymbol{\mu}) \leq r_{\alpha}^{2}\right\}$ with $\mathbb{P}(E_{\alpha})=\alpha$. 
\end{lem}

Given an $n \times p$ data matrix $\boldsymbol{X}=(\boldsymbol{x}_1,\dots,\boldsymbol{x}_n)'$, with its
estimated center $\hat{\boldsymbol{\mu}}$ and scatter matrix $\hat{\boldsymbol{\Sigma}}$, we
denote with $\mathcal{D}\left(\boldsymbol{x}_{i};\hat{\boldsymbol{\mu}} , \hat{\boldsymbol{\Sigma}}\right)=\sqrt{\left(\boldsymbol{x}_{i}-\hat{\boldsymbol{\mu}}\right)^{T} \hat{\boldsymbol{\Sigma}}^{-1}\left(\boldsymbol{x}_{i}-\hat{\boldsymbol{\mu}}\right)}$ the Mahalanobis distance of $\boldsymbol{x}_i$. 
The C-step described in Algorithm~\ref{alg} is crucial for MCD-type algorithms.
\begin{algorithm}[h]
	\caption{The C-step} 
	\hspace*{0.02in}{\bf Input: initial subset $H_{old}$ or the estimates ($\hat{\boldsymbol{\mu}}_{\text {old}}, \hat{\boldsymbol{\Sigma}}_{\text {old}}$), subset size $h$}.\\
	\hspace*{0.02in}{\bf Output: $H_{new}$ or ($\hat{\boldsymbol{\mu}}_{\text {new}}, \hat{\boldsymbol{\Sigma}}_{\text {new}}$)}
	\begin{algorithmic}[1]
\State Compute the distances $d_{i,{\rm old}}=\mathcal{D}\left(\boldsymbol{x}_{i};\hat{\boldsymbol{\mu}}_{\text {old}}, \hat{\boldsymbol{\Sigma}}_{\text {old}}\right)$ for $i=1, \ldots, n$.

\State Sort these distances, yielding a permutation $\pi$ for which $d_{\pi(1),\text {old}} \leqslant d_{\pi(2),\text {old}} \leqslant \ldots \leqslant d_{\pi(n), \text {old}}$, and set $H_{new}=\{\pi(1), \pi(2), \ldots, \pi(h)\}$.

\State Update $\hat{\boldsymbol{\mu}}$ and $\hat{\boldsymbol{\Sigma}}$ as
$$\hat{\boldsymbol{\mu}}_{\text {new}}=\frac{1}{h}\sum_{i \in H_{new}} \boldsymbol{x}_{i}\quad {\rm and}\quad \hat{\boldsymbol{\Sigma}}_{\text {new}}=\frac{1}{h-1}\sum_{i \in H_{new}}\left(\boldsymbol{x}_{i}-\hat{\boldsymbol{\mu}}_{\text {new}}\right)\left(\boldsymbol{x}_{i}-\hat{\boldsymbol{\mu}}_{\text {new}}\right)^{T}.$$
\end{algorithmic}\label{alg}
\end{algorithm}

\cite{rousseeuw1999fast} proposed the first computationally feasible algorithm, termed FASTMCD, for approximating the MCD subset. Specifically, they randomly constructed several initial subsets and applied two C-steps for each subset, yielding the ten subsets with the lowest determinant. Then, they took C-step iteratively for these ten subsets until the determinant sequence converged and eventually chose the MCD subset as the one leading to the smallest determinant. 
Given this, the computational efficiency of FASTMCD is thus roughly proportional to the number of the initial subsets. 
\cite{hubert2012deterministic} proposed an alternative algorithm DetMCD, which replaces random initial subsets (of which there could be many) in FASTMCD with six well-designed deterministic estimators of $(\boldsymbol{\mu}, \boldsymbol{\Sigma})$, and also involves the C-step in a similar way. Denote the estimates $\left(\hat{\boldsymbol{\mu}}, \hat{\boldsymbol{\Sigma}}\right)$ as the location and scatter matrix estimates of the $h$-subset for which the determinant of the sample covariance matrix is as small as possible. Further, an additional reweighted step is employed in both algorithms to improve the efficiency of the estimators. To be more specific, the estimators are renewed as trimmed estimates for location and scatter,
\begin{equation}\label{rew}
\hat{\boldsymbol{\mu}}_{\rm re}=\frac{1}{\sum_{i=1}^{n} W_i}\sum_{i=1}^{n} W_i \boldsymbol{x}_{i}\quad {\rm and}\quad  \hat{\boldsymbol{\Sigma}}_{\rm re}= \frac{1}{\sum_{i=1}^{n} W_i-1}\sum_{i=1}^{n} W_i\left(\boldsymbol{x}_{i}-\hat{\boldsymbol{\mu}}_{re}\right)\left(\boldsymbol{x}_{i}-\hat{\boldsymbol{\mu}}_{re}\right)^{T},
\end{equation}
where $W_i=1$ when $ \mathcal{D}\left(\boldsymbol{x}_{i};\hat{\boldsymbol{\mu}}, c_0\hat{\boldsymbol{\Sigma}}\right) \le \sqrt{\chi_{p,0.975}^2}$  and 0 otherwise,  $\chi_{p,\alpha}^2$ is the $\alpha$-quantile of the $\chi_{p}^2$ distribution and $c_0=\operatorname{med}_{i} \mathcal{D}^2\left(\boldsymbol{x}_{i}, \hat{\boldsymbol{\mu}}, \hat{\boldsymbol{\Sigma}}\right)/{\chi_{p, 0.5}^{2}}$.

\section{A Depth-based Alternative}
The idea behind the premier step of MCD-type algorithms is to construct outlier-free subsets as the initial values for the C-step. This motivates us to 
approach such a purpose using statistical depth, a popular tool for robust multivariate data analysis. More importantly, we find the equivalence between the eventual MCD subset and the depth-trimmed region, which avoids the implementation of the iterative C-steps and hence improves the computational efficiency dramatically.

In general, a statistical depth notion is a function $D:(\boldsymbol{x}, P) \mapsto[0,1]$, for $\boldsymbol{x} \in \mathbb{R}^p$ and $P$ from some class $\mathcal{P}$ of $p$-variate probability distributions, that 
provides a center-outward order for a collection of data. 
Taking into account the robustness as well as the computational efficiency, to be discussed later, we specifically consider the following two depth notions for the proposed method.

\textbf{Projection depth} \citep{zuo2000general}:
$$
D_{\operatorname{Proj}}(\boldsymbol{x}; P)=\left(1+\sup _{\|\boldsymbol{u}^{\prime}\|=1} \frac{\left|\boldsymbol{u}^{\prime}\boldsymbol{x}-\operatorname{med}\left(\boldsymbol{u}^{\prime} \boldsymbol{y}\right)\right|}{\operatorname{MAD}\left(\boldsymbol{u}^{\prime} \boldsymbol{y}\right)}\right)^{-1},
$$
where $P$ is the distribution of $\boldsymbol{y}$, $\operatorname{med}(V)$ denotes the median of a univariate random variable $V$, and $\operatorname{MAD}(V)=$ $\operatorname{med}(|V-\operatorname{med}(V)|)$ its median absolute deviation from the median. Practically, one may choose a finite number of random directions to approximate the projection depth values.

\textbf{$\mathbb{L}_2$ depth} \citep{zuo2000general}:
$$
D_{\mathbb{L}_{\mathrm{2}}}(\boldsymbol{x};P)=\left(1+E\left[\|\boldsymbol{y}-\boldsymbol{x}\|_2\right]\right)^{-1},
$$
where $\|\cdot\|_2$ is the $L^2$ norm.

Consider random samples $\boldsymbol{x}_1,\dots,\boldsymbol{x}_n$ independently generated from the elliptically symmetric (ES) distribution. For a given depth function $D(\cdot ; {\rm ES})$ and for $\alpha>0$, we call
$$
E_{\alpha,\text{depth}} = \left\{\boldsymbol{x} \in \mathbb{R}^{p} \mid D(\boldsymbol{x} ; {\rm ES}) \geq D_{\alpha}\right\},
$$
and the sample version is
$$
\hat{E}_{\alpha_n, \mathrm{depth}}=\left\{x \in \mathbb{R}^p\mid \hat{D}_n(\boldsymbol{x} ; {\rm ES}) \geq D_{\alpha_n}\right\},$$
the corresponding $\alpha$-trimmed region with $\mathbb{P}(E_{\alpha,\text{depth}})=\alpha$.

\begin{lem}[\cite{zuo2000structural}]\label{thm2}
Assume that the random samples $\boldsymbol{x}_i \sim {\rm ES}(f; \boldsymbol{\mu},\boldsymbol{\Sigma})$, $i=1,\dots ,n$. Then for the projection depth, the depth trimmed region (subset) $E_{\alpha_n,\text{\rm depth}}$ satisfies 
$$
\mathbb{P}\left(\Delta\left(\hat{E}_{\alpha_n, \text{\rm depth}}, E_{\alpha}\right)\right) \rightarrow 0 \text { as } n \rightarrow \infty,
$$ 
for any sequence $\alpha_n\to \alpha>0$ as $n\to \infty$. 
\end{lem}

Other depth notions could also lead to the same conclusion except for the projection depth. We omit those notions since they are either not robust enough or computationally demanding. Also note that the result does not necessarily hold for the $L_2$ depth, although $\texttt{FDB}_{\rm L_2}$ indeed provides satisfactory results in the simulation. Combining Lemmas \ref{thm1} and \ref{thm2}, it is straightforward that the two subsets are asymptotic equivalent. We state this result formally in the following theorem.

\begin{thm}\label{thm3}
Assume that the random samples $\boldsymbol{x}_i \sim {\rm ES}(f; \boldsymbol{\mu},\boldsymbol{\Sigma})$, $i=1,\dots , n$. Under the conditions of Lemmas \ref{thm1} and \ref{thm2}, we have
$$\mathbb{P}(\Delta(E_{\alpha_n,\text{\rm depth}},E_{\alpha_n,\text{\rm MCD}}))\to 0, \text{ as } n \rightarrow \infty.$$ 
\end{thm}

\begin{figure}[!t] 
	\centering
	\subfigure[]{\label{a}
		\centering
		\includegraphics[width=0.4\linewidth]{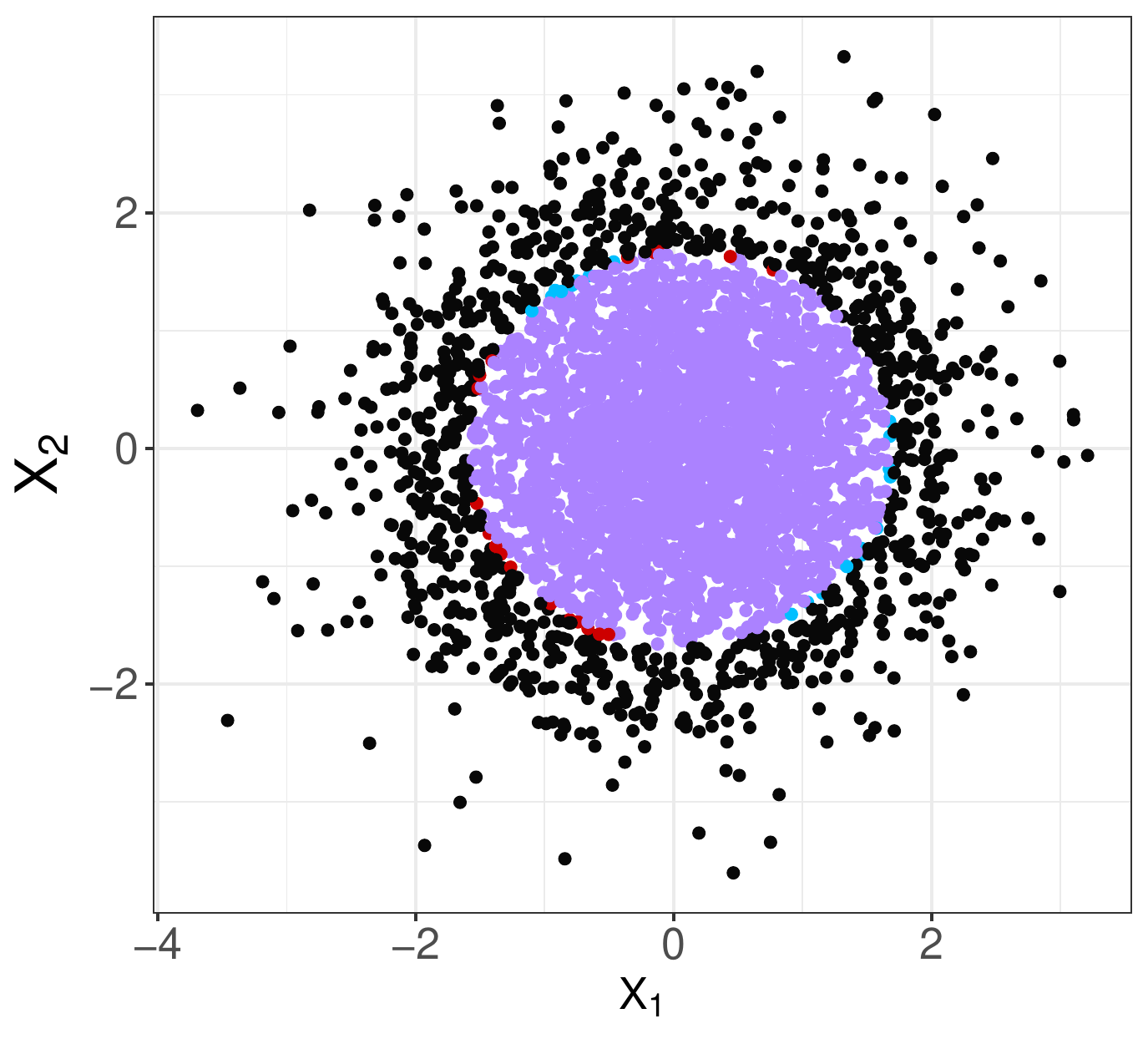}}
	\subfigure[]{\label{b}
		\centering
		\includegraphics[width=0.4\linewidth]{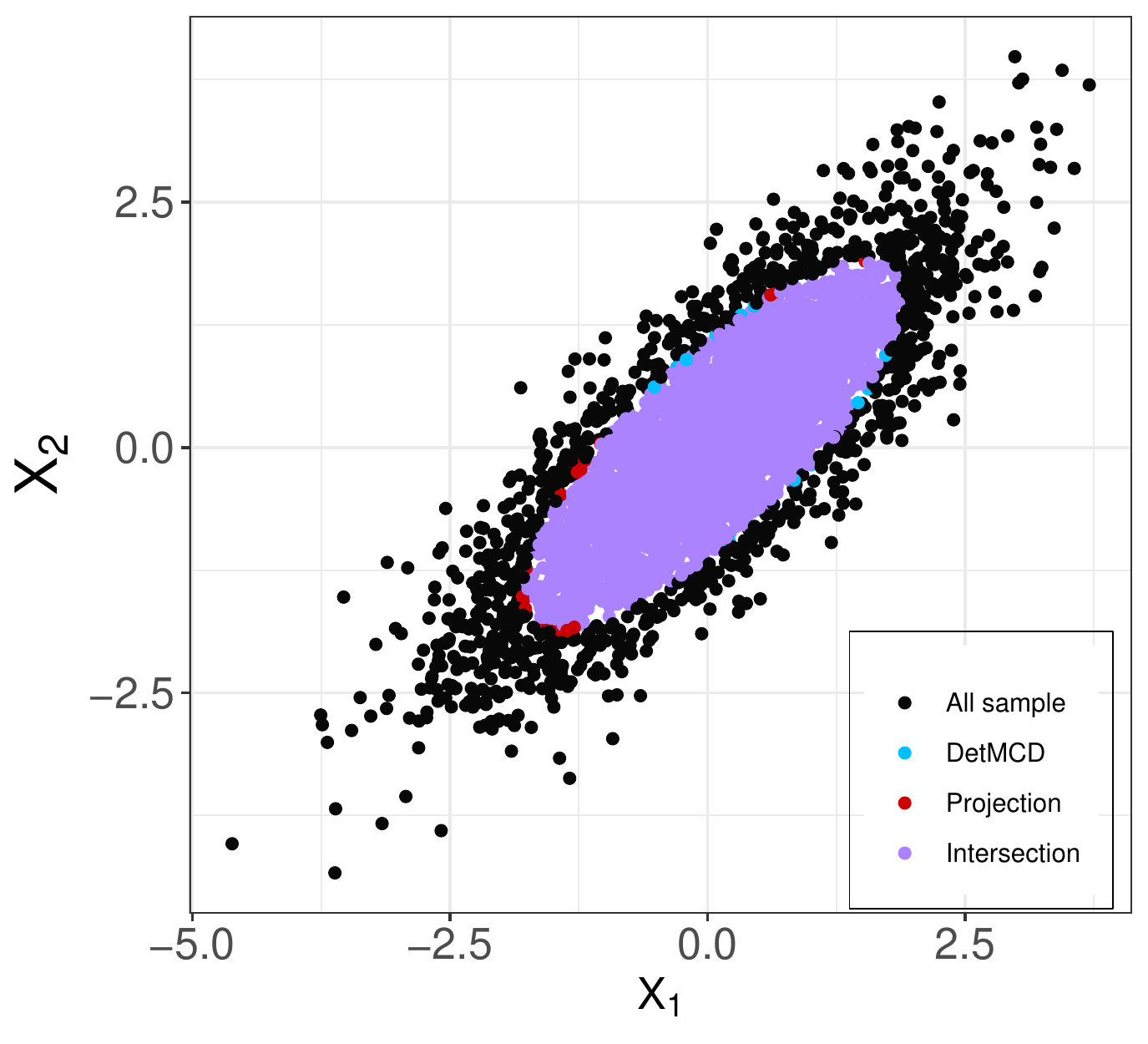}}
\caption{The subsets induced by the projection depth and DetMCD. The black points represent the samples that are outside both subsets; blue points indicate samples in DetMCD subset only; red points indicate samples in depth-based subset only; purple points are their intersection. Plots (a) and (b) are for independent and correlated data, respectively.}\label{s1}
\end{figure}

Proof of Theorem \ref{thm3} is provided in Section S1.2 of the Supplementary Material. Motivated by the above result, we propose to approximate the eventual MCD subset with the depth-trimmed region and avoid the iterative implementation of the C-step. We provide two toy examples in Figure \ref{s1} to illustrate such a coincidence. Specifically, we generate data from bivariate normal distributions with unit variance and correlation coefficients 0 and 0.5 for Figure \ref{a} and \ref{b}, respectively. We consider $n=4000$ and $h=3000$. In both cases, the two subsets match quite well, such that the proportions of the common elements are no less than 97\%, which well supports the high effectiveness of the proposed method.

For data of low dimensions, both MCD subsets and depth-based trimmed regions can be computed efficiently, and the result matches well, as shown in Figure \ref{s1}. However, for high-dimensional data, the MCD algorithms are severely challenged by the cubically increased computational complexity, and hence the approximation will be less efficient. To alleviate this challenge, we consider replacing the MCD subsets with the trimmed regions induced by some computationally efficient depth notions. By doing so, we may not only reduce the computational time significantly but also attain comparable or better robust estimation for both the location and scatter matrix, especially in high-dimensional cases. In what follows, we introduce the \texttt{FDB} algorithm.

\begin{algorithm}[h]
	\caption{\texttt{FDB}} 
	\hspace*{0.02in}{\bf Input: $\boldsymbol{x_i},i=1,\dots ,n$, subset size $h$, selected depth notion. }\\
	\hspace*{0.02in}{\bf Output:} $\hat{\boldsymbol{\mu}}_{\text{FDB}}$, $\hat{\boldsymbol{\Sigma}}_{\text{FDB}}$.
	\begin{algorithmic}[1]
		\State Calculate the depth value for each observation $\boldsymbol{x}_{i}$, denoted as $\text{GD}(i)$.
		\State Sort these values, yielding a permutation $\pi$ of $1,\dots, n$, for which $\text{GD}(\pi(1)) \geq,\dots,\geq \text{GD}(\pi(n))$, and a set $H_{sub}=\{\pi(1),\dots,\pi(h)\}$.
		\State Get the location and the scatter matrix estimates as
		$$\hat{\boldsymbol{\mu}}_{\rm raw}=\frac{1}{h}\sum_{i \in H_{sub}} \boldsymbol{x}_{i}\quad {\rm and}\quad \hat{\boldsymbol{\Sigma}}_{\rm raw}=\frac{c_1}{h} \sum_{ i \in H_{sub}}\left(\boldsymbol{x}_{i}-\boldsymbol{\mu}_{\rm raw}\right)\left(\boldsymbol{x}_{i}-\boldsymbol{\mu}_{\rm raw}\right)^{T},$$
		where $c_1=\underset{i}{\operatorname{med}} \mathcal{D}^2\left(\boldsymbol{x}_{i}, \hat{\boldsymbol{\mu}}_{\rm raw}, \hat{\boldsymbol{\Sigma}}_0\right)/{\chi_{p, 0.5}^{2}}$ with $ \hat{\boldsymbol{\Sigma}}_{0}=\frac{1}{h} \sum_{ i \in H_{sub}}\left(\boldsymbol{x}_{i}-\boldsymbol{\mu}_{\rm raw}\right)\left(\boldsymbol{x}_{i}-\boldsymbol{\mu}_{\rm raw}\right)^{T}$.
		\State Apply the reweighted step (\ref{rew}) to the raw estimates, yielding the final FDB estimates, $\hat{\boldsymbol{\mu}}_{\mathrm{FDB}}$ and $\hat{\boldsymbol{\Sigma}}_{\mathrm{FDB}}$.
	\end{algorithmic}\label{algfdb}
\end{algorithm}

Algorithm \ref{algfdb} considers the case of $h>p$, which is the condition to guarantee the invertibility of estimated matrix \citep{rousseeuw1990unmasking}. All algorithms for the original MCD require that the dimension $p$ be lower than $h$ to obtain an invertible covariance matrix. It is recommended that $n > 5p$ in practice \citep{rousseeuw1999fast}.

According to Lemmas 1 and 2, for data from an elliptically symmetric distribution, MCD and \texttt{FDB} algorithms both approximate the optimal subset, though from different perspectives. MCD approaches the solution by combining well selected (or random) initial subset and the iterative implementation of the C-step, which could be computationally demanding for high dimensional data. In contrast, \texttt{FDB} relies on ordering the data from the center outward, and hence its computational complexity is mainly determined by the cost of assigning depth values to each sample.

The idea of incorporating depth (outlyingness) to construct MCD estimators has been considered in the literature. 
For example, the Stahel–Donoho outlyingness \citep{don0ho_1982breakdown}, equivalent to the projection depth, is applied to determine an $h$-subset consisting of the $h$ points with the lowest outlyingness, and the corresponding sample mean and covariance matrix are used as one initial value for the C-step \citep{Hubert2005ROBPCAAN,Schreurs2021OutlierDI}. \cite{debruyne2009influence} studied the influence function and asymptotic relative efficiency of the estimators obtained directly based on such a subset (without the reweighted step). For the first time, we establish the equivalence of the two subsets, which indicates that the depth-based subset is a reasonable approximation to the optimal subset rather than just one option of the initial value for the C-step.

\section{Properties of FDB}
This section focuses on the properties of the \texttt{FDB} estimators. 
Specifically, We discuss three types of properties, that are of main interest for such methods \citep{maronna2002robust,hubert2012deterministic}, invariance, robustness and computational complexity. We show that the proposed estimators are quite satisfactory in these aspects.

\subsection{Invariant Properties}

 {\bf Affine equivariance} makes the analysis independent of any affine transformation of the data. 
 For any nonsingular $p \times p$ matrix $\boldsymbol{A}$ and $p \times 1$ vector $\boldsymbol{v}$, the estimators $\hat{\boldsymbol{\mu}}$ and $\hat{\boldsymbol{\Sigma}}$ are affine equivariant if they satisfy 
 $$\hat{\boldsymbol{\mu}}\left(\boldsymbol{X} \boldsymbol{A}+\mathbf{1}_{n} \boldsymbol{v}^{T}\right) =\hat{\boldsymbol{\mu}}(\boldsymbol{X}) \boldsymbol{A}+\boldsymbol{v} {\rm~and~} \hat{\boldsymbol{\Sigma}}\left(\boldsymbol{X} \boldsymbol{A}+\mathbf{1}_{n} \boldsymbol{v}^{T}\right) =\boldsymbol{A}^{T} \hat{\boldsymbol{\Sigma}}(\boldsymbol{X}) \boldsymbol{A},$$ 
 where $\mathbf{1}_{n}=(1,1, \ldots, 1)^{T}$. 
The projection depth has been shown affine equivariant \citep{zuo2000general,zuo2006multidimensional}, that is the depth value does not vary through affine transformation for any sample, and hence the indexes of samples forming the trimmed region remain the same. 
Consequently, $\texttt{FDB}_{\rm pro}$ is obviously affine equivariant. 
For $\texttt{FDB}_{\rm L_2}$, a similar property holds for rigid transformation \citep{mosler2022choosing}, which is a bit more restrictive than the affine transformation. For high dimensional situations, the affine equivariance may be less important under nonstandard data contamination such as componentwise outliers \citep{alqallaf2009propagation}.

{\bf Permutation invariance} provides an effective way to guaranteeing the robustness of analysis to the perturbation of observations. An estimator $T(\cdot)$ is said to be permutation invariant if $T(\boldsymbol{P} \boldsymbol{X})=T(\boldsymbol{X})$ for any permutation matrix $\boldsymbol{P}$. 
Permutation invariance holds for both projection depth and $L_2$ depth since they do not involve any random subsets, and hence the depth values remain the same through permutation and so will the $h$-subset.

\subsection{Robustness}
	
Robustness is the property of main interest when outlier contamination of the data is suspected. As aforementioned, the MCD estimator is highly robust that it achieves the highest possible asymptotic breakdown point, about 1/2, with $h\approx n/2$. The robustness of \texttt{FDB} is determined by the property of the employed depth notion. According to \cite{zuo2006multidimensional} and \cite{lopuhaa1991breakdown}, the breakdown points of the trimmed regions induced by the projection depth and $L_2$ depth are both 1/2, with $\alpha \approx 0.5$. That is, $\texttt{FDB}_{\rm pro}$ and $\texttt{FDB}_{\rm L_2}$ both have a breakdown point as high as that of the MCD estimator. 
Another indicator is the influence function, which captures the local robustness of estimators. \cite{zuo2006multidimensional} and \cite{niinimaa1995influence} showed the influence functions of depth regions induced by projection depth and $L_2$ depth are both bounded. Therefore, $\texttt{FDB}_{\rm pro}$ and $\texttt{FDB}_{\rm L_2}$ are highly robust locally as well as globally.

\begin{figure}[!t] 
	\centering
	\subfigure[10\% cluster]{\label{cluster0.1}
		\centering
		\includegraphics[width=0.4\linewidth]{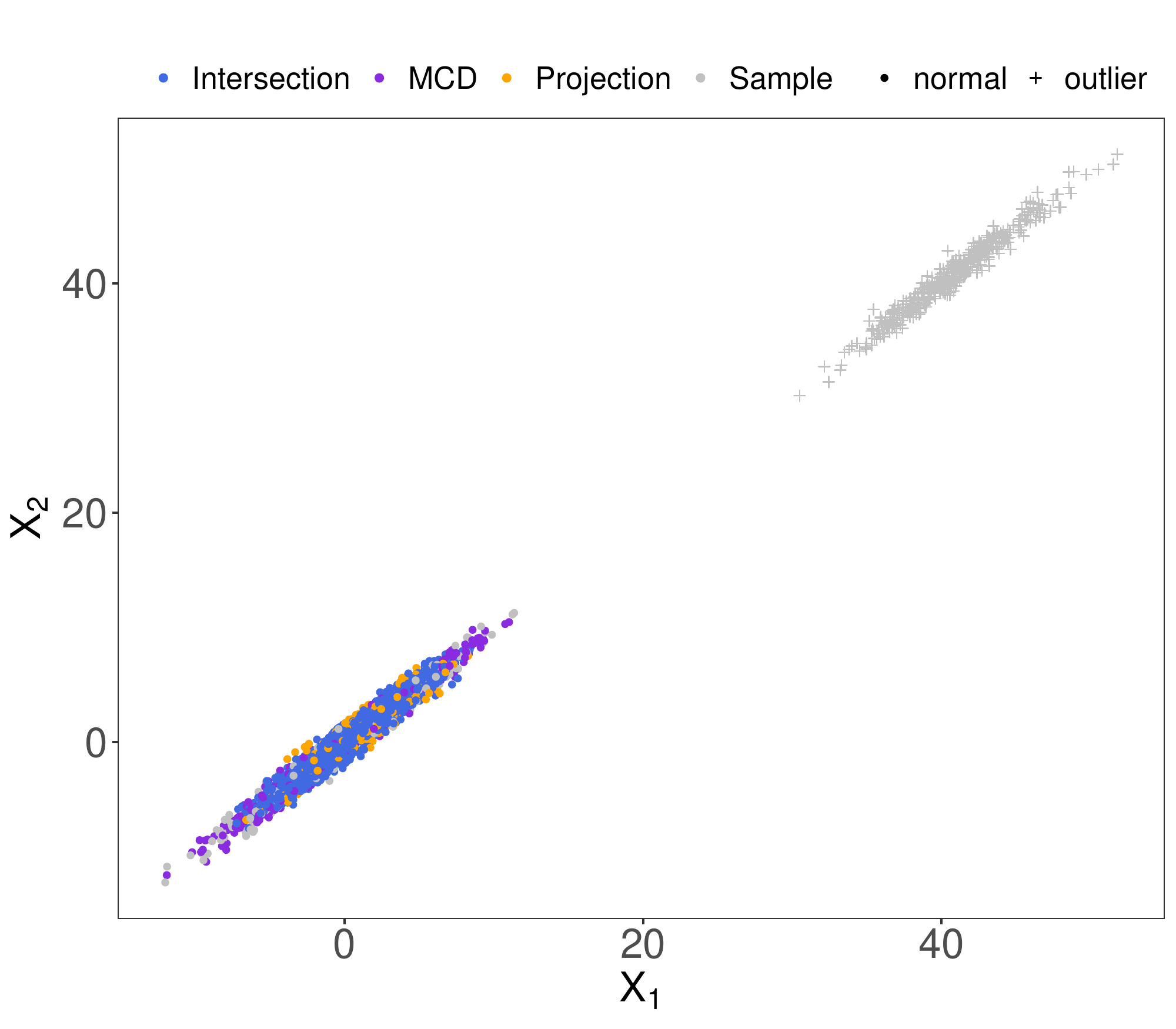}}
	\subfigure[40\% cluster]{\label{cluster0.4}
		\centering
		\includegraphics[width=0.4\linewidth]{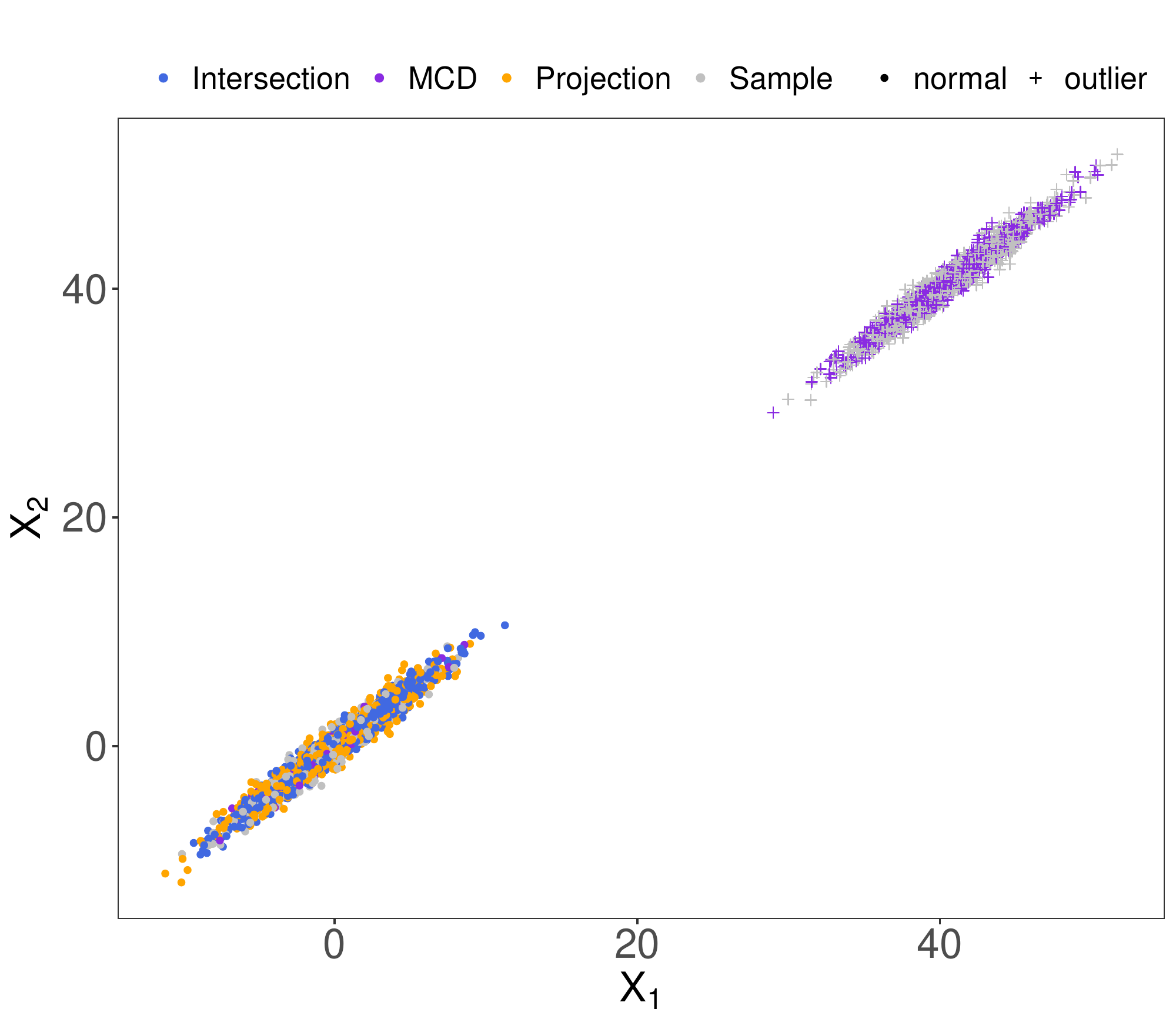}}
	\subfigure[10\% point ]{\label{point0.1}
		\centering
		\includegraphics[width=0.4\linewidth]{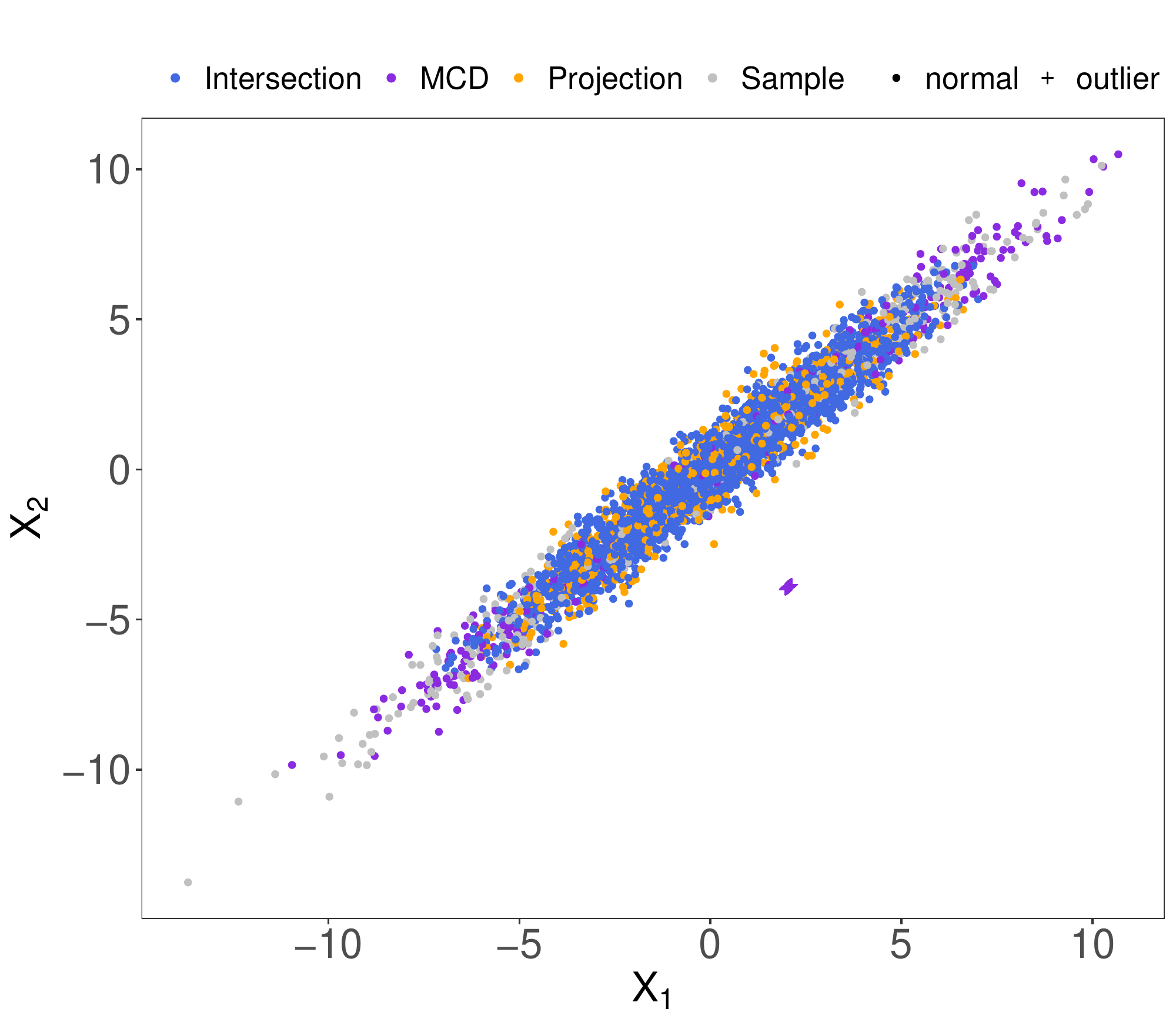}}
  \subfigure[40\% point ]{\label{point0.4}
		\centering
		\includegraphics[width=0.4\linewidth]{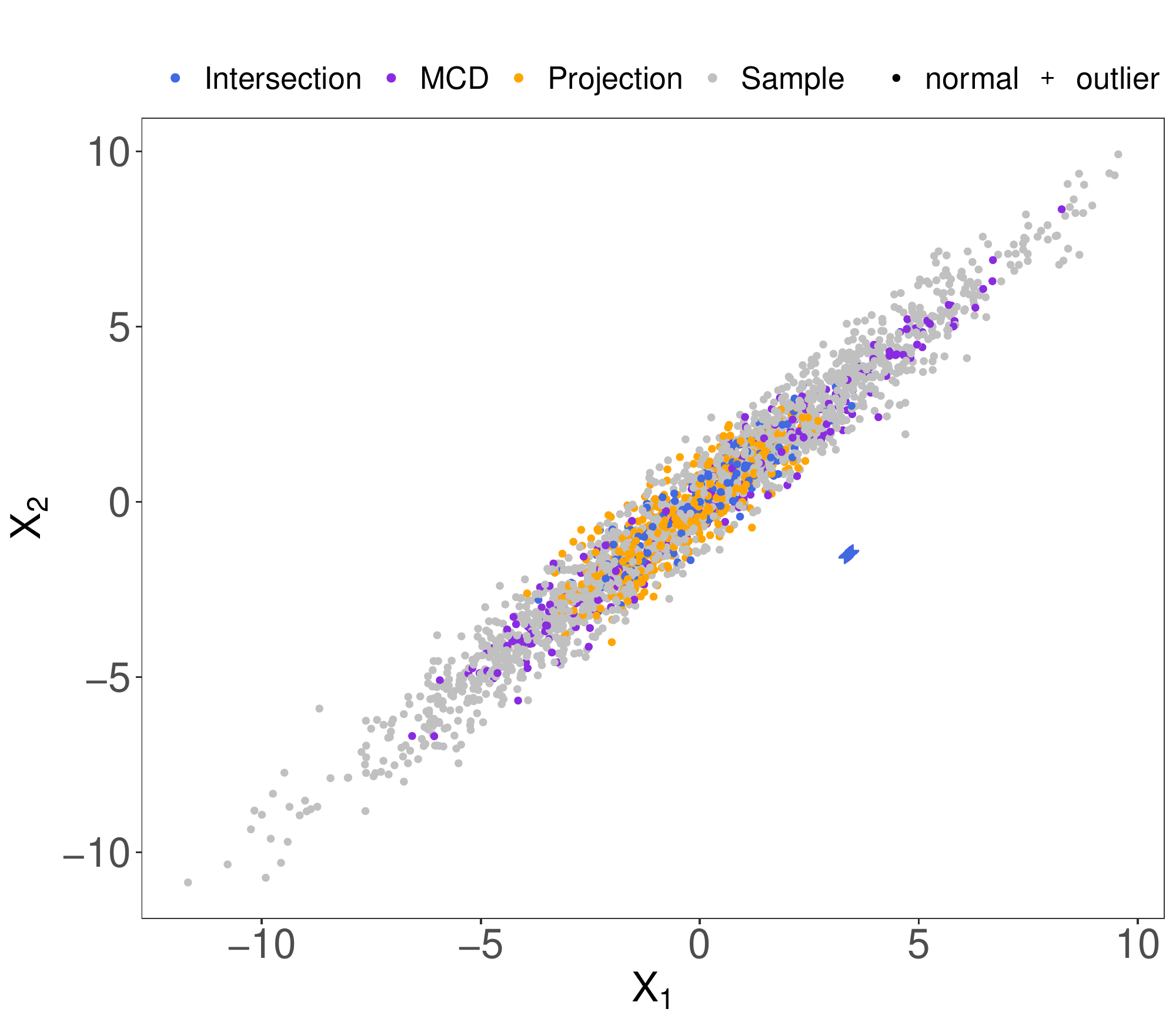}}
\caption{The subsets induced by $\texttt{FDB}_{\rm pro}$ and DetMCD.
Dots denote normal sample and crosses denote outliers. Blue ones form the intersection of two subsets; orange (purple) ones indicate the samples unique to the $\texttt{FDB}_{\rm pro}$ (DetMCD) subset; blue ones are samples dropped by both subsets.}\label{s2}
\end{figure}

The robustness discussed above is from the theoretical aspect. Practically, both \texttt{FDB} and MCD approximate the theoretically optimal subsets, and their empirical performances do not necessarily match the theoretical results under all the scenarios. This means that the subset selected by the two methods under finite samples may still contain outliers. To show this point, we provide an example in Figure \ref{s2}, where we generate $n=4000$ samples from a 40-dimensional normal distribution with standard normal marginal distributions and a correlation coefficient of 0.5. We consider two levels of contamination, 10\% and 40\%, and two types of outliers, point and cluster (see for details in Section \ref{sym}), respectively. For the first column with 10\% outliers, we set $h=\lfloor 75\%n \rfloor$; for the second column with 40\% outliers, $h=\lfloor 50\%n \rfloor$.
$\texttt{FDB}_{\rm pro}$ performs perfectly for the first three cases and fails for the last scenario; in contrast, DetMCD is only satisfactory for the first case but fails for the rest. This indicates that for high-dimensional data, $\texttt{FDB}_{\rm pro}$ provides more reliable approximations to the optimal subset.

\subsection{Computational complexity}\label{cc}

\begin{figure}[!b]
	\centering
	\subfigure[$n=1000$]{
		\centering
		\includegraphics[width=0.4\linewidth]{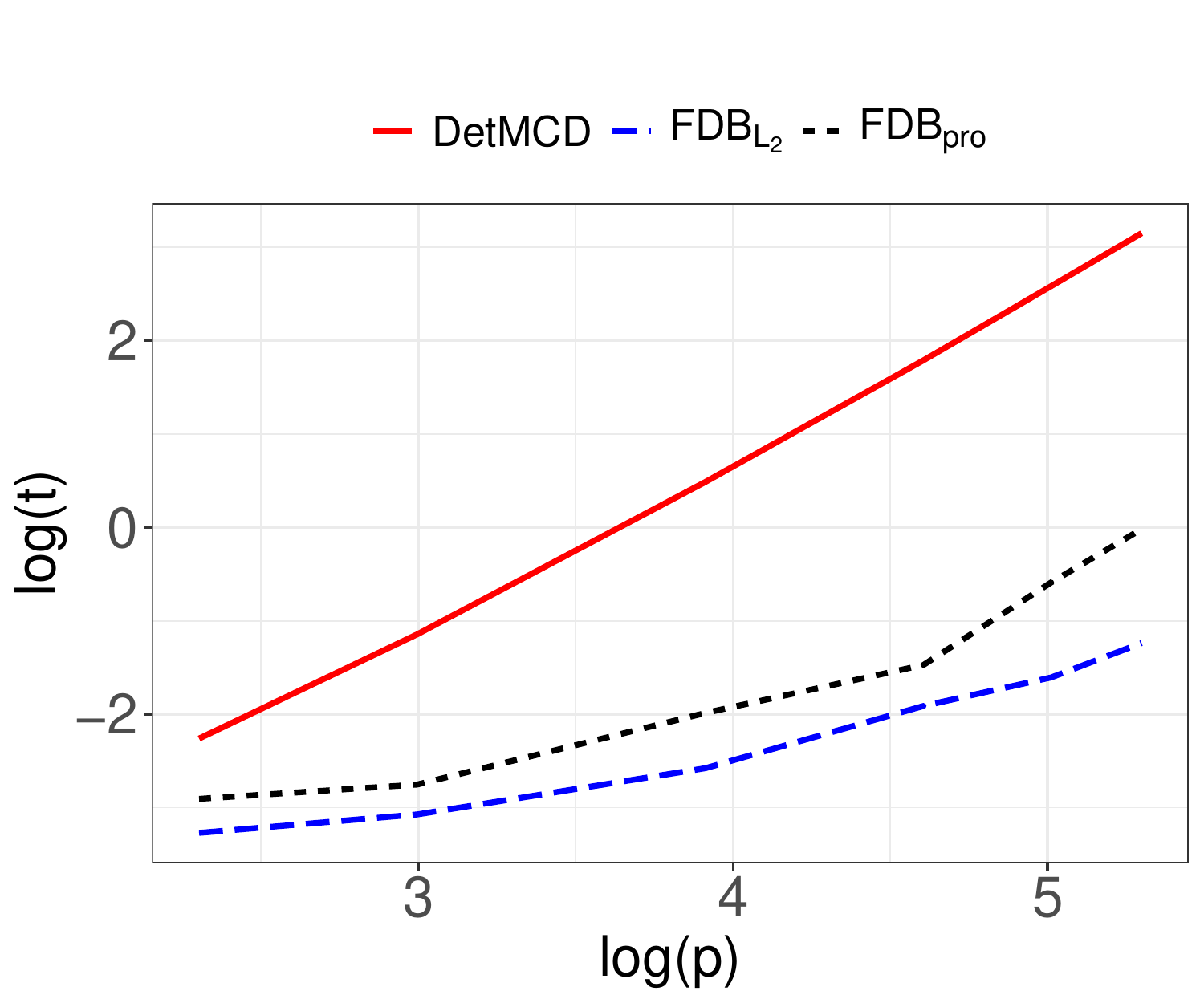}}
	\subfigure[$p=200$]{
		\centering
		\includegraphics[width=0.4\linewidth]{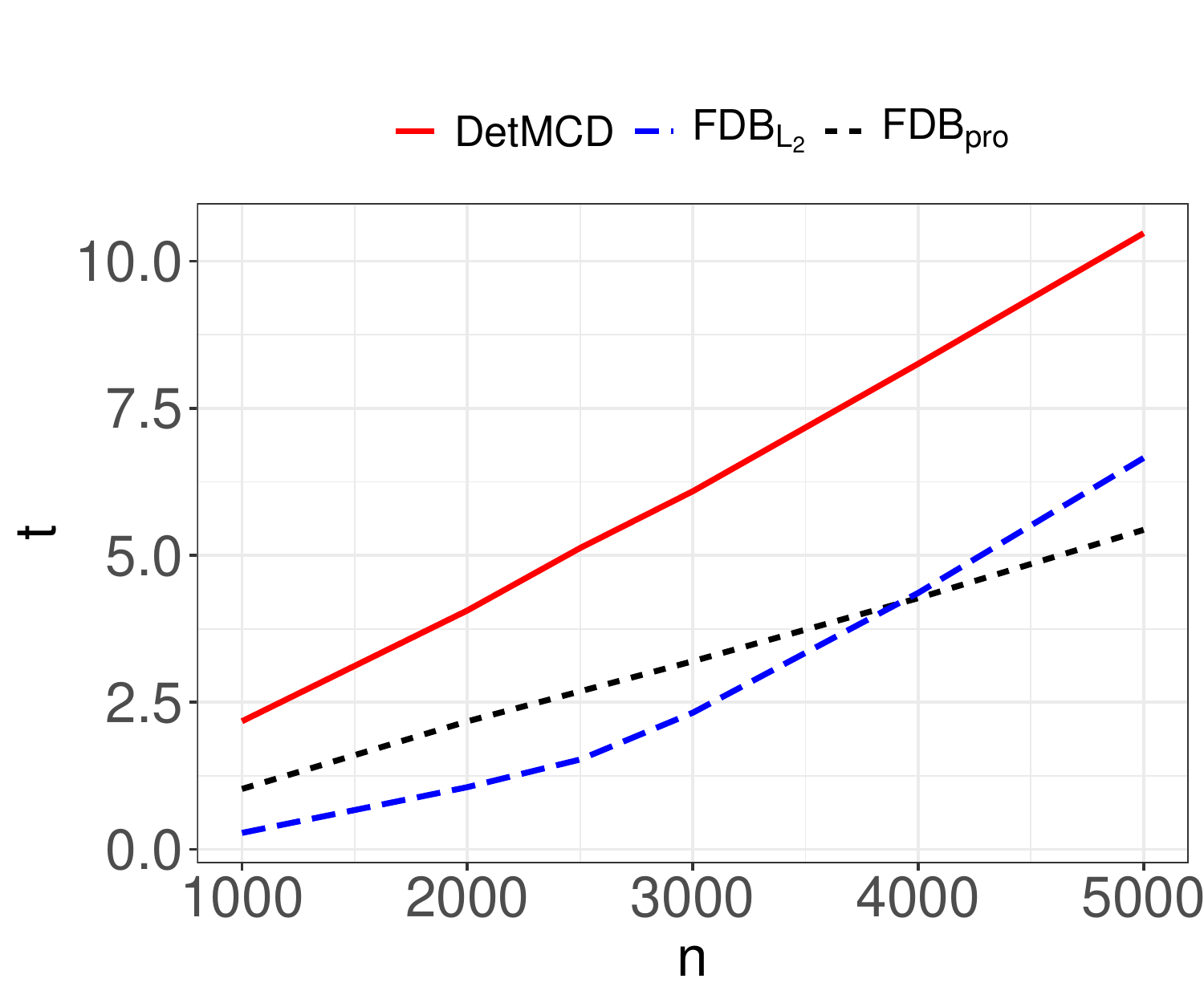}}
	\caption{The average computation time (seconds) for different settings over 20 replicates. (a): $\log(t)$ versus $\log(p)$ with $n=1000$; (b): $t$ versus $n$ with $p=200$.}\label{time1-2}
\end{figure}

The computational complexity for finding the $h$-subset by MCDs is $O(\psi(np^2+p^3))$. Specifically, for each C-step, it requires computing the covariance matrix and Mahalanobis distances, with complexities $O(np^2)$ and $O(p^3+np^2)$ respectively, and $\psi$ depends on the number of initial estimates and the times of C-step iteration. For FASTMCD, the number of initial estimates defaults to 500; for DetMCD, it defaults to 6; and for RT-DetMCD, it is further reduced to 2. However, these efforts only reduced the value of $\phi$ but the order term still remains the same. 

In contrast, the computational complexity of $\texttt{FDB}_{\rm pro}$ for finding the subset is $O(knp)$ with $k$ as the number of projection directions. According to our numerical experiments, The performance of $\texttt{FDB}_{\rm pro}$ is quite stable when the number of random directions is set around $1000$; see Figure S2 of the Supplement. Hence, $\texttt{FDB}_{\rm pro}$
leads to a significant improvement over the MCD estimators. We remark that it is possible to further reduce the number of projection directions according to some elaborate generative algorithms \citep{DYCKERHOFF2021107166}. However, these algorithms may instead lengthen the total computational time due to the tedious procedure for searching ``better'' directions, and hence we stick to selecting the directions randomly. For the case of ultra-high dimensional data, we suggest an adaptive rule, $k=\max (1000, 10p)$.
As for $\texttt{FDB}_{\rm L_2}$, the computational complexity is $O(n^2p)$, which scales linearly with the dimension of data.

To show the improvement, we provide some numerical results for computation time in Figure \ref{time1-2}. 
Notably, the speed of DetMCD is also influenced by the way of constructing the initial estimates. Specifically, $Q_n$ \citep{rousseeuw1993alternatives} is applied to construct initial estimates for DetMCD, which is computationally demanding. To improve speed, \cite{hubert2012deterministic} suggested substituting $Q_n$ with the $\tau$-scale of \cite{yohai1988high} when $n >1000$. We follow this suggestion by using the $Q_n$ estimator for DetMCD in Figure \ref{time1-2}(a), and the $\tau$-estimator in Figure \ref{time1-2}(b), respectively. 
For $\texttt{FDB}_{\rm pro}$, we let $k=1000$. All experiments are run using \texttt{R-package ddalpha} for DetMCD on an Intel(R) Xeon(R) with 3.10GHz and 192 GB memory processor.

\texttt{FDB}s show significant improvement over DetMCD under all the settings. Specifically, in Figure \ref{time1-2}(a), the line of DetMCD is steeper than those of the other two methods, which matches well with different orders of dimension $p$ in their theoretical computational complexities aforementioned. In Figure \ref{time1-2}(b), both DetMCD and $\texttt{FDB}_{\rm pro}$ reveal linear trends with the increasing sample size, while $\texttt{FDB}_{\rm L_2}$ shows a quadratic trend, though its computation time is the least when $n<4000$.

\section{Simulations}

We conduct extensive simulations with data from symmetric distributions to assess the performance of our proposed algorithms, $\texttt{FDB}_{\rm pro}$ and $\texttt{FDB}_{L_2}$, and make a comparison with DetMCD \citep{hubert2012deterministic}. Besides, we also provide some exploration for the scenarios of asymmetric distributions in Section S3.1 of the Supplement.
To evaluate the estimation results, we use the following five measures (the smaller the better). 
\begin{itemize}
	\item An error measure of the location estimator, given by $e_{\mu}=\|\hat{\boldsymbol{\mu}}-\boldsymbol{\mu_0}\|$, where $\boldsymbol{\mu_0}$ denotes the true sample mean.
	\item An error measure of the scatter estimator, defined as the logarithm of the condition number of $\hat {\mathbf{\Sigma}} \mathbf{\Sigma^{-1}}$, $e_{\Sigma}=\log _{10}(\operatorname{cond}(\hat{\mathbf{\Sigma}}\mathbf{\Sigma}^{-1})).$
		\item The mean squared error (MSE) of $\mathbf{\Sigma}$, ${\rm MSE}=\frac{1}{Sp^2}\sum_{s=1}^S||\hat{\mathbf{\Sigma}}-\mathbf{\Sigma}||_F^2.$	
	\item The Kullback Leibler (KL) divergence between $\hat{\mathbf{\Sigma}}$ and ${\mathbf{\Sigma}}$,
	${\rm KL}\left(\hat{\mathbf{\Sigma}},\mathbf{\Sigma} \right)=\operatorname{trace}\left(\hat{\mathbf{\Sigma}} \mathbf{\Sigma}^{-1}\right)-\log \left(\operatorname{det}\left(\hat{\mathbf{\Sigma}} \mathbf{\Sigma}^{-1}\right)\right)-p,$, which is identical the KL divergence between the two Gaussian distributions with the same mean.
	\item The computation time $t$ (in seconds) of the whole procedure, including the optimal subset pursuit and the reweighted step.
\end{itemize}

\subsection{Estimation performance}\label{sym}

In this subsection, we generate the bulk of non-outlying samples as $\boldsymbol{x}_{i}=\boldsymbol{G}\boldsymbol{y}_{i}$, where $\boldsymbol{y}_{i}$ are from $N_{p}(\mathbf{0},\mathbf{I})$ and $\boldsymbol{G}$ is a $p\times p$ matrix with unit diagonal elements and off-diagonal elements equal to 0.75. 
The number of outliers is $m=\lfloor n \varepsilon\rfloor$ and $\varepsilon$ denotes the level of contamination.
Four contamination types are considered: point, random, cluster, and radial outliers. 
{\bf Point outliers} are obtained by generating $\boldsymbol{y}_i \sim N_p\left(r \boldsymbol{a} \sqrt{p}, 0.01^2 \mathbf{I}\right)$, where $\boldsymbol{a} $ is a unit vector generated orthogonal to $\boldsymbol{a_{0}}=(1,1, \ldots, 1)^{T}$. 
{\bf Random outliers} are obtained by generating $\boldsymbol{y}_{i} \sim N_{p}\left( \boldsymbol{\mu}_{ir},  \mathbf{I}\right)$, where $\boldsymbol{\mu}_{ir}=rp^{1/4}\boldsymbol{\nu}/\|\boldsymbol{\nu}\|$ with $\boldsymbol{\nu}$ a random vector from $N_p(\boldsymbol{0}, \mathbf{I})$. 
{\bf Cluster outliers} are obtained by generating $\boldsymbol{y}_{i} \sim N_{p}\left(rp^{-1/4}\boldsymbol{a_0},  \mathbf{I}\right)$, where $r$ is a constant. 
{\bf Radial outliers} are obtained by generating $\boldsymbol{y}_{i} \sim N_{p}(\mathbf{0}, 5 \mathbf{I})$. 
Except for the random outliers, the other three types have been considered by~\cite{hubert2012deterministic}.

Different contamination levels are considered, namely $\varepsilon=0 \%, 10 \%$, and $40\%$. Let $h$ be $\lfloor0.75n\rfloor$ when $\varepsilon=0 \%$ or $10\%$, and $\lfloor0.5n\rfloor$ when $\varepsilon=40 \%$ for each method under investigation. 
The number of directions for projection depth is set as $k=\max (1000, 10p)$ as suggested in Section 4.3. For each method, we compute the reweighted location vectors $\hat{\boldsymbol{\mu}}_{\boldsymbol{X}}$ and the reweighted scatter matrices  $\hat{\boldsymbol{\Sigma}}_{\boldsymbol{X}}$. The corresponding
	estimators for the data set $\boldsymbol{Y}$ are obtained as $\hat{\boldsymbol{\mu}}_{\boldsymbol{Y}}=\boldsymbol{G}^{-1} \hat{\boldsymbol{\mu}}_{\boldsymbol{X}}$ and $\hat{\boldsymbol{\Sigma}}_{\boldsymbol{Y}}=\boldsymbol{G}^{-1} \hat{\mathbf{\Sigma}}_{\boldsymbol{X}} \boldsymbol{G}^{-1}$, which are compared to the true values using the aforementioned measures. In this part, the true covariance matrix of $\boldsymbol{Y}$ is ${\mathbf{\Sigma}}=\mathbf{I}_p$.

\begin{figure}[h!]
	\centering
	\subfigure[point]{\label{r_point_0.1}
		\centering
		\includegraphics[width=0.3\linewidth]{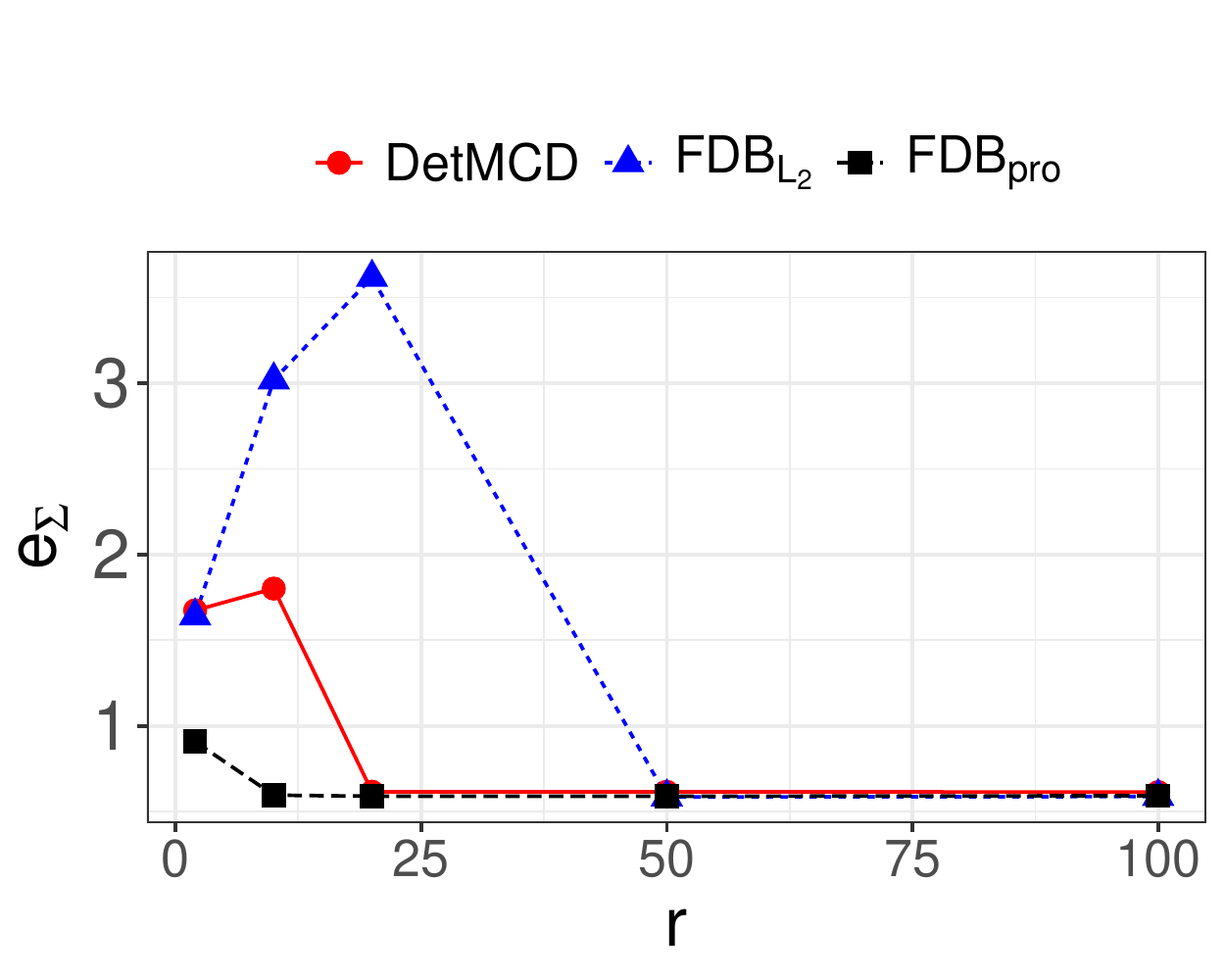}}
			\centering
		\subfigure[random]{\label{r_random_0.1}
		\centering
		\includegraphics[width=0.3\linewidth]{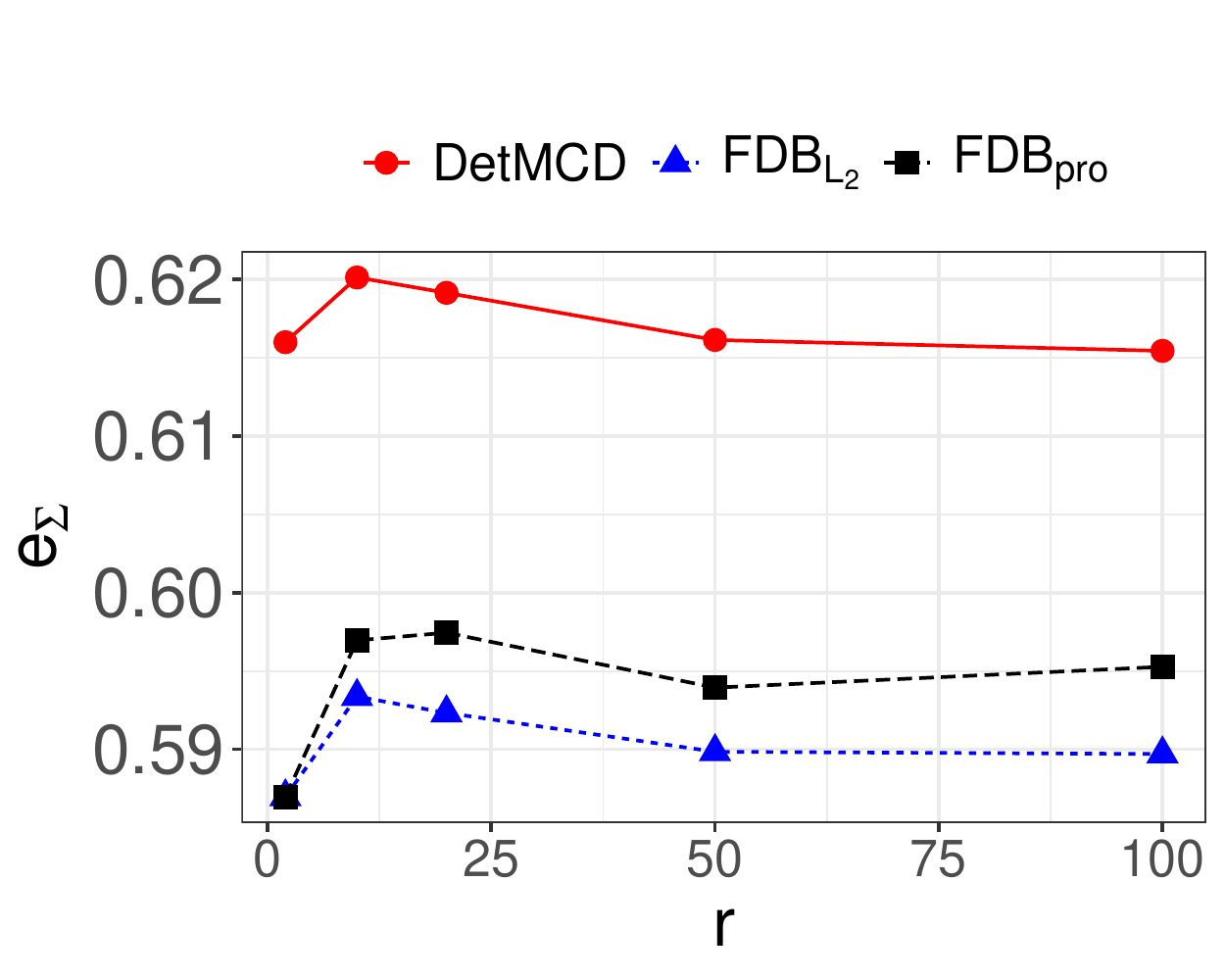}}
		\subfigure[cluster]{\label{r_clu_0.1}
		\centering
		\includegraphics[width=0.3\linewidth]{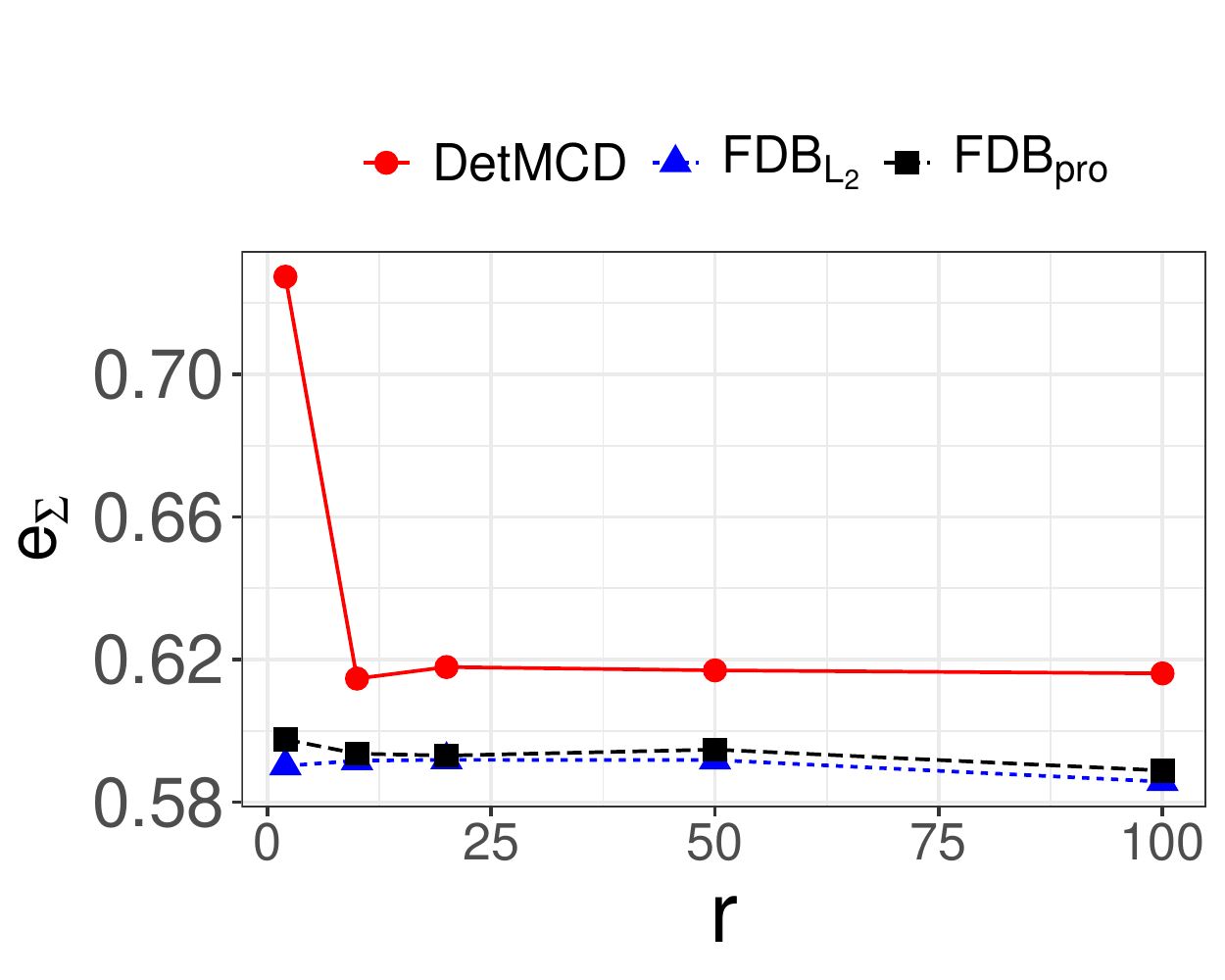}}
			\centering
		\subfigure[point]{\label{r_point-0.4}
		\includegraphics[width=0.3\linewidth]{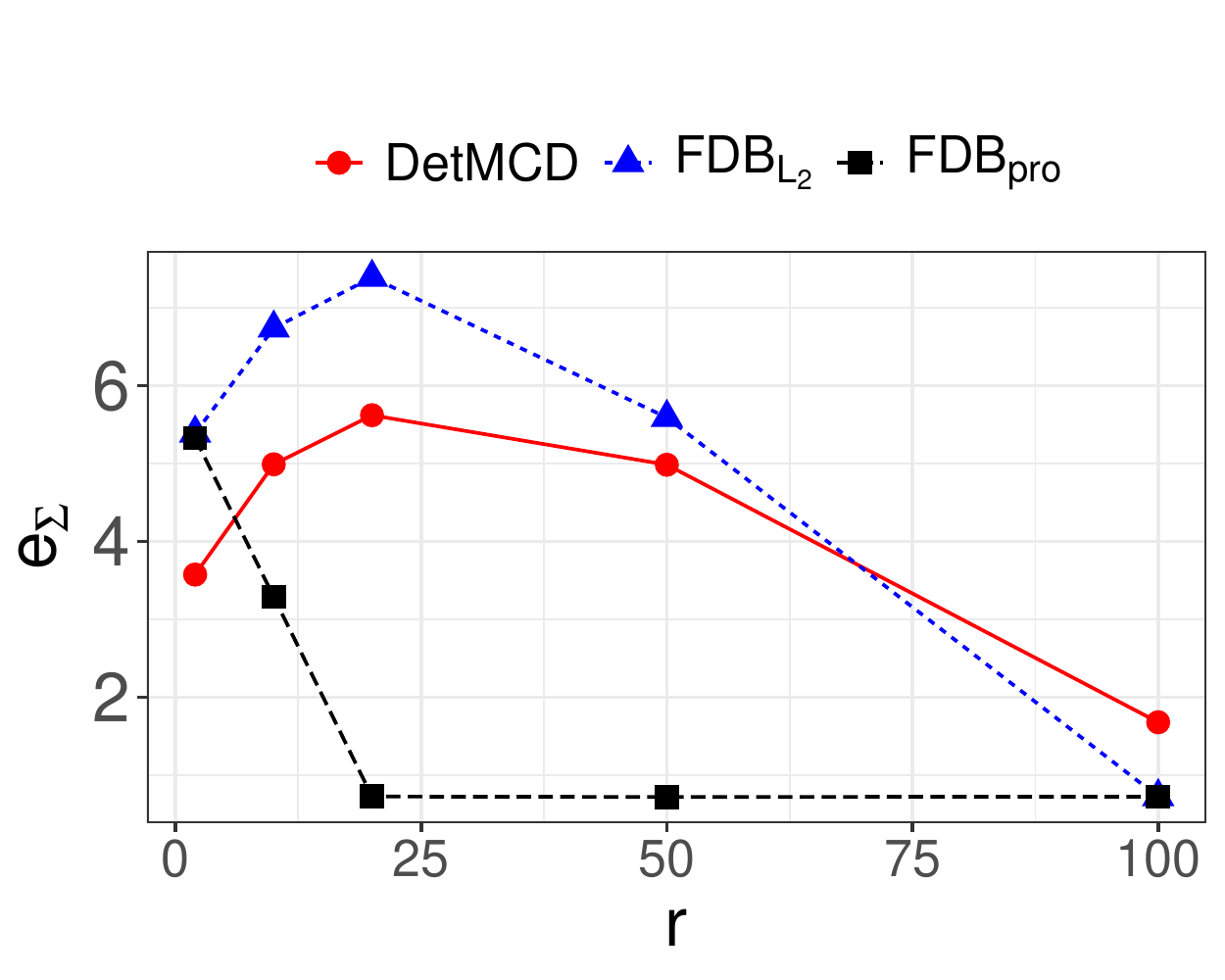}}
			\centering
		\subfigure[random]{\label{r_random_0.4}
		\centering
		\includegraphics[width=0.3\linewidth]{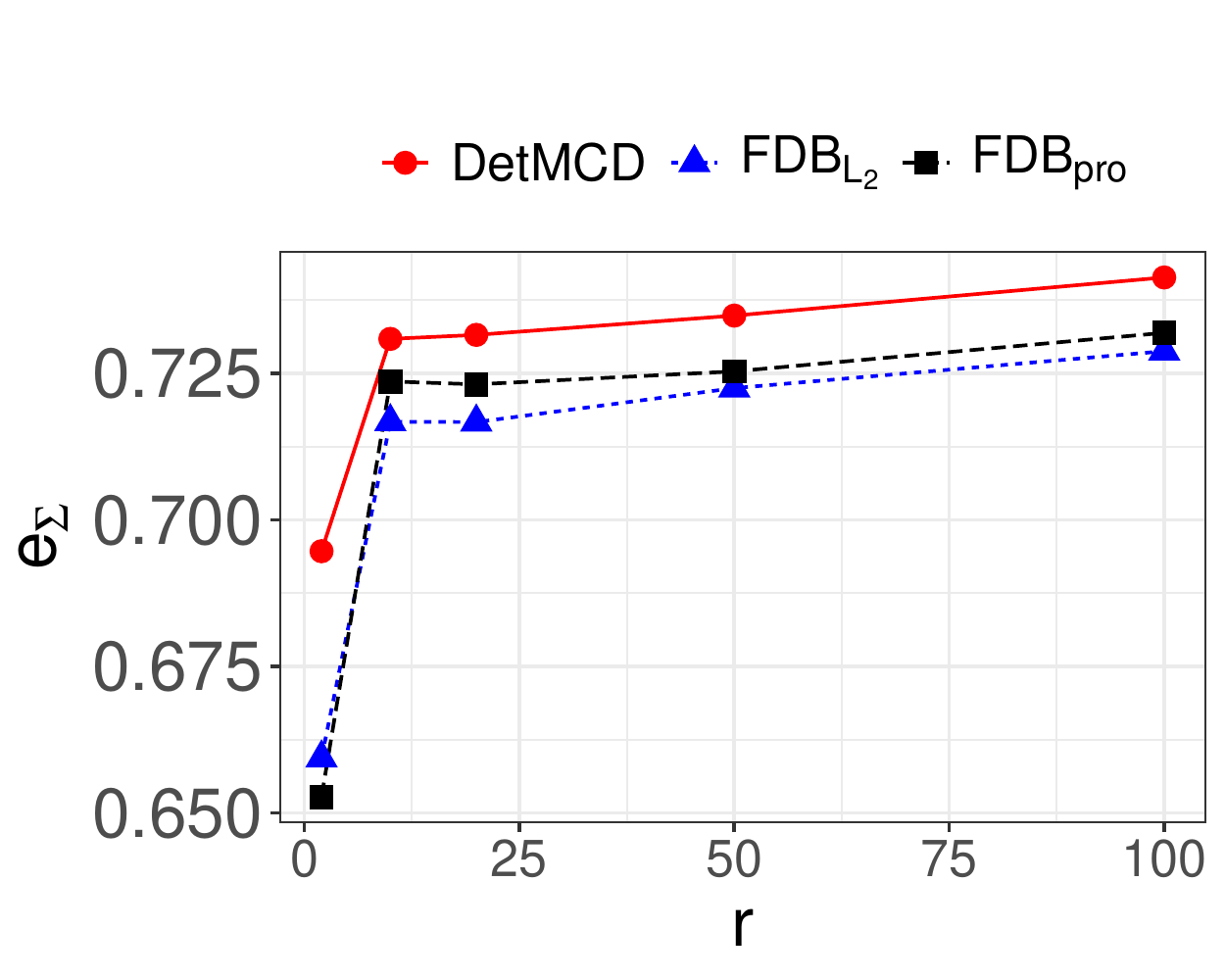}}
		\subfigure[cluster]{\label{r_clu_0.4}
		\centering
		\includegraphics[width=0.3\linewidth]{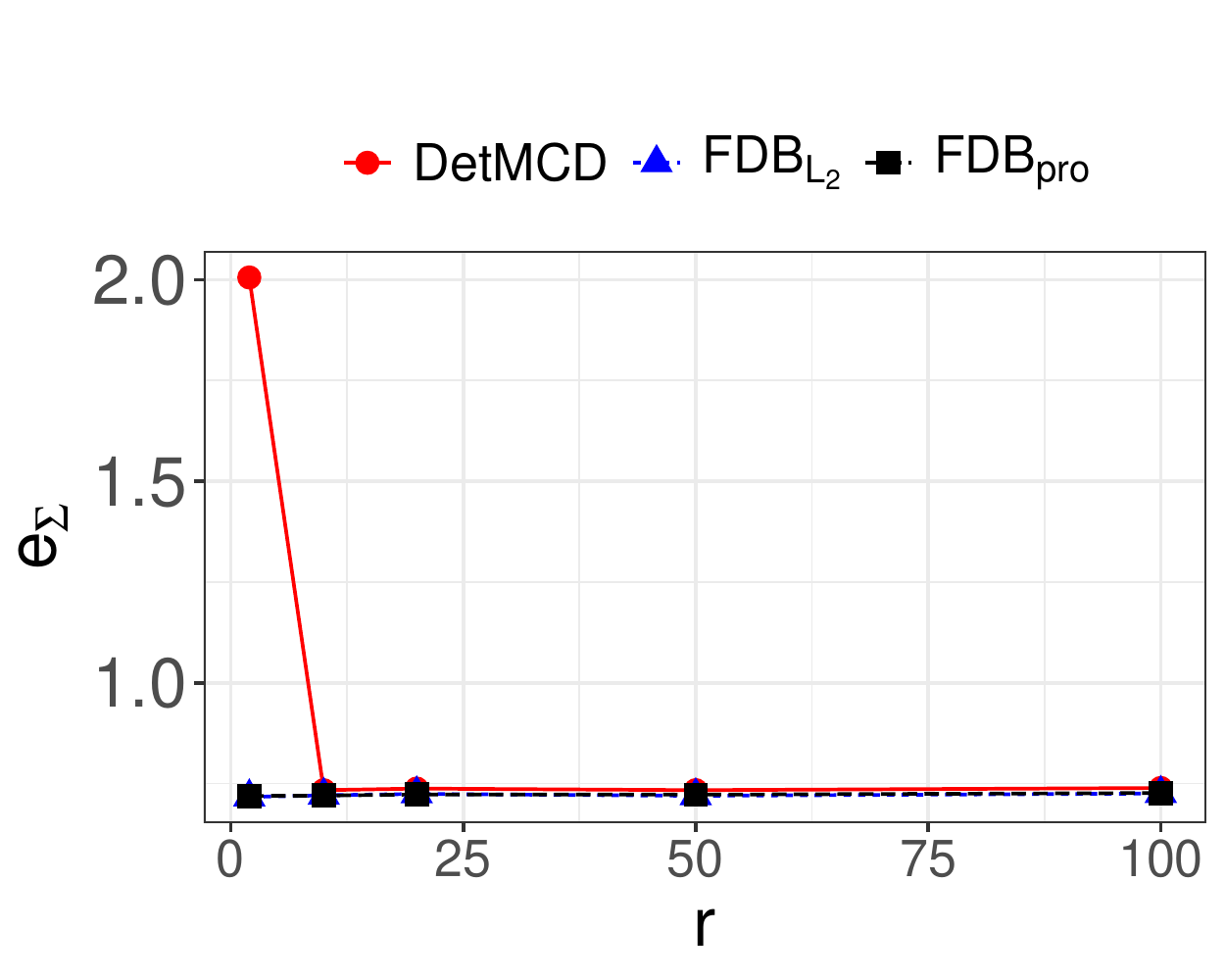}}
	\caption{The average $e_\Sigma$ for levels of abnormality ($r$), with $n=400$ and $p=40$, the first row is for $\varepsilon=0.1$ and $\alpha=0.75$, and the second row is for $\varepsilon=0.4$ and $\alpha=0.5$.} \label{change_with_r}
\end{figure}

\begin{figure}[h!]
	\centering
	\subfigure[point]{\label{point_0.1_r2}
		\centering
		\includegraphics[width=0.22\linewidth]{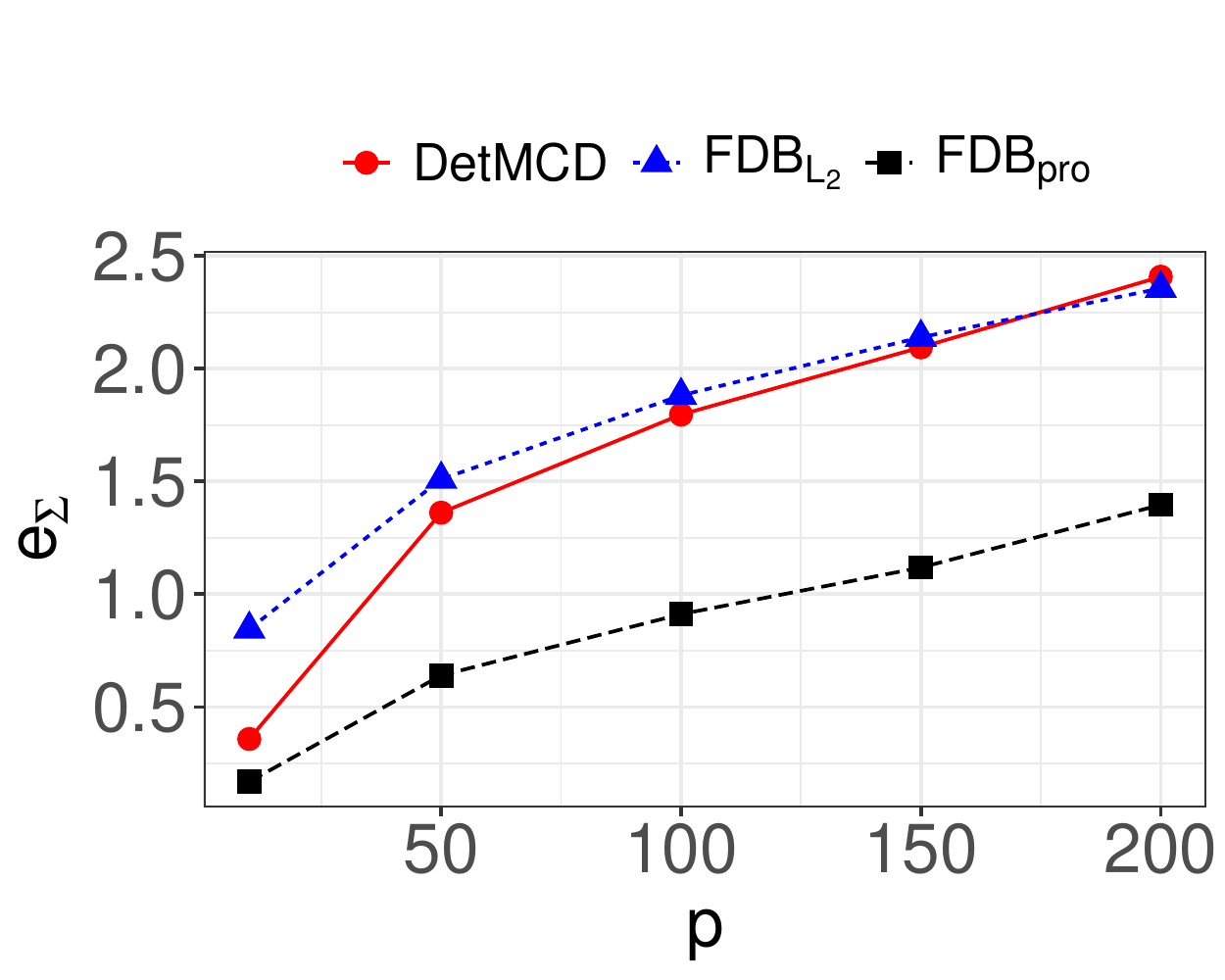}}
			\centering
		\subfigure[cluster]{\label{cluster_0.1_r2}
		\centering
		\includegraphics[width=0.22\linewidth]{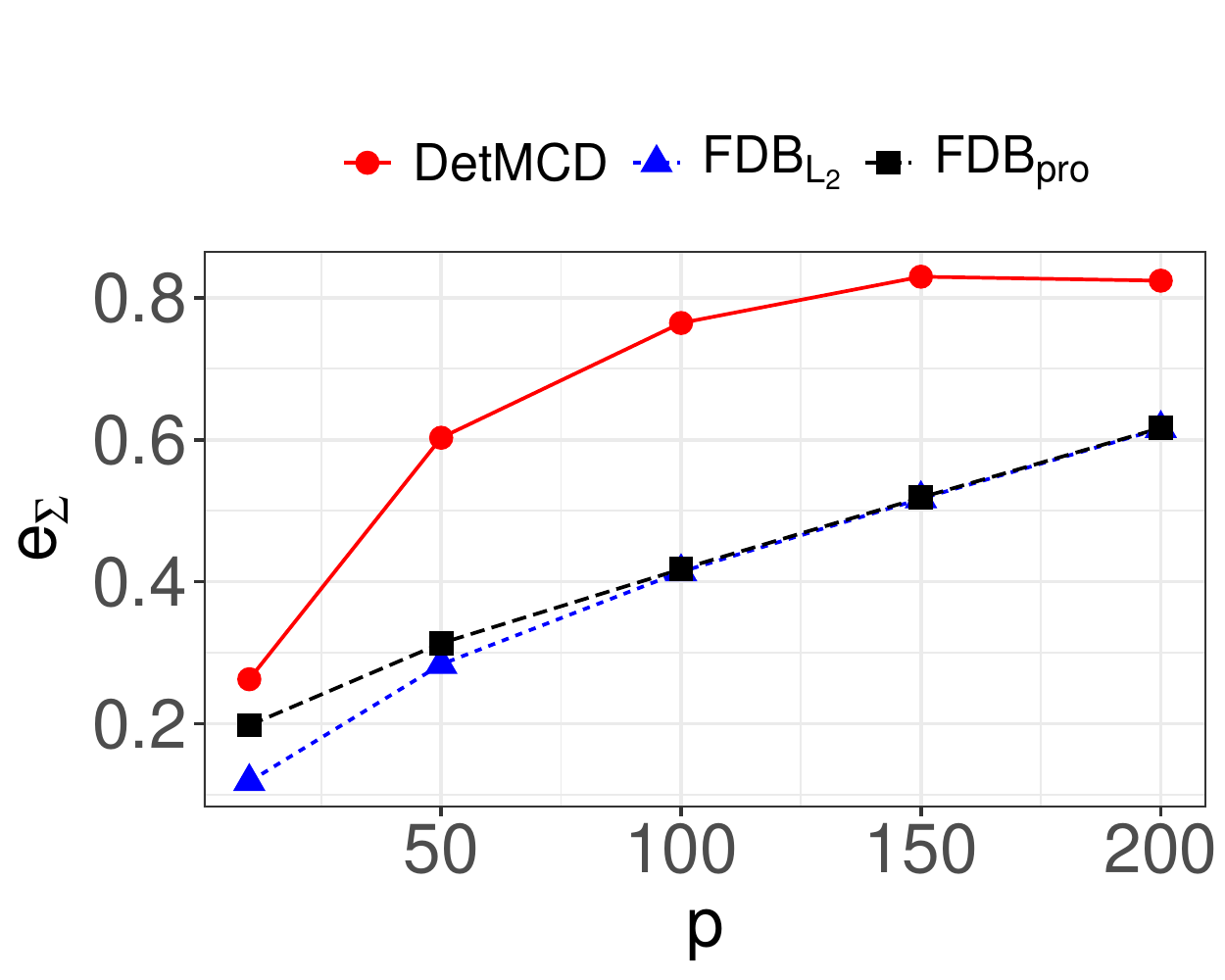}}
		\subfigure[point]{\label{point_0.1_r20}
		\centering
		\includegraphics[width=0.22\linewidth]{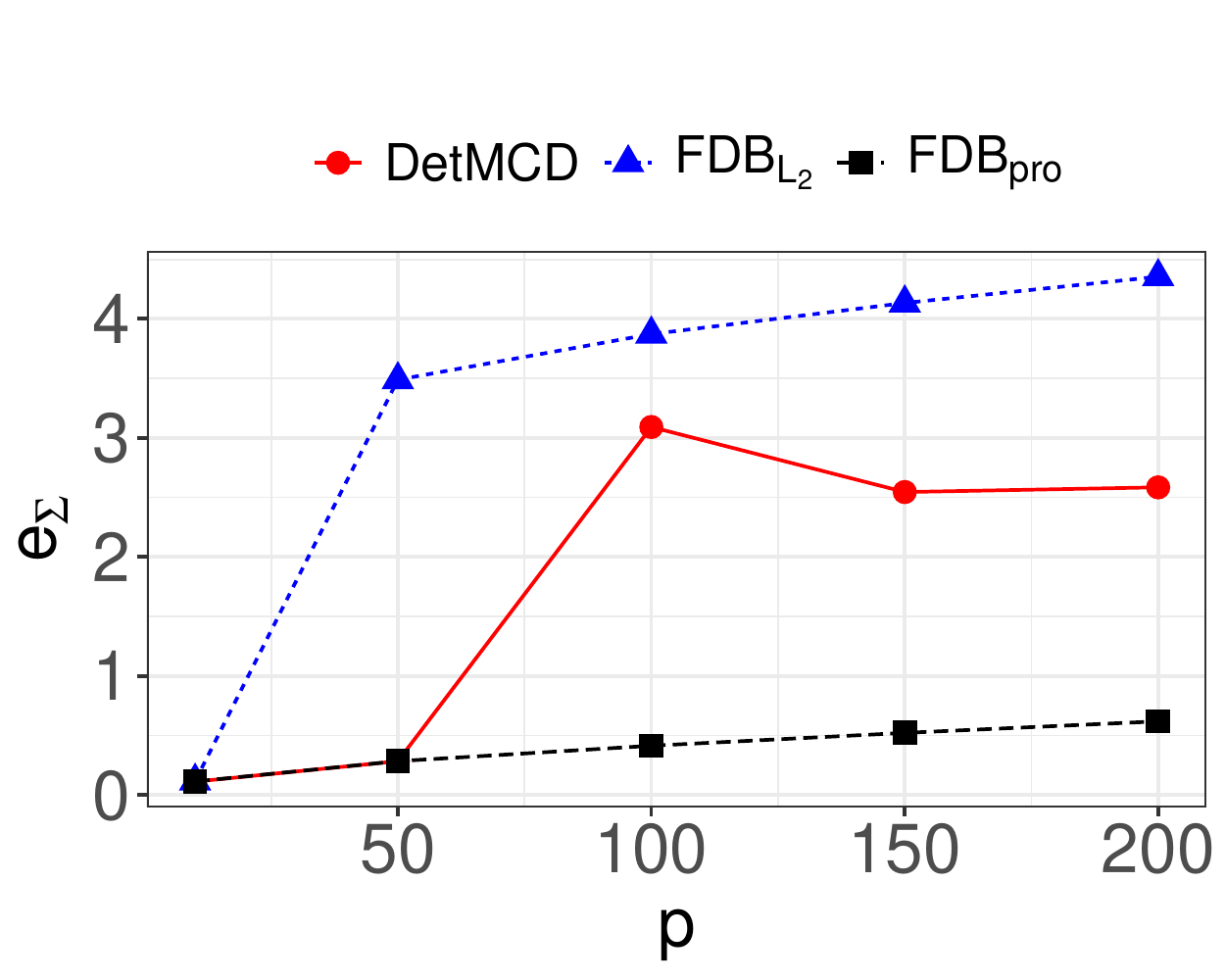}}
	\subfigure[cluster]{{\label{cluster_0.1_r20}}
		\centering
		\includegraphics[width=0.22\linewidth]{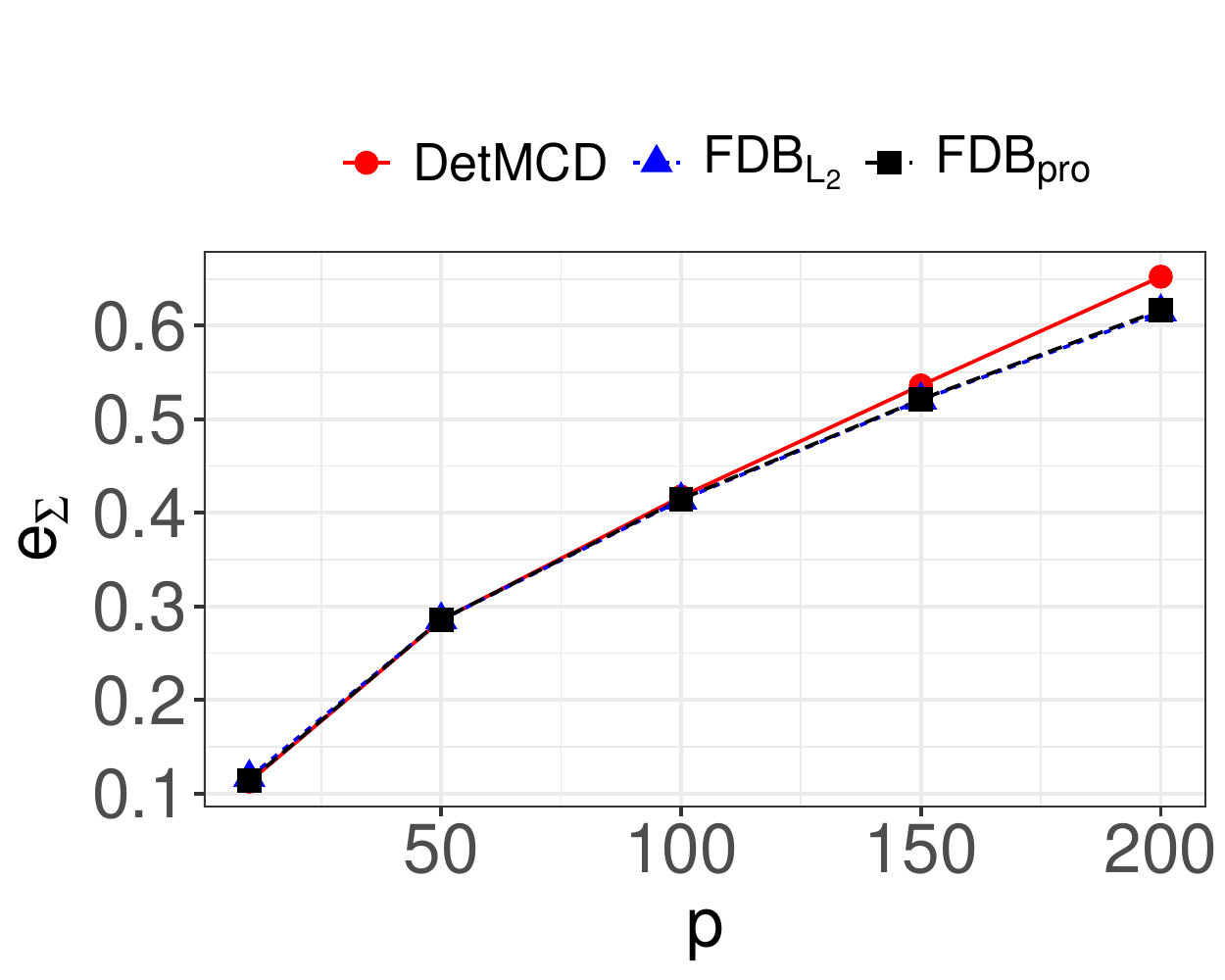}}
		\centering
	\subfigure[point]{\label{point_0.4_r2}
		\centering
		\includegraphics[width=0.22\linewidth]{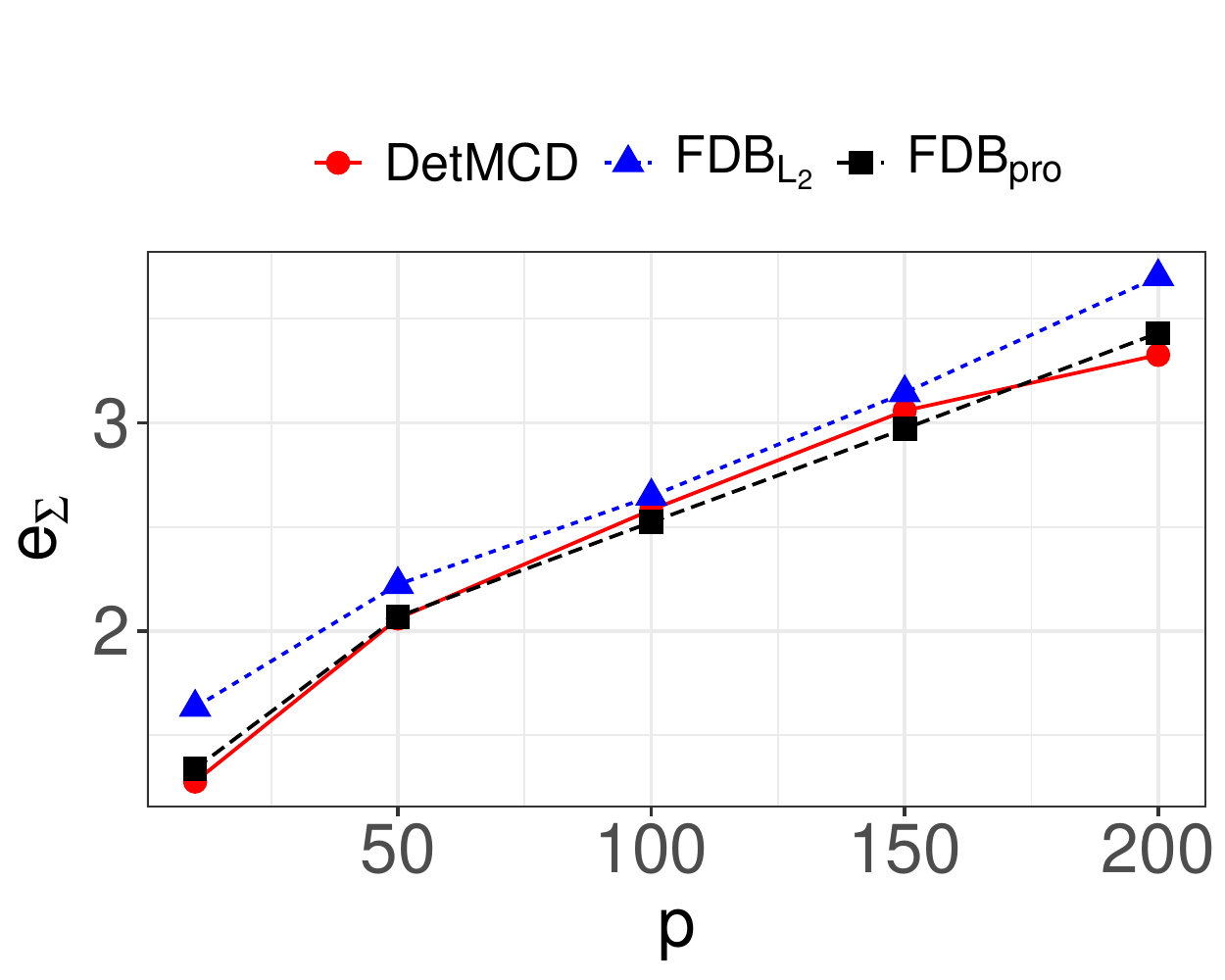}}
			\centering
		\subfigure[cluster]{\label{cluster_0.4_r2}
		\centering
		\includegraphics[width=0.22\linewidth]{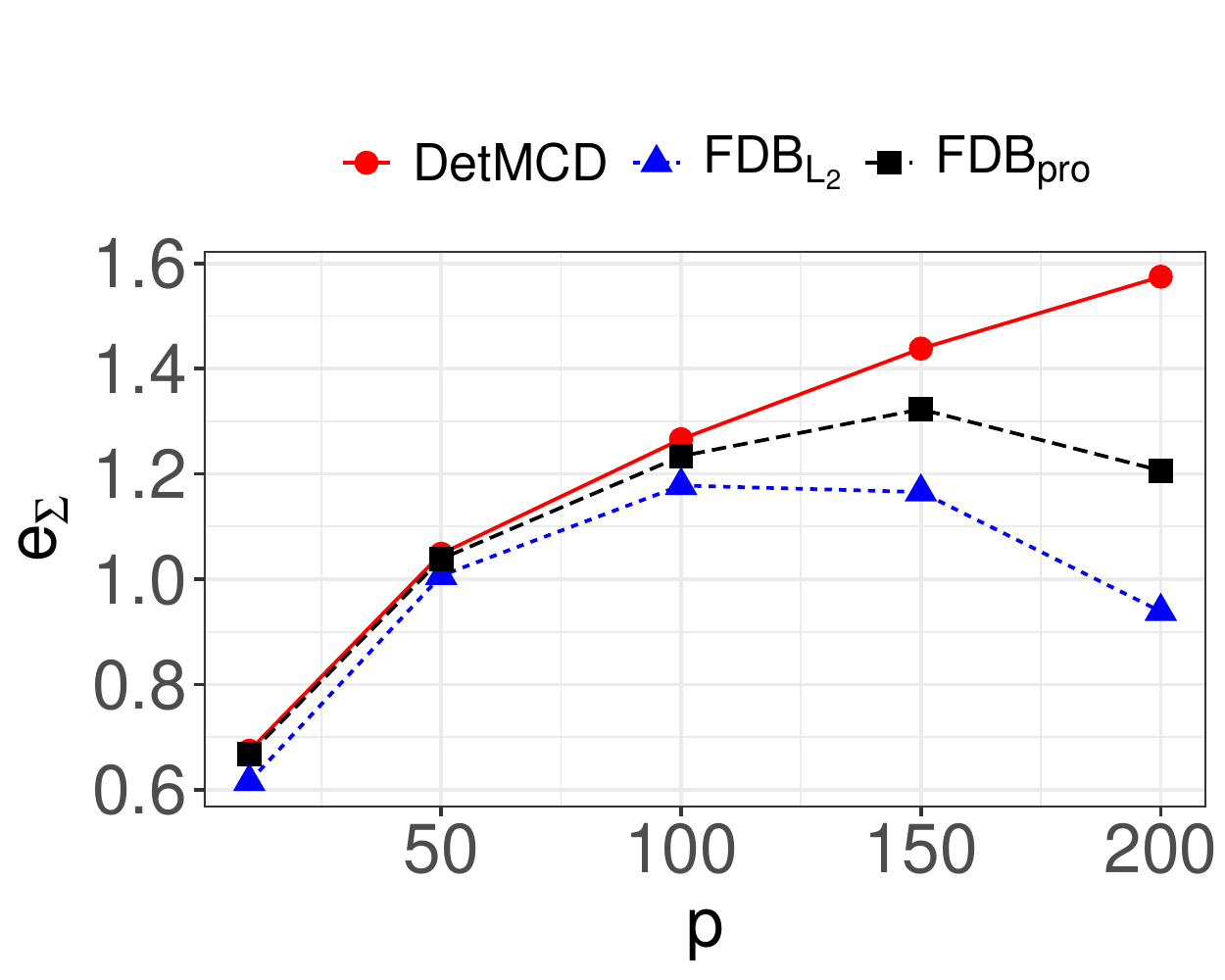}}
		\subfigure[point]{\label{point_0.4_r20}
		\centering
		\includegraphics[width=0.22\linewidth]{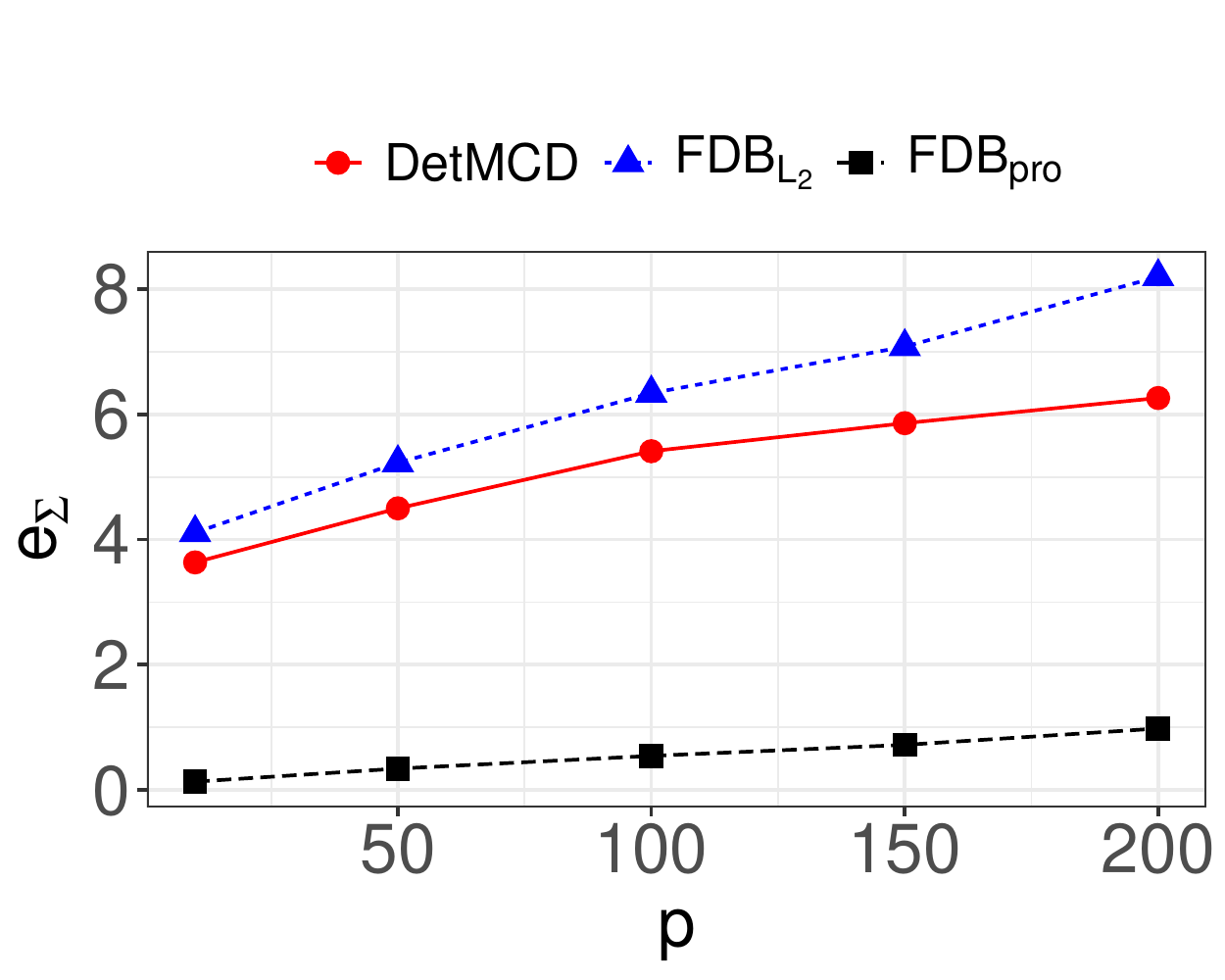}}
	\subfigure[cluster]{\label{cluster_0.4_r20}
		\centering
		\includegraphics[width=0.22\linewidth]{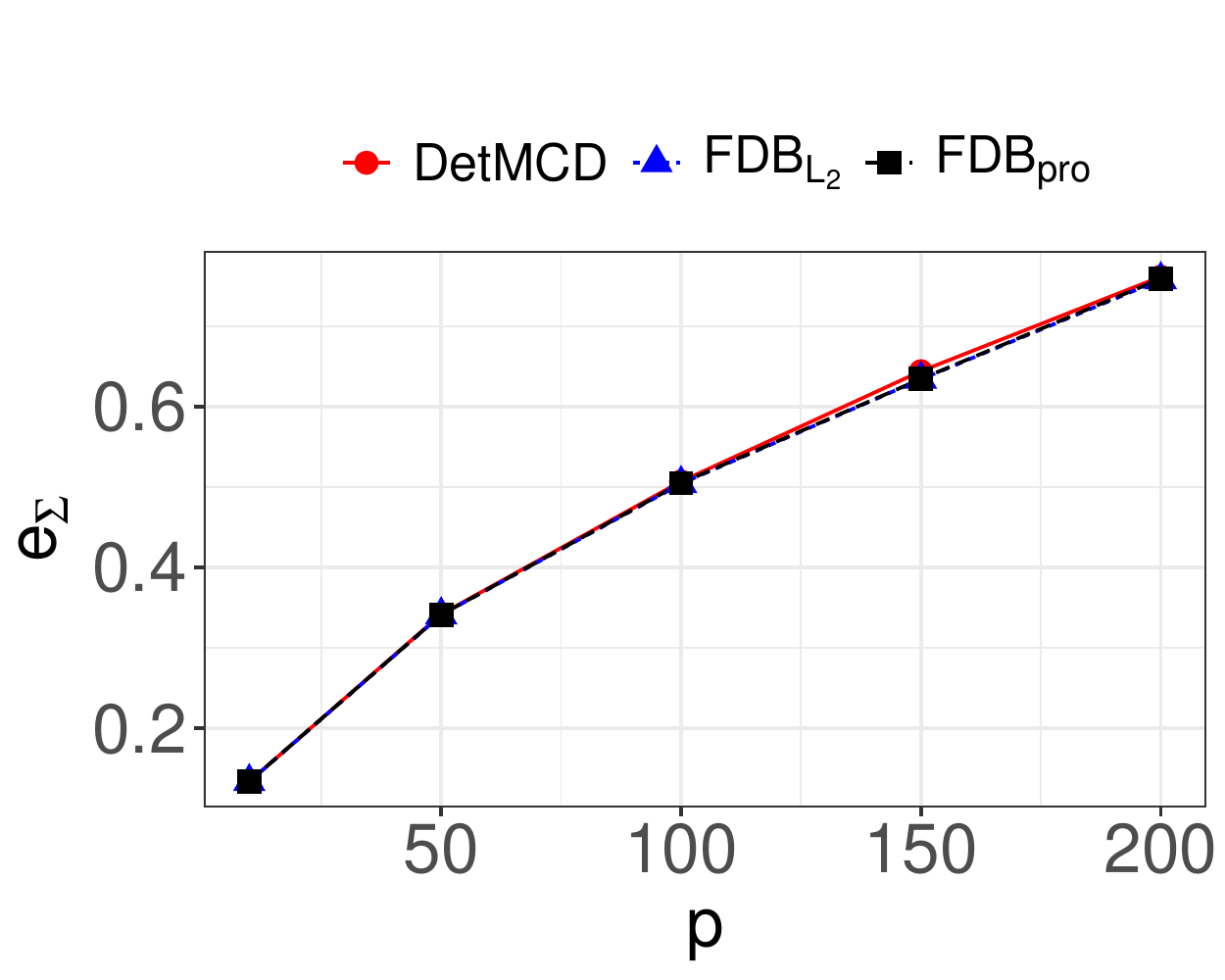}}
	\caption{The average $e_\Sigma$ for different $p$, with $n=2000$, the first row is for $\varepsilon=0.1$ and $\alpha=0.75$; the second row is for $\varepsilon=0.4$ and $\alpha=0.5$; the first two columns are for $r=2$, and the last two columns are for $r=20$, respectively. The cases of random contamination is omitted since all methods perform similarly.} \label{change with p}
\end{figure}

We first provide a full picture for the performance of each method by conducting simulation studies under a broad range of settings. To be more specific, we generated $1000$ data sets consisting of different types of contamination under a broad range of $r$ and $p$, and computed the average $e_{\Sigma}$ with the three methods. 
The results are illustrated in Figures \ref{change_with_r} and \ref{change with p}. 
Other measures, $e_{\mu}$, MSE, and KL divergence, all reveal similar patterns as $e_{\Sigma}$ and hence are omitted.
As shown in Figure \ref{change_with_r}, $\texttt{FDB}_{\rm pro}$ and $\texttt{FDB}_{\rm L_2}$ outperform DetMCD across all values of $r$ for both random and cluster outliers under either low or high contamination levels. For the case of point contamination, $\texttt{FDB}_{\rm pro}$ still holds the upper hand, with $\texttt{FDB}_{\rm L_2}$ slightly worse than DetMCD, especially for small $r$ values.

Figure \ref{change with p} shows that $\texttt{FDB}s$ is among the best estimators for all the settings. However, in general, DetMCD's performance deviates more seriously with the increase of dimension $p$, suggesting that it is less suitable for high-dimensional data.
Overall, $\texttt{FDB}_{\rm pro}$ provides the best performance among the three options, that both $\texttt{FDB}_{\rm L_2}$ and DetMCD produce large $e_{\Sigma}$ under point contamination since their induced subsets may contain outliers, which is consistent with the results in Figure \ref{s2}. 
Between $\texttt{FDB}_{\rm L_2}$ and DetMCD, the later is better for point contamination while the former is better (or even the best) for cluster contamination.

Next, we provide more detailed numerical outputs for typical simulation settings under study. Following \cite{hubert2012deterministic}, we consider three options, A: $n=200$ and $p=5$, B: $n=400$ and $p=40$, and C: $n=2000$ and $p=200$, representing low, moderate and high dimensions, respectively. Other settings remain the same as those in Figure \ref{change_with_r} except that $r$ is fixed at $5$ for point, random, and cluster outliers. 
We report the average measures over 1000 runs in Tables \ref{tab:cl}, \ref{tab:0.1} and \ref{tab:0.4}, corresponding to 0\% (clean data), 10\% and 40\% contamination, respectively. 
\begin{table}[!t]  
\renewcommand\arraystretch{1.5}
	\centering  
	\caption{Simulation results for clean data.}  
	\begin{tabular}{|c|c|c|c|c|c|c|}
	\cline{3-7}
		\multicolumn{2}{c|}{}&$e_{\mu}$&$e_{\Sigma}$&MSE&KL&t\\
		\hline
		\multirow{3}{*}{A} &DetMCD&0.158 (0.050) & 0.238 (0.051) &0.008 (0.003) & 0.119 (0.046)&0.022\\
		%\cline{2-6}
		&$ \texttt{FDB}_{\rm L_2}$&0.157 (0.051)& 0.245 (0.054) & 0.008 (0.003) & 0.122 (0.049) &0.003\\
			&$ \texttt{FDB}_{\rm pro}$&0.157 (0.050)& 0.232 (0.049) & 0.007 (0.003) & 0.112 (0.044) &0.011\\
		\hline
		\multirow{3}{*}{B} &DetMCD&0.344 (0.040) & 0.615 (0.029) &0.003 (0.000) & 2.771 (0.179)&0.388\\
		&$ \texttt{FDB}_{\rm L_2}$&0.329 (0.038)& 0.571 (0.025) & 0.003 (0.000) & 2.404 (0.135) &0.014\\
			&$ \texttt{FDB}_{\rm pro}$&0.331 (0.038)& 0.573 (0.025) & 0.003 (0.000) & 2.435 (0.137) &0.039\\
			\hline
			\multirow{3}{*}{C} &DetMCD&0.355 (0.016) & 0.656 (0.011) &6e-4 (0.000) & 14.202 (0.249)&3.876\\
		&$ \texttt{FDB}_{\rm L_2}$&0.331 (0.016)& 0.596 (0.008) & 5e-4 (0.000) & 11.796 (0.166) &1.043\\
			&$ \texttt{FDB}_{\rm pro}$&0.334 (0.016)& 0.598 (0.008) &5e-4 (0.000) & 11.876 (0.177) &1.109\\
				\hline
	\end{tabular}
	\label{tab:cl}
\end{table}

\begin{table}[t!]  
\renewcommand\arraystretch{1.3}
\setlength\tabcolsep{1.2pt}
	\centering  
	\caption{Simulation results for data with $\varepsilon=10\%$ and $r=5$.}  
%	\begin{adjustwidth}{-1cm}{-0.1cm}
	\begin{tabular}{|c|c|ccc|ccc|ccc|ccc|}
	\cline{3-14}
		\multicolumn{2}{c}{}&\multicolumn{3}{|c|}{Point}&\multicolumn{3}{c|}{Random}&\multicolumn{3}{c|}{Cluster}&\multicolumn{3}{c|}{Radial}\\
	\cline{3-14}
		\multicolumn{2}{c|}{} &Det & $\texttt{FDB}_{\rm L_2}$&$ \texttt{FDB}_{\rm pro}$ &Det & $\texttt{FDB}_{\rm L_2}$&$ \texttt{FDB}_{\rm pro}$ &Det & $\texttt{FDB}_{\rm L_2}$&$ \texttt{FDB}_{\rm pro}$ &Det & $\texttt{FDB}_{\rm L_2}$&$ \texttt{FDB}_{\rm pro}$\\
		\hline
		\multirow{8}{*}{\text{ A }} &\multirow{2}{*}{$e_{\mu}$}&0.166&1.188&0.165&0.164&0.166&0.164&0.163&0.163&0.163&0.165&0.171&0.163\\
		&&(0.054)&(0.071)&(0.054)&(0.054)&(0.056)&(0.054)&(0.053)&(0.053)&(0.053)&(0.051)&(0.053)&(0.051)\\
		\cline{2-14}
		&\multirow{2}{*}{$e_{\Sigma}$}&0.236&1.333&0.236&0.228&0.236&0.230&0.228 &0.227&0.229&0.233 &0.251&0.233\\
		&&(0.052)&(0.070)&(0.052)&(0.045)&(0.047)&(0.046)&(0.045) &(0.045)&(0.045)&(0.049) &(0.054)&(0.048)\\
		\cline{2-14}
		&\multirow{2}{*}{MSE}&0.008&5.526&0.008&0.008&0.008&0.008 &0.007 & 0.007&0.007&0.008 & 0.009&0.008\\
		&&(0.003)&(0.268)&(0.003)&(0.003)&(0.003)&(0.003) &(0.03) &(0.003)&(0.003)&(0.003) & (0.003)&(0.003)\\
		\cline{2-14}
		&\multirow{2}{*}{KL}&0.107&9.375&0.112&0.101&0.112&0.105&0.101&0.101&0.102&0.103&0.117&0.104\\
		&&(0.041)&(0.314)&(0.043)&(0.038)&(0.042)&(0.039)&(0.038)&(0.038)&(0.038)&(0.038)&(0.046)&(0.037)\\
		\hline
		\multirow{8}{*}{B} &\multirow{2}{*}{$e_{\mu}$}&2.403&3.456&0.339&0.347&0.339 &0.340&0.349&0.340&0.341&0.347&0.337&0.338\\
		&&(1.597)&(0.67)&(0.037)&(0.040)&(0.038) &(0.039)&(0.037)&(0.036)&(0.038)&(0.039)&(0.037)&(0.037)\\
		\cline{2-14}
		&\multirow{2}{*}{$e_{\Sigma}$}&1.787&2.420&0.591&0.622&0.595 &0.596&0.616&0.588&0.592&0.616&0.593&0.594\\
		&&(0.911)&(0.028)&(0.026)&(0.031)&(0.026) &(0.027)&(0.026)&(0.025)&(0.024)&(0.027)&(0.026)&(0.026)\\
		\cline{2-14}
		&\multirow{2}{*}{MSE}&4.049&5.912&0.003&0.003&0.003 &0.003&0.003&0.003&0.003&0.003&0.003&0.003\\
		&&(3.145)&(0.147)&(0.000)&(0.000)&(0.000) &(0.000)&(0.000)&(0.000)&(0.000)&(0.000)&(0.000)&(0.000)\\
		\cline{2-14}
		&\multirow{2}{*}{KL}&64.051&95.728&2.591&2.789&2.591 &2.597&2.781&2.558&2.591&2.801&2.574&2.587\\
		&&(47.593)&(1.304)&(0.131)&(0.172)&(0.141) &(0.138)&(0.168)&(0.144)&(0.140)&(0.162)&(0.147)&(0.142)\\
		\hline
		\multirow{8}{*}{C} &\multirow{2}{*}{$e_{\mu}$}&7.824&8.056&0.348&0.355&0.340 &0.341&0.359&0.344&0.346&0.362&0.345&0.347\\
		&&(2.507)&(0.082)&(0.018)&(0.015)&(0.015) &(0.015)&(0.017)&(0.016)&(0.017)&(0.018)&(0.018)&(0.018)\\
		\cline{2-14}
		&\multirow{2}{*}{$e_{\Sigma}$}&2.965&3.151&0.618&0.654&0.617 &0.618&0.654&0.617&0.618&0.653&0.613&0.616\\
		&&(0.773)&(0.012)&(0.009)&(0.010)&(0.009) &(0.009)&(0.010)&(0.008)&(0.008)&(0.012)&(0.010)&(0.009)\\
		\cline{2-14}
		&\multirow{2}{*}{MSE}&6.488&6.366&6e-04&6e-04&6e-04 &6e-04&6e-04&6e-04&6e-04&e-04&e-04&e-04\\
		&&(2.185)&(0.103)&(8e-06)&(7e-06)&(7e-06)&(7e-06) &(8e-06)&(8e-06)&(8e-06)&(8e-06)&(8e-06)&(8e-06)\\
		\cline{2-14}
		&\multirow{2}{*}{KL}&495.2&513.9&12.61&14.08&12.648&12.92&14.02&12.57&12.65&14.06&12.58&12.65\\
		&&(161.4)&(4.320)&(0.174)&(0.167)&(0.151) &(0.150)&(0.195)&(0.141)&(0.142)&0.174&0.156&0.142\\
		\hline
	\end{tabular}
%	\end{adjustwidth}
	\label{tab:0.1}
\end{table}

\begin{table}[h!]  
\renewcommand\arraystretch{1.3}
\setlength\tabcolsep{1.2pt}
	\centering  
	\caption{Simulation results for data with $\varepsilon=40\%$ and $r=5$.}  
%	\begin{adjustwidth}{-1cm}{-0.1cm}
	\begin{tabular}{|c|c|ccc|ccc|ccc|ccc|}
	\cline{3-14}
		\multicolumn{2}{c}{}&\multicolumn{3}{|c|}{Point}&\multicolumn{3}{c|}{Random}&\multicolumn{3}{c|}{Cluster}&\multicolumn{3}{c|}{Radial}\\
	\cline{3-14}
		\multicolumn{2}{c|}{} &Det & $\texttt{FDB}_{\rm L_2}$&$ \texttt{FDB}_{\rm pro}$ &Det & $\texttt{FDB}_{\rm L_2}$&$ \texttt{FDB}_{\rm pro}$ &Det & $\texttt{FDB}_{\rm L_2}$&$ \texttt{FDB}_{\rm pro}$ &Det & $\texttt{FDB}_{\rm L_2}$&$ \texttt{FDB}_{\rm pro}$\\
		\hline
		\multirow{8}{*}{\text{ A }} &\multirow{2}{*}{$e_{\mu}$}&7.226&6.946&6.335&0.193&0.280 &0.191&0.604&0.179&0.240&0.216&0.233&0.215\\
		&&(0.982)&(0.354)&(0.376)&(0.066)&(0.100) &(0.066)&(0.215)&(0.061)&(0.106)&(0.072)&(0.071)&(0.072)\\
		\cline{2-14}
		&\multirow{2}{*}{$e_{\Sigma}$}&2.989&2.981&2.563&0.279&0.623 &0.278&0.421&0.263&0.284&0.299&0.447&0.291\\
		&&(0.290)&(0.208)&(0.184)&(0.062)&(0.126) &(0.062)&(0.386)&(0.058)&(0.170)&(0.058)&(0.081)&(0.053)\\
		\cline{2-14}
		&\multirow{2}{*}{MSE}&31.34&33.45&36.43&0.011&0.474 &0.011&1.037&0.009&0.071&0.041&0.120&0.045\\
		&&(3.592)&(2.941)&(2.329)&(0.006)&(0.360) &(0.005)&(2.017)&(0.004)&(0.111)&(0.019)&(0.054)&(0.020)\\
		\cline{2-14}
		&\multirow{2}{*}{KL}&29.66&29.46&29.91&0.142&2.363 &0.141&1.673&0.130&0.349&0.365&0.963&0.400\\
		&&(0.918)&(0.894)&(0.795)&(0.060)&(1.453) &(0.058)&(3.523)&(0.053)&(0.483)&(0.142)&(0.324)&(0.152)\\
		\hline
		\multirow{8}{*}{B} &\multirow{2}{*}{$e_{\mu}$}&22.98&25.30&25.10&0.408&0.408 &0.407&4.796&0.411&0.409&0.412&0.413&0.410\\
		&&(0.079)&(0.041)&(1.138)&(0.047)&(0.048) &(0.048)&(0.371)&(0.045)&(0.045)&(0.044)&(0.046)&(0.045)\\
		\cline{2-14}
		&\multirow{2}{*}{$e_{\Sigma}$}&4.373&6.161&6.003&0.732&0.729 &0.724&2.000&0.718&0.720&0.737&0.739&0.726\\
		&&(0.112)&(0.151)&(0.470)&(0.033)&(0.036) &(0.035)&(0.093)&(
		0.031)&(0.032)&(0.037)&(0.036)&(0.033)\\
		\cline{2-14}
		&\multirow{2}{*}{MSE}&24.51&15.98&16.55&0.004&0.004 &0.004&0.875&0.004&0.004&0.004&0.004&0.004\\
		&&(0.568)&(0.366)&(3.196)&(0.000)&(0.000) &(0.000)&(0.074)&(0.000)&(0.000)&(0.000)&(0.000)&(0.000)\\
		\cline{2-14}
		&\multirow{2}{*}{KL}&242.1&229.1&231.1&3.868&3.815 &3.796&36.654&3.731&3.747&3.877&3.890&3.799\\
		&&(2.489)&(2.716)&(5.105)&(0.217)&(0.217) &(0.216)&(2.466)&(0.201)&(0.210)&(0.218)&(0.236)&(0.204)\\
		\hline
		\multirow{8}{*}{C} &\multirow{2}{*}{$e_{\mu}$}&51.43&56.56&55.64&0.420&0.409 &0.413&6.345&0.418&0.417&0.421&0.410&0.415\\
		&&(0.016)&(0.012)&(3.242)&(0.020)&(0.020) &(0.021)&(0.161)&(0.019)&(0.020)&(0.021)&(0.020)&(0.021)\\
		\cline{2-14}
		&\multirow{2}{*}{$e_{\Sigma}$}&5.061&6.996&6.756&0.769&0.745 &0.753&2.317&0.757&0.758&0.769&0.747&0.760\\
		&&(0.039)&(0.027)&(0.674)&(0.012)&(0.011) &(0.011)&(0.009)&(0.012)&(0.011)&(0.011)&(0.009)&(0.011)\\
		\cline{2-14}
		&\multirow{2}{*}{MSE}&24.57&16.01&17.33&9e-04&8e-04 &4e-04&0.155&8e-04&8e-04&9e-04&8e-04&8e-04\\
		&&(0.085)&(0.055)&(4.426)&(1e-05)&(1e-05) &(1e-05)&(0.004)&(1e-05)&(1e-05)&(1e-05)&(9e-06)&(1e-05)\\
		\cline{2-14}
		&\multirow{2}{*}{KL}&1204&1151&1140&19.20& 18.09&18.45&88.81&18.63&18.66&19.14&18.12&18.70\\
		&&(2.094)&(2.722)&(31.32)&(0.238)&(0.199) &(0.224)&(0.975)&(0.233)&(0.237)&(0.240)&(0.198)&(0.222)\\
		\hline
	\end{tabular}
%	\end{adjustwidth}
	\label{tab:0.4}
\end{table}

For clean data (Table \ref{tab:cl}), the three estimators are comparable for low-dimensional cases; $\texttt{FDB}$s achieve slightly smaller values for $e_{\mu}$, $e_{\Sigma}$ and KL when the data dimension is moderate or high. More importantly, the running time of DetMCD is reduced in all settings, with the relative computational efficiency, defined as $t_{\rm MCD}/t_{\rm FDB}$, ranging between 2 and 10 for $\texttt{FDB}_{\rm pro}$, and between 3 and 27 for $\texttt{FDB}_{\rm L_2}$. Such a comparison of computation time holds similarly when contamination presents, and hence is omitted in the remaining tables.

When the amount of contamination is 10\% (Table \ref{tab:0.1}), the performances of three methods are all satisfying under settings with random, cluster or radial contamination, and the comparison is similar to that for clean data. 
For point contamination, $\texttt{FDB}_{\rm L_2}$ performs the worse for each of the three settings, that it generates the largest values for the four measures of estimation accuracy; DetMCD gets problematic for moderate and high dimensional data (settings B and C), which indicates its deficiency for such cases; in contrast, $\texttt{FDB}_{\rm pro}$ remains robust under all three settings and provides values of measures quite close to corresponding ones for the clean data in Table \ref{tab:cl}.

When the amount of contamination increases to 40\% (Table \ref{tab:0.4}), the three methods all work well and are comparable under settings with random or radial contamination. However, none of them is satisfactory when the contamination presents as point outliers, and this weak performance was also observed for both DetMCD and FASTMCD under similar settings in \cite{hubert2012deterministic}. 
For cluster contamination, DetMCD leads to larger estimation errors in each cell, especially when the dimension of data is moderate or high; however, both $\texttt{FDB}_{\rm L_2}$ and $\texttt{FDB}_{\rm pro}$ instead remains very stable across all three settings. 
Additional results for $r=2$ and $20$ are provided in Tables S1--S4 of the Supplementary Material, from which we may draw similar conclusions for the comparison of the three methods.

\begin{figure}[t!]
	\centering
	\subfigure[point]{\label{hpo}
		\centering
		\includegraphics[width=0.4\linewidth]{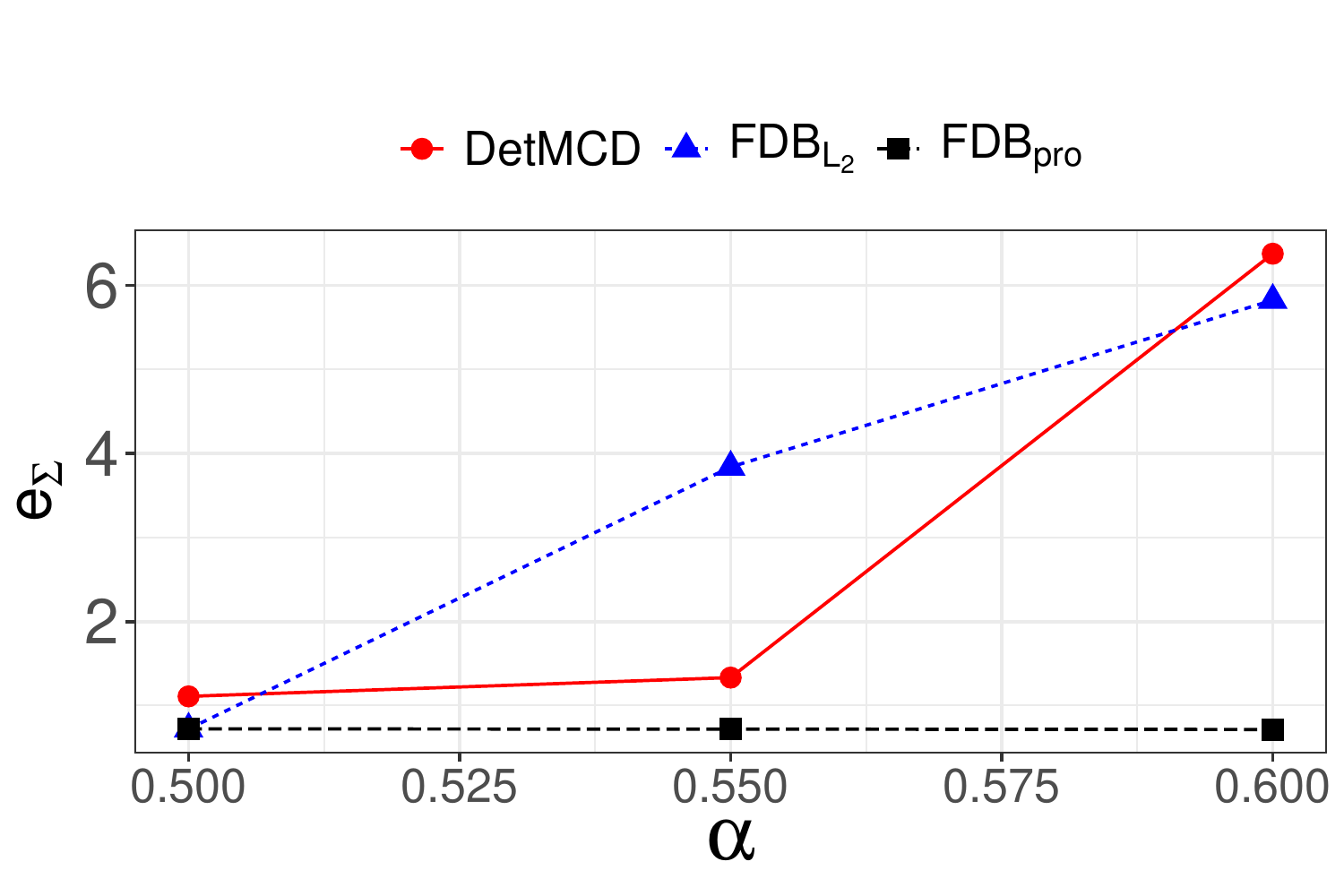}}
			\centering
		\subfigure[random]{\label{hra}
		\centering
		\includegraphics[width=0.4\linewidth]{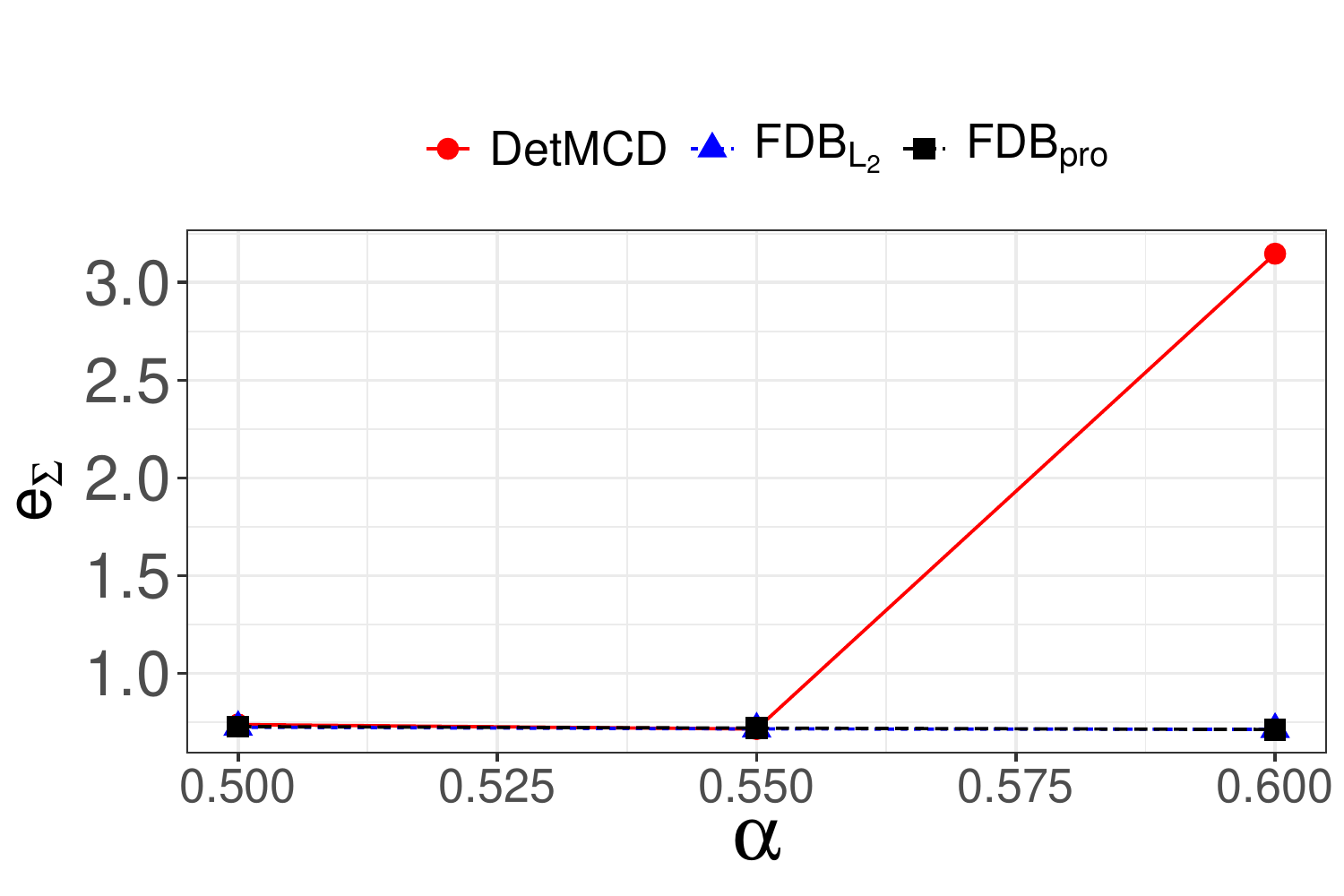}}
		\subfigure[cluster]{\label{hcl}
		\centering
		\includegraphics[width=0.4\linewidth]{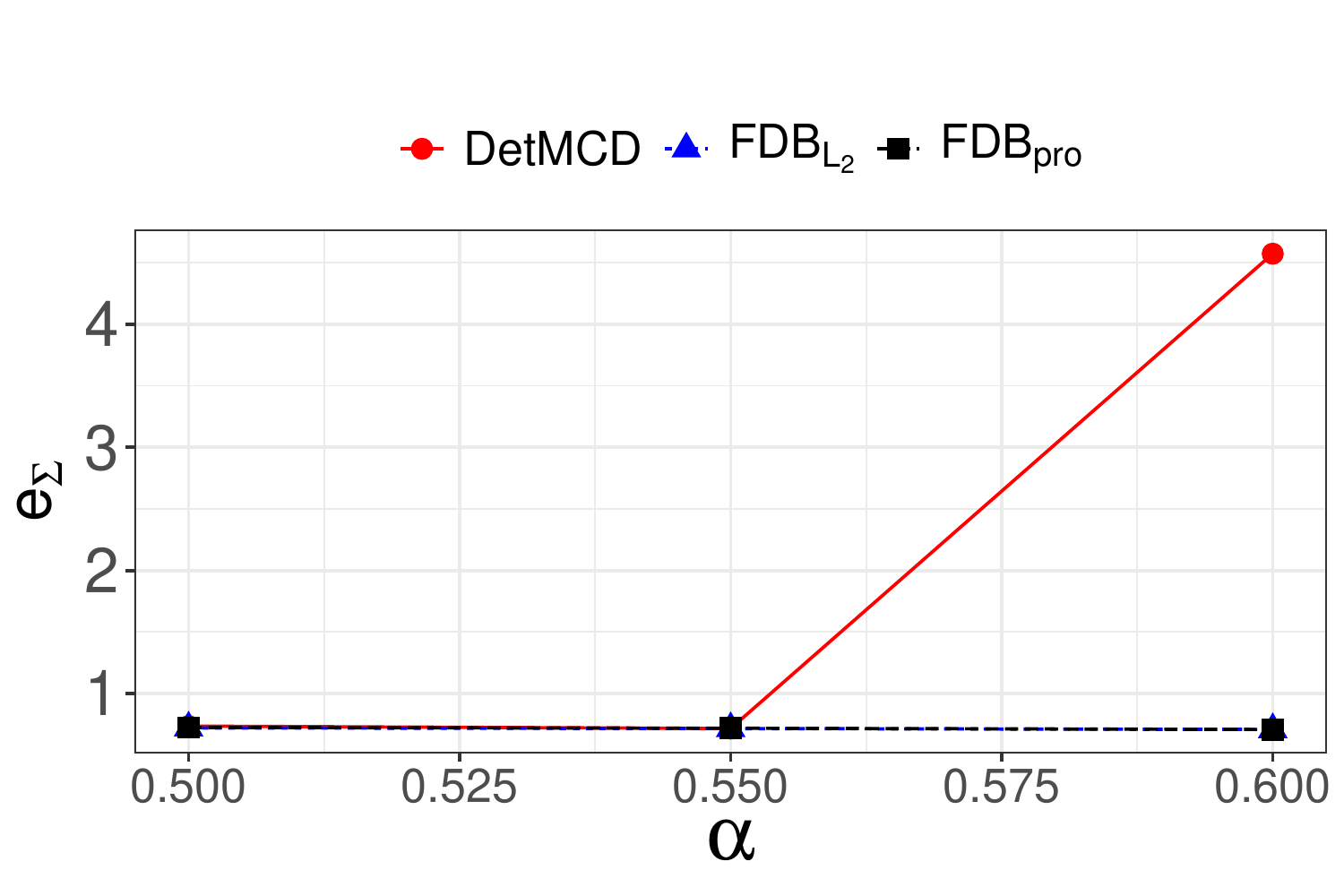}}
			\subfigure[radial]{
		\centering
		\includegraphics[width=0.4\linewidth]{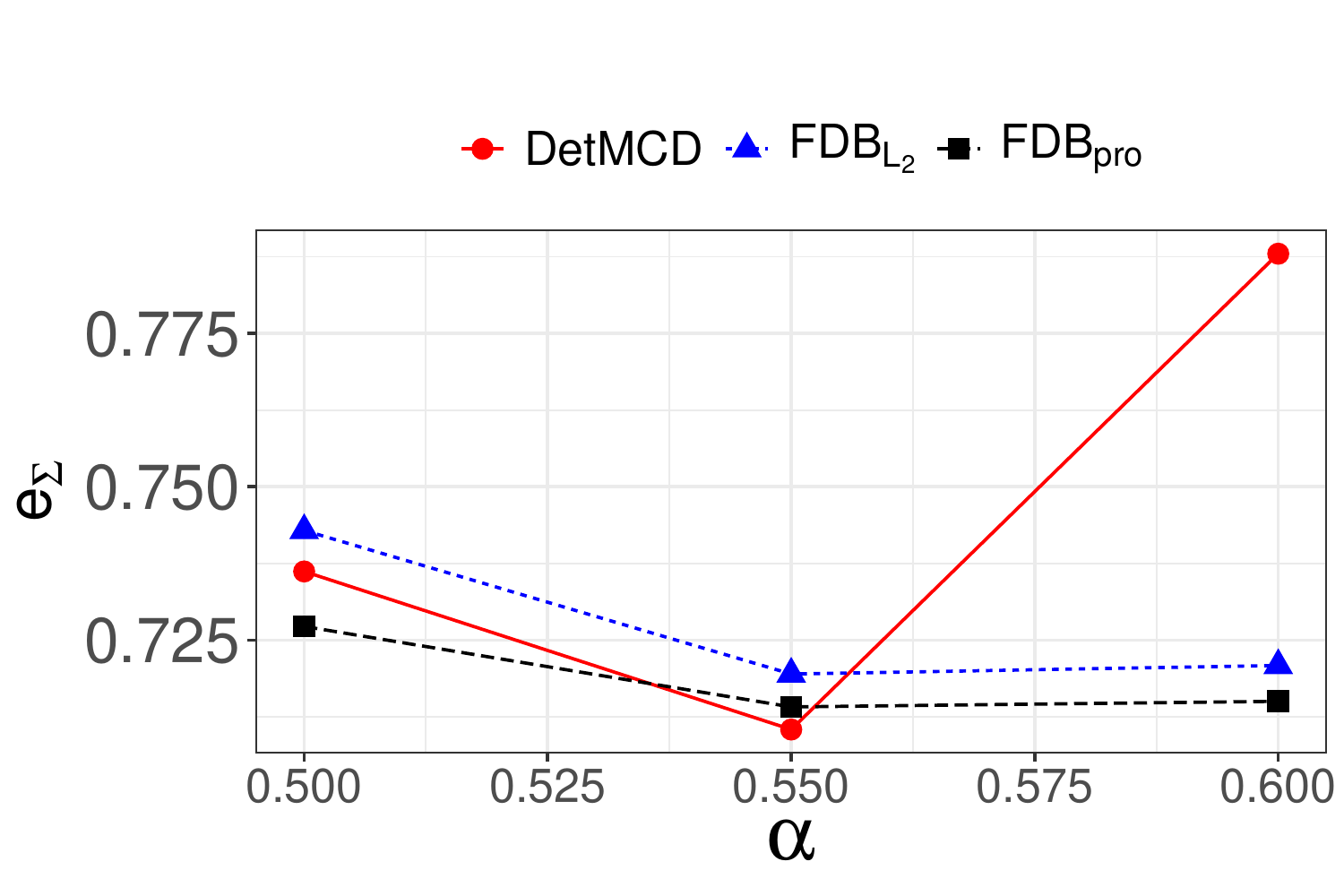}}
	\caption{The average $e_\Sigma$ for different $\alpha$, with $n=400$, $p=40$, $r=100$, and $\varepsilon=0.4$.} \label{hes}
\end{figure}

\subsection{Robustness assessment}

In addition, we assess the tolerance of each method to the core-set size $h=\lfloor\alpha n\rfloor$. Specifically, we generate data from Setting B with 40\% contamination, which is high enough for most practical implementations, and we set $r=100$ for point, random and cluster outliers. Meanwhile, we consider $\alpha$ ranging from 0.5 to 0.6, which is the highest value that possibly produces a clean subset. The average $e_{\rm \Sigma}$ from 1000 replications are reported in Figure \ref{hes}. 
$\texttt{FDB}_{\rm pro}$ remain robust for different values of $\alpha$ under all the investigated settings; $\texttt{FDB}_{\rm L_2}$ is satisfactory for random, cluster and radial contamination, while its estimation error grows substantially with $\alpha$ for point contamination; DetMCD generates table estimation when $\alpha=0.5$ and $0.55$ but becomes deficient when $\alpha$ raises up to 0.6 in each plot of Figure \ref{hes}. 
The other three measures show similar patterns as illustrated in Figure S1 of the Supplementary Material. In conclusion, $\texttt{FDB}_{\rm pro}$ reaches the strongest tolerance to the core-set size, when the proportion of outliers in the data is very high.

To sum it up, $\texttt{FDB}$s improves the computational efficiency significantly, which is the main motivation of this work, and $\texttt{FDB}_{\rm L_2}$ generally achieves the highest computational efficiency. Besides, we surprisingly find that $\texttt{FDB}_{\rm pro}$ shows superiority in terms of both estimation accuracy and robustness, especially in high-dimensional cases. One may safely substitute DetMCD with $\texttt{FDB}_{\rm pro}$ for practical implementation.

\section{Real data examples}
In this section, we apply the $\texttt{FDB}$ methods to four real datasets of various dimensions and sizes. The resultant robust multivariate location and scatter estimates are evaluated via some typical tasks in multivariate analysis such as outlier detection, linear discriminant analysis (LDA), principal component analysis (PCA), and image denoise. 
The same three methods from Section 5 are utilized.
The computations are performed in R (R Core Team 2021) on a laptop with a 10-core and 32GB memory processor.

\subsection{Robust PCA for forged bank notes data}
\label{sec:subsec:fb}
\begin{figure}[!b]
\centering
\subfigure[$\texttt{FDB}_{\rm L_2}$]{
\label{fig:subfig:FDpcab} 
\includegraphics[scale=0.33]{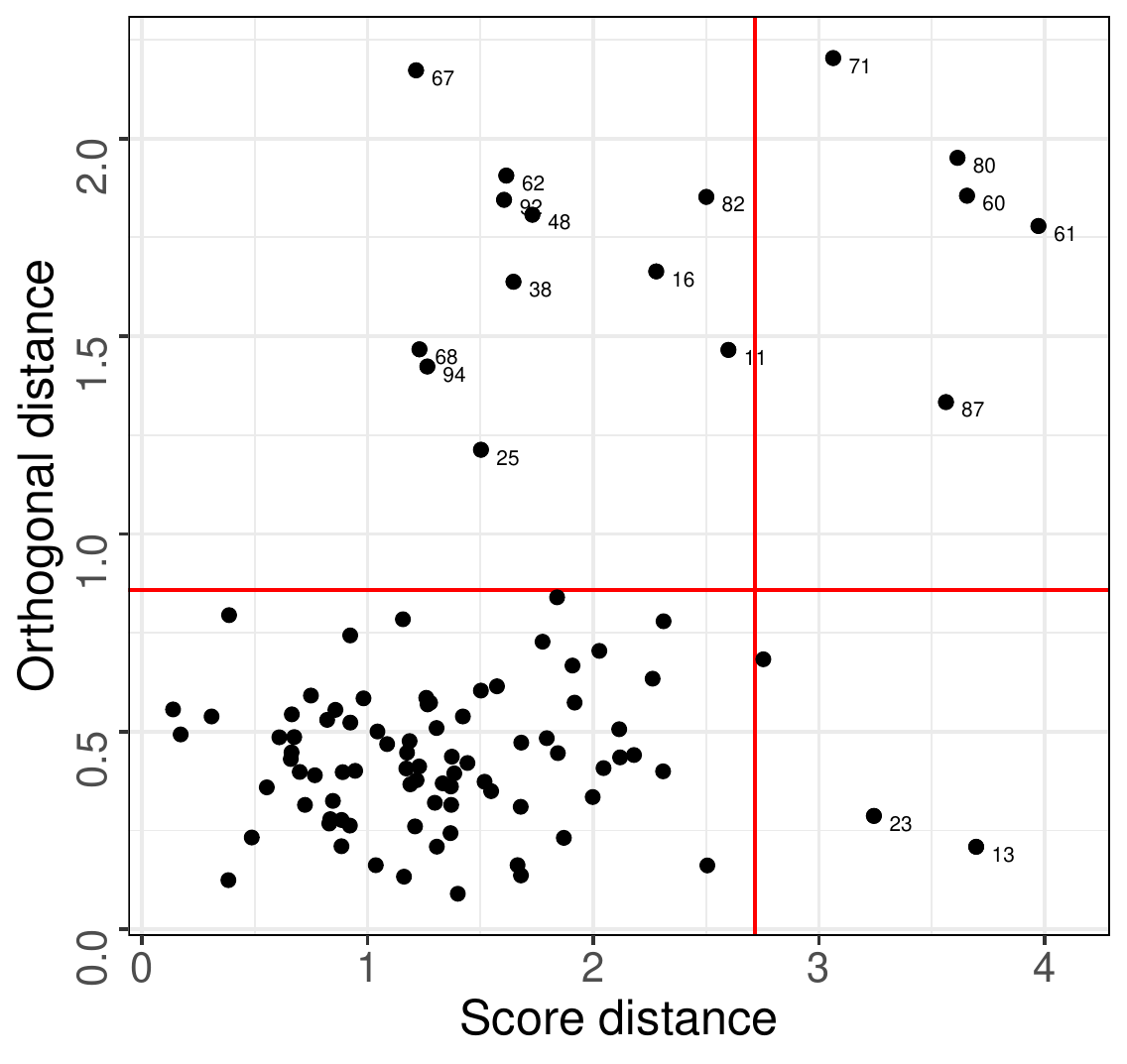}}
\subfigure[$\texttt{FDB}_{\rm pro}$]{
\label{fig:subfig:FDpcap} 
\includegraphics[scale=0.33]{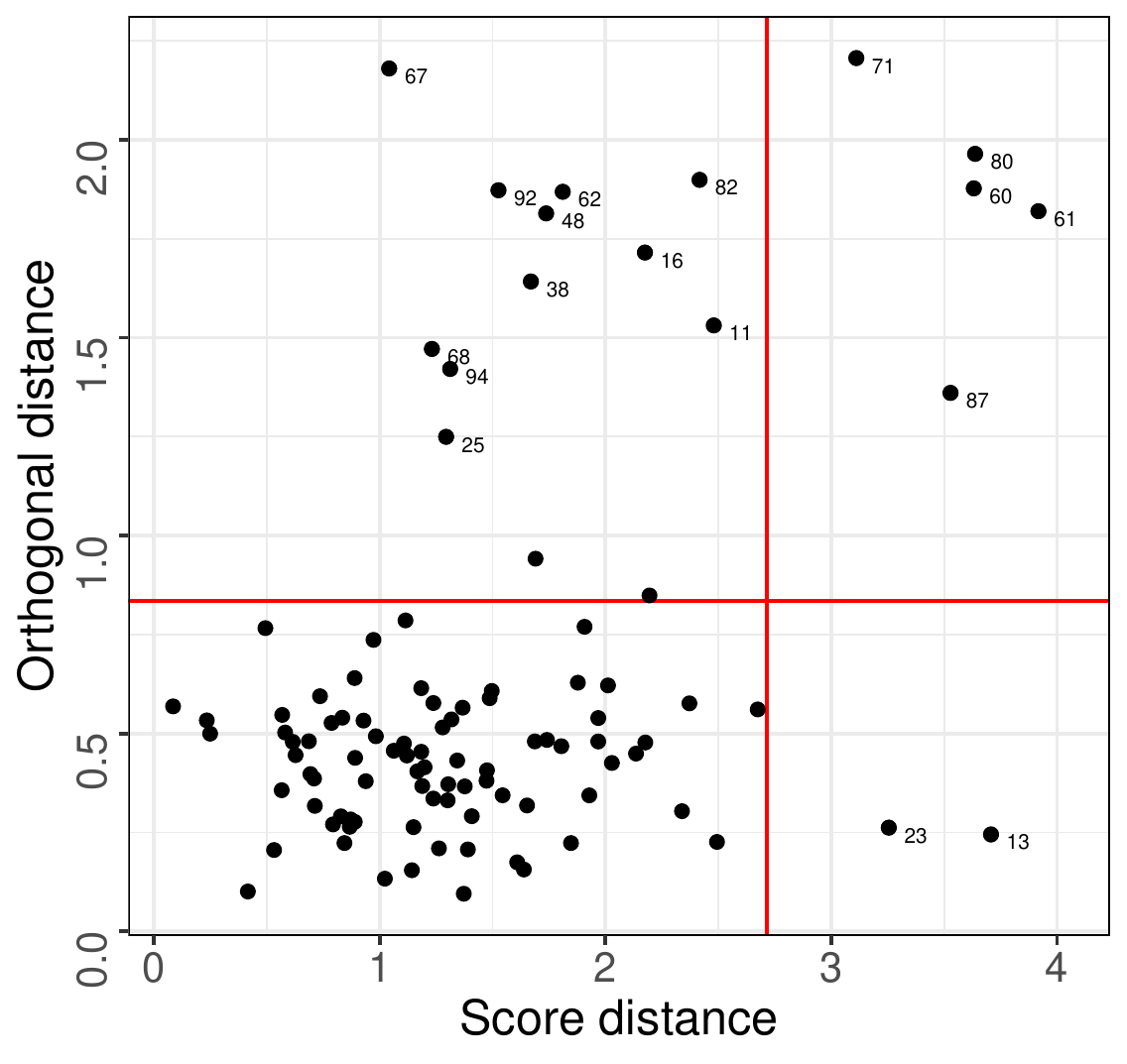}}
\subfigure[DetMCD]{
\label{fig:subfig:FDpcad} 
\includegraphics[scale=0.33]{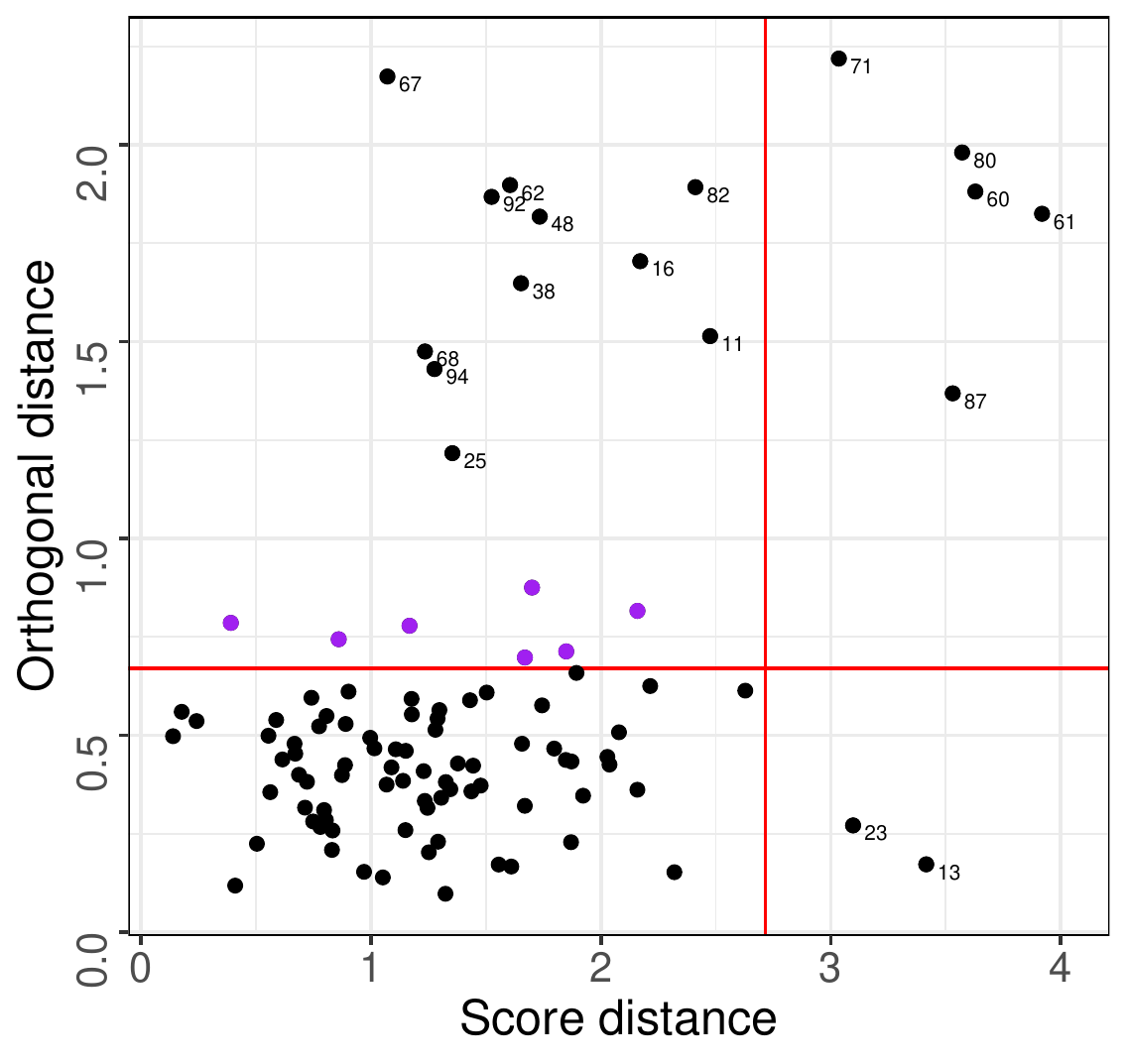}}
\subfigure[Sample]{
\label{fig:subfig:FDpcan} 
\includegraphics[scale=0.33]{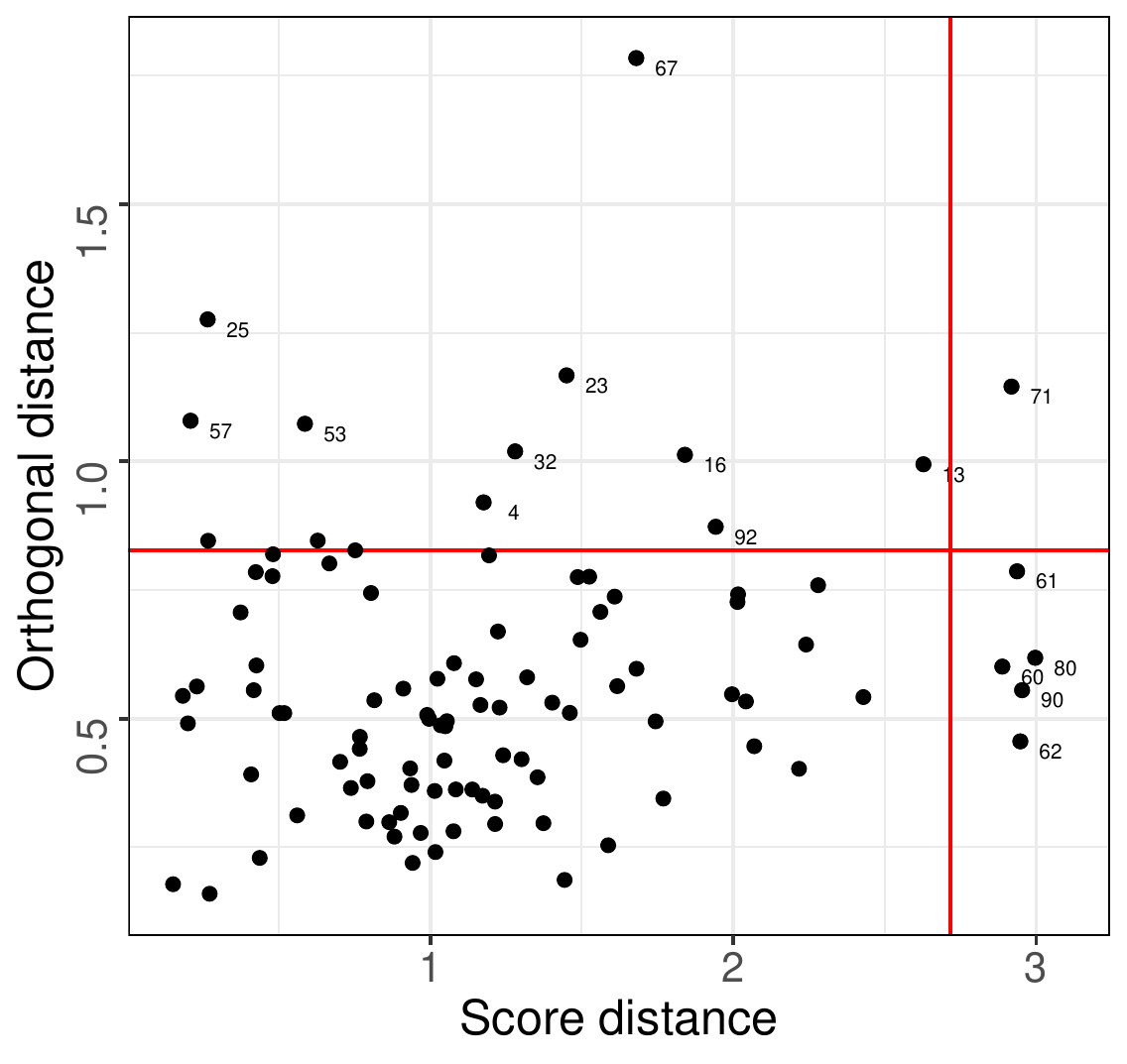}}\\
\subfigure[$\texttt{FDB}_{\rm L_2}$]{
\label{fig:subfig:FDb} 
\includegraphics[scale=0.33]{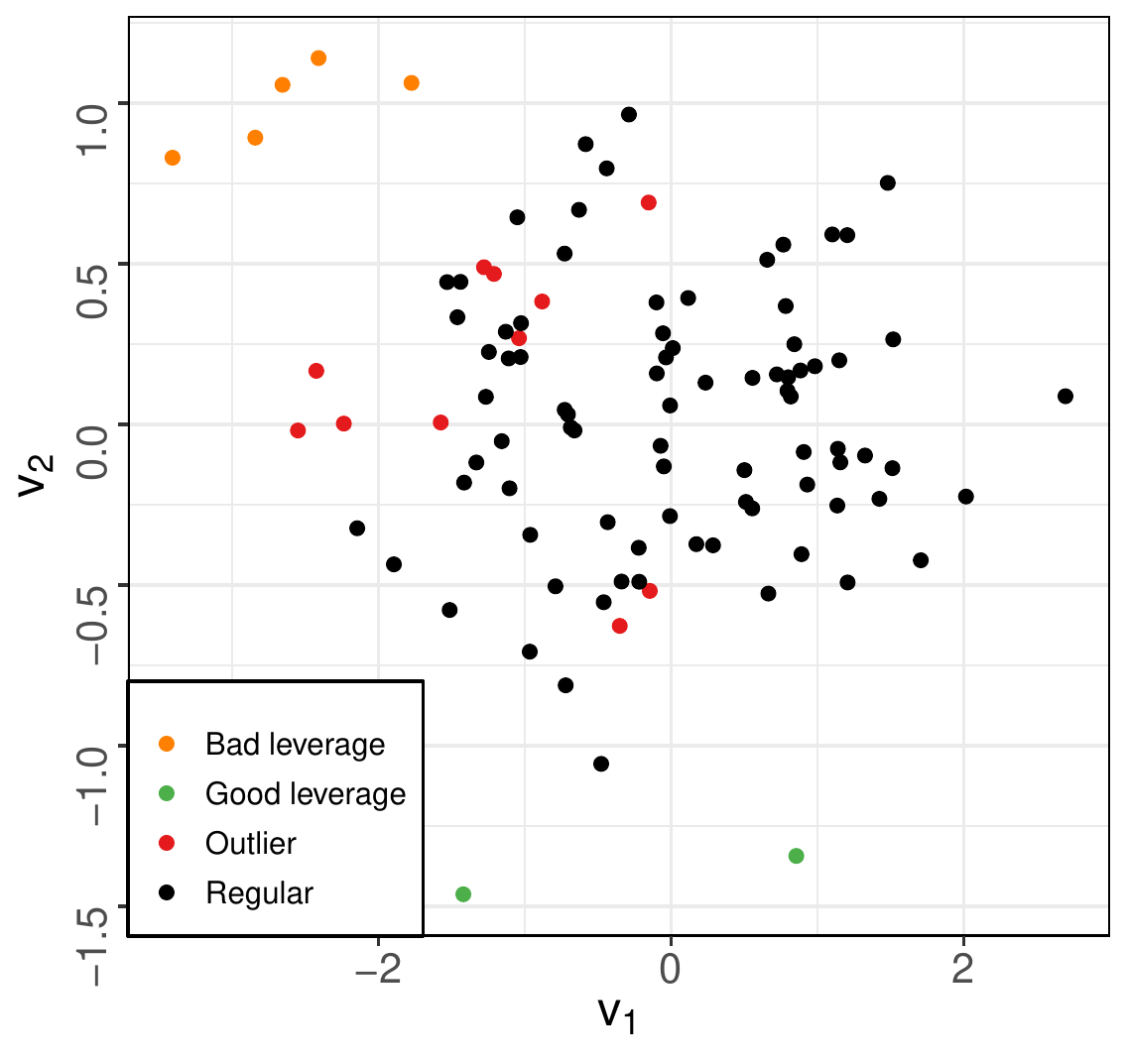}}
\subfigure[$\texttt{FDB}_{\rm pro}$]{
\label{fig:subfig:FDp} 
\includegraphics[scale=0.33]{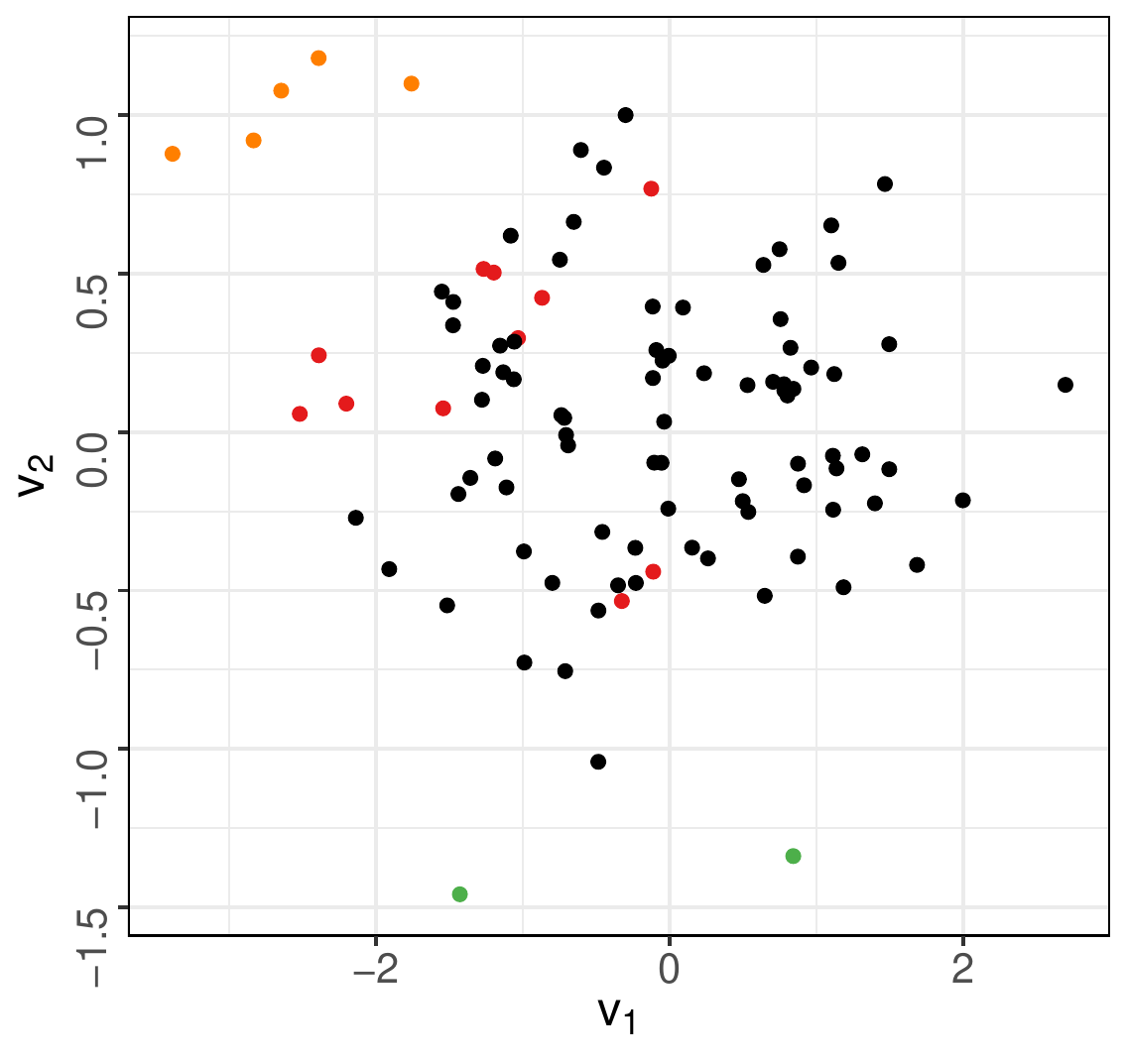}}
\subfigure[DetMCD]{
\label{fig:subfig:FDd} 
\includegraphics[scale=0.33]{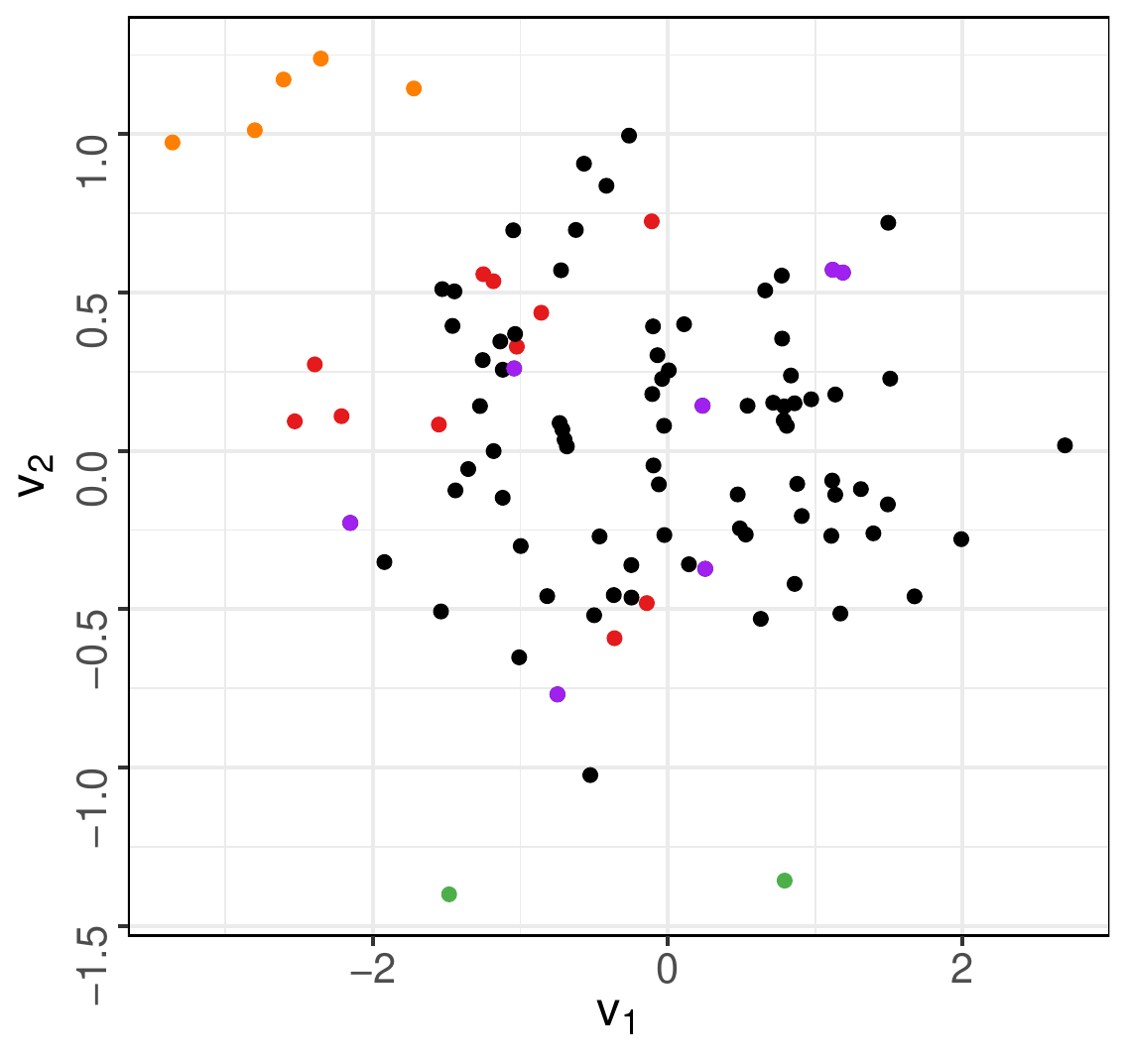}}
\subfigure[Sample]{
\label{fig:subfig:FDn} 
\includegraphics[scale=0.33]{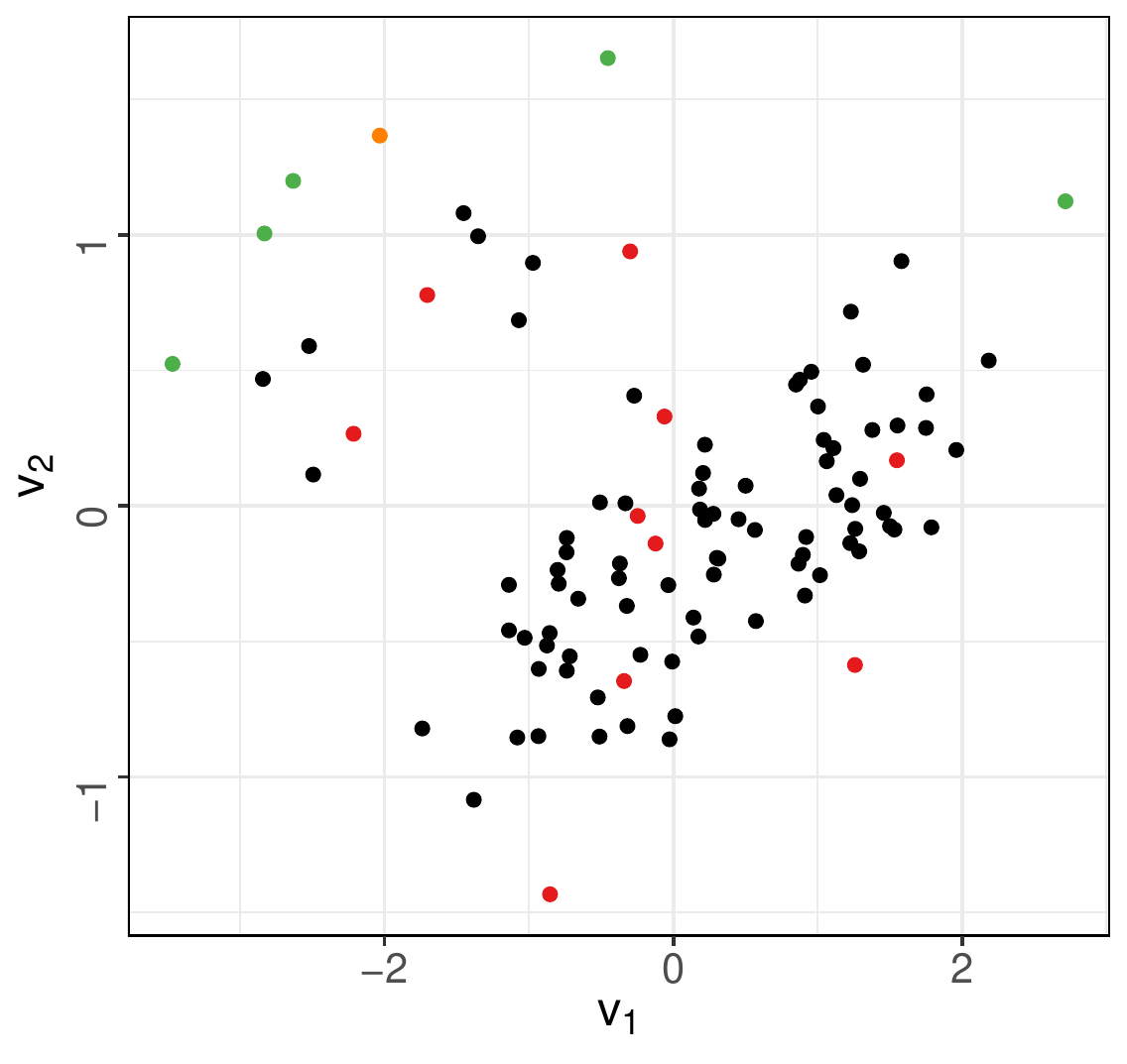}}
\caption{Diagnostic plots of the forged bank notes using robust PCAs based on (a) $\texttt{FDB}_{\rm L_2}$, (b) $\texttt{FDB}_{\rm pro}$ and (c) DetMCD and classical PCA based on (d) Sample. Projection of data on 2-dimensional subspace and diagnostic results obtained by (e) $\texttt{FDB}_{\rm L_2}$, (f) $\texttt{FDB}_{\rm pro}$, and (g) DetMCD and (h) classical PCA. Purple points in (c) and (g) represent samples that may be misclassified by using DetMCD.}
\label{fig:FD}
\end{figure}

The first dataset is the \href{https://www.tandfonline.com/doi/abs/10.1198/jcgs.2009.0005}{forged Swiss bank notes} data \citep{Forgedbanknotes}, which is also used in  \citet{hubert2012deterministic}. The data are of size $n=100$ and dimension $p=6$, denoted as $\boldsymbol{X}\in \mathbb{R}^{n\times p}$. Since this dataset includes outliers and highly correlated variables \citep{FD1,FD2}, we employ the proposed algorithms and DetMCD to get robust estimation $(\hat{\boldsymbol\mu},\hat{\boldsymbol\Sigma})$ first and conduct robust PCA based on these estimates. The classical PCA obtained by sample location and scatter is also demonstrated for comparison. We use the first two principal components $\boldsymbol{P}\in\mathbb{R}^{p\times 2}$, which explain over $80\%$ of the total variance. The projections of data on the $2$-dimensional PCA subspace, $\boldsymbol{T}=\{t_{ik}\}=(\boldsymbol{X}-\mathbf{1}_n\hat{\boldsymbol{\mu}}^T)\boldsymbol{P}$, are shown in Fig.~\ref{fig:subfig:FDb}--\ref{fig:subfig:FDn}. 

The score distance (SD) and orthogonal distance (OD) represent the robust distance of samples in the two-dimensional PCA subspace and the orthogonal distance of samples to the PCA subspace, respectively. For each sample $\boldsymbol{x}_i$, we have   $\mathbf{SD}_i=\sum_{k=1}^2 t_{ik}^2/\lambda_k$, where $\lambda_1$ and $\lambda_2$ are the first two eigenvalues of $\hat{\boldsymbol{\Sigma}}$, and $\mathbf{OD}_i=\sum_{j=1}^p e_{i,j}^2$, where $\boldsymbol{X}-\boldsymbol{TP}^T=\{e_{i,j}\}$.
Figure~\ref{fig:subfig:FDpcab}--\ref{fig:subfig:FDpcan} illustrate diagnostic plots. By two cutoff values in each diagnostic plot, we categorize samples into four types and 
assign different colors to them in Fig.~\ref{fig:subfig:FDb}--\ref{fig:subfig:FDn}. 
Regular samples gather in the bottom-left region of diagnostic plots with both score distances and orthogonal distances relatively small, which form the main body of the data cloud. 
%They gather in the bottom-left region of diagnostic plots. 
Good leverage samples are close to the PCA subspace but far from the regular samples, e.g., the samples $13$ and $23$ in the bottom-right region of Fig.~\ref{fig:subfig:FDpcab}--\ref{fig:subfig:FDpcad}, and the green points in Fig.~\ref{fig:subfig:FDb}--Fig.~\ref{fig:subfig:FDd}. 
Orthogonal outliers are far from the PCA subspace but not distinguishable by only observing their projections. With larger orthogonal distances but smaller score distances, they locate on the top-left region of diagonal plots, e.g., samples $11$, $62$ and $67$ in Fig.~\ref{fig:subfig:FDpcab}--\ref{fig:subfig:FDpcad}, and red points in Fig.~\ref{fig:subfig:FDb}--Fig.~\ref{fig:subfig:FDd}. 
For bad leverage samples, both score and orthogonal distances are large. They lie on the top-right region of diagonal plots and are represented as orange points in projection plots. 
In general, the three robust methods lead to comparable analysis results and all significantly improve the performance of the classical PCA, which agrees with Theorem~\ref{thm3}. 
However, DetMCD may identify some regular points as outliers according to its specific cutoff value for the orthogonal distance; see the purple points in the third column of Figure \ref{fig:FD}.

\subsection{Outlier detection for phoneme data}
In the second example, we detect outliers for a \href{https://web.stanford.edu/~hastie/ElemStatLearn/}{phoneme} dataset, which comes from a speech recognition database TIMIT and has been discussed in \citet{hastie2009elements}. The data includes $1050$ speech frames, $1000$ of which are ``ao'' and $50$ of which are ``iy''. Each data frame has been transformed to be a log-periodogram of length $256$. First, we reduce the dimensions by smoothing splines. For each sample, we replace the original variables $\boldsymbol{x}\in\mathbb{R}^{256}$ with $50$-dimensional variables $\Tilde{\boldsymbol{x}}=\boldsymbol{N}^T\boldsymbol{x}$, where $\boldsymbol{N}\in\mathbb{R}^{256\times 50}$ is the basis matrix of natural cubic splines. We use $50$ basis functions with knots uniformly placed over $1,\ldots,256$. To this end, we are dealing with data of $n=1050$ and $p=50$.

\begin{table}[!b] 
\renewcommand\arraystretch{1.5}
	\centering  
	\caption{Performance of various methods in outlier detection for Phoneme data.}  
	\vspace{0.5em}
	\begin{tabular}{|c|c|c|c|c|}
		\hline
		Method & $\texttt{FDB}_{\rm L_2}$ & $\texttt{FDB}_{\rm pro}$ & DetMCD & Sample\\
		\hline
		Number & $49$ & $49$ & $49$ & $13$\\
		\hline
          AUC & $0.9895$ & $0.9895$ & $0.9895$ & $0.6115$\\
		\hline
		Time & $0.5390$ & $0.5225$ & $1.7911$ &  $0.2865$\\
		\hline
	\end{tabular}
	\label{tab:phoneme}
\end{table}

We take ``ao'' and ``iy'' to be regular cases and outliers, respectively, and perform outlier detection. To be more specific, we calculate the robust Mahalanobis distances based on FDBs and DetMCD, and classical Mahalanobis distances based on Sample. Then, we treat cases with the top $50$ largest distances as outliers, and the remaining ones as regular cases. Table~\ref{tab:phoneme} records the number of ``iy'' that are flagged as outliers, the area under the ROC curve (AUC), and the average computational time (second) of $100$ replicates for these methods. Again, the proposed methods and DetMCD perform similarly and they all outperform Sample with higher AUCs. In addition, the proposed methods reveal advantages in computation time.

\subsection{Denoise for MNIST data}
The \href{https://archive.ics.uci.edu/ml/machine-learning-databases/mnist-mld/}{Modified National Institute of Standards and Technology} (MNIST) database is widely used for training various image processing systems. It contains a large set of handwritten images representing the digits zero through nine. Each digit is stored as a gray-scale image with a size $p=28\times 28$. Training data $\boldsymbol{X}$ and testing data $\boldsymbol{X}_t$ of $n=10000$ are randomly selected from MNIST. Noises $\bm\epsilon\sim\mathcal{N}(\mathbf{0}_p,60^2\mathbf{I}_p)$ are generated and added to $20\%$ of the images in the training set and all images in the testing set. Our task is to denoise the testing images, which has been considered in \cite{Schreurs2021OutlierDI}.

\begin{figure}[!t]
\centering
\includegraphics[scale=0.4]{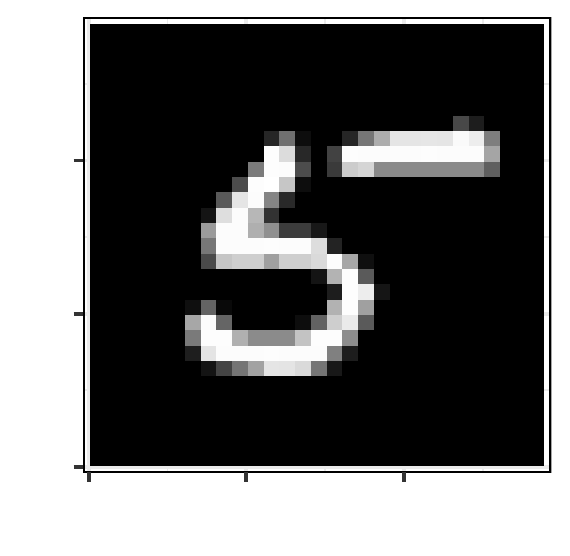}
\includegraphics[scale=0.4]{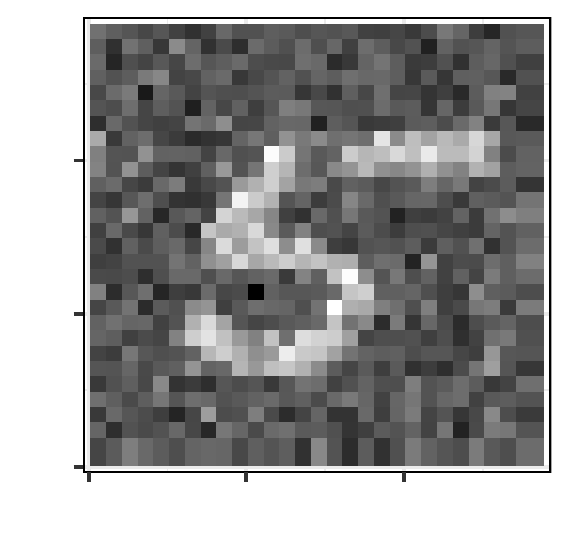}
\includegraphics[scale=0.4]{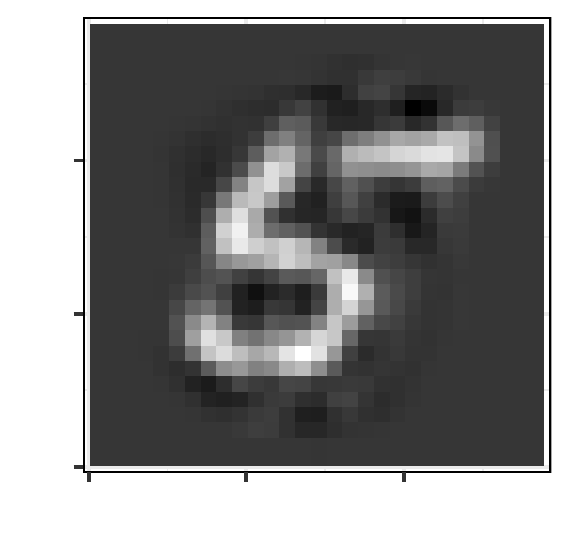}
\includegraphics[scale=0.4]{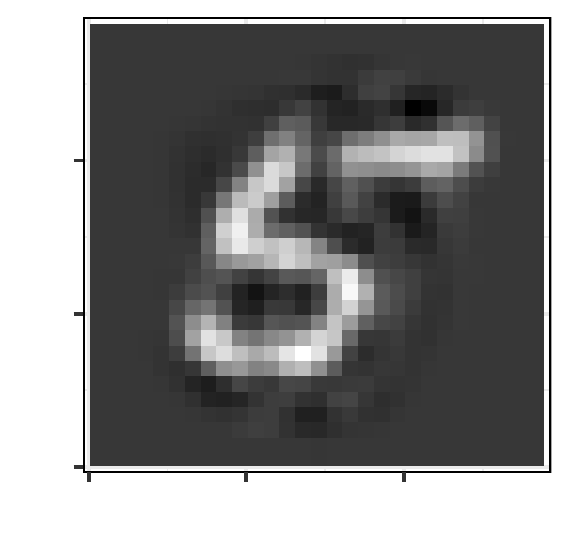}
\includegraphics[scale=0.4]{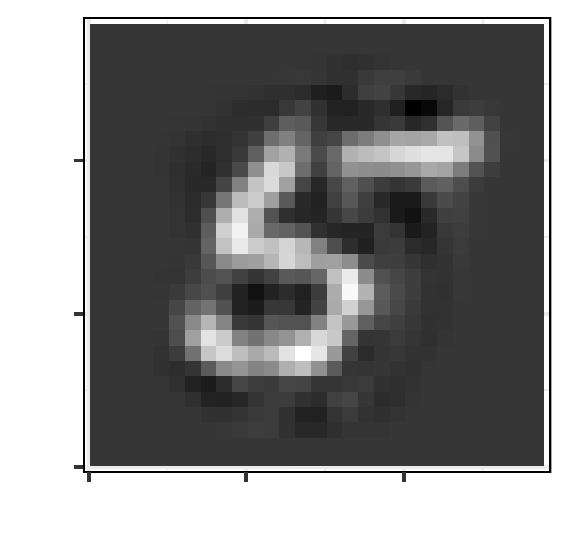}
\includegraphics[scale=0.4]{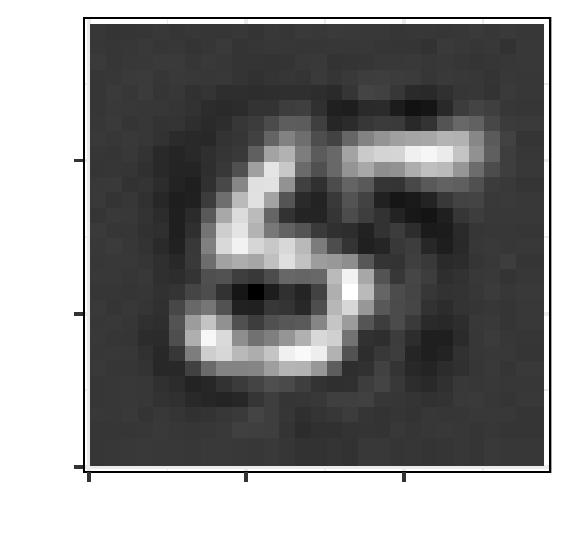}\\
\includegraphics[scale=0.4]{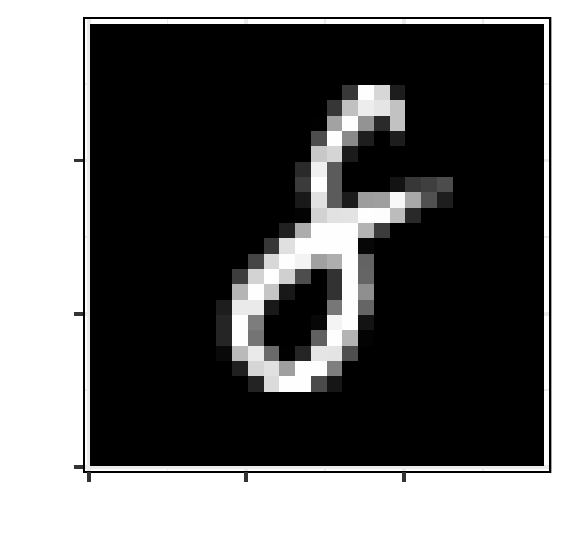}
\includegraphics[scale=0.4]{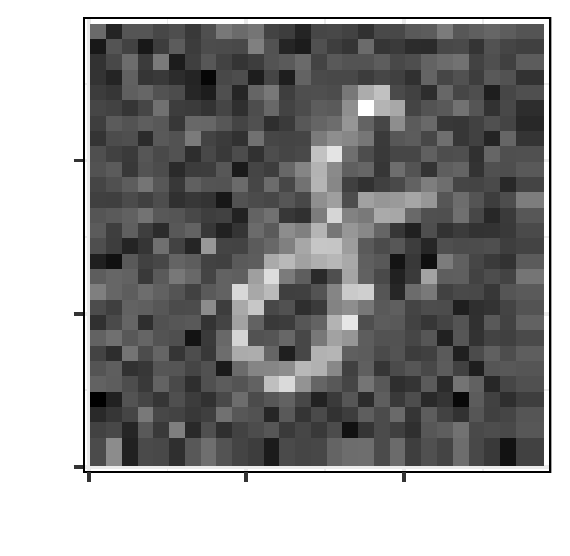}
\includegraphics[scale=0.4]{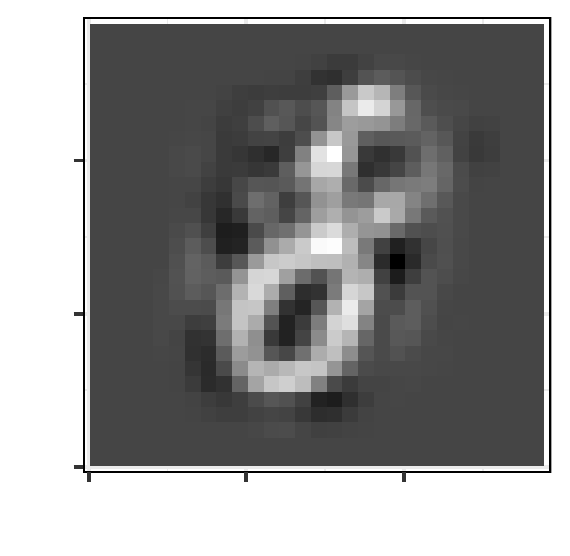}
\includegraphics[scale=0.4]{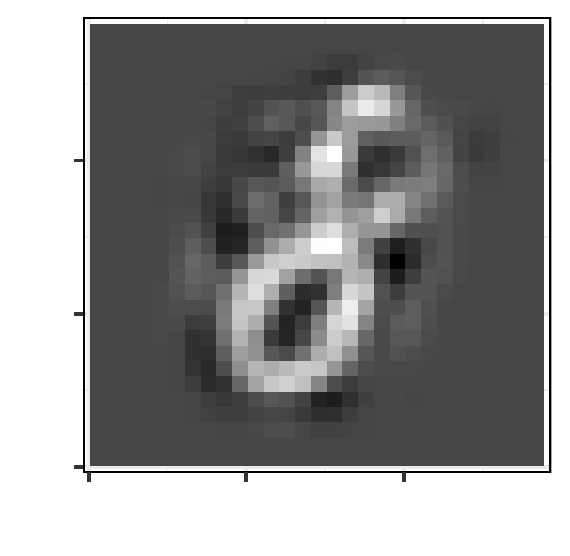}
\includegraphics[scale=0.4]{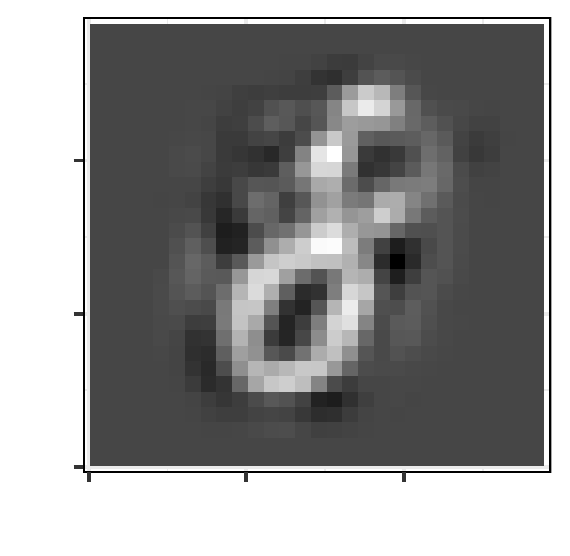}
\includegraphics[scale=0.4]{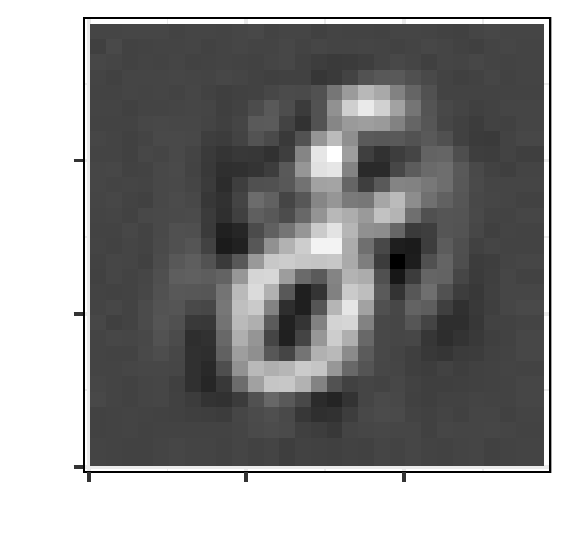}\\
\includegraphics[scale=0.4]{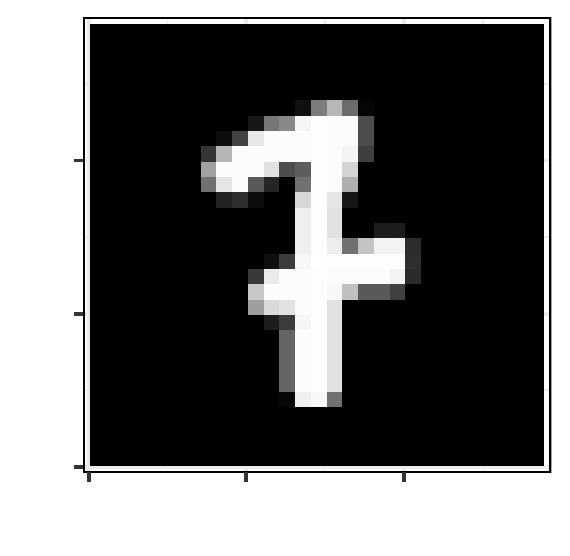}
\includegraphics[scale=0.4]{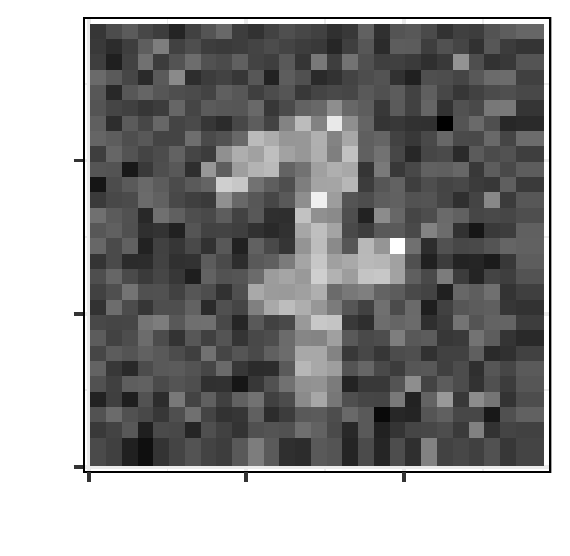}
\includegraphics[scale=0.4]{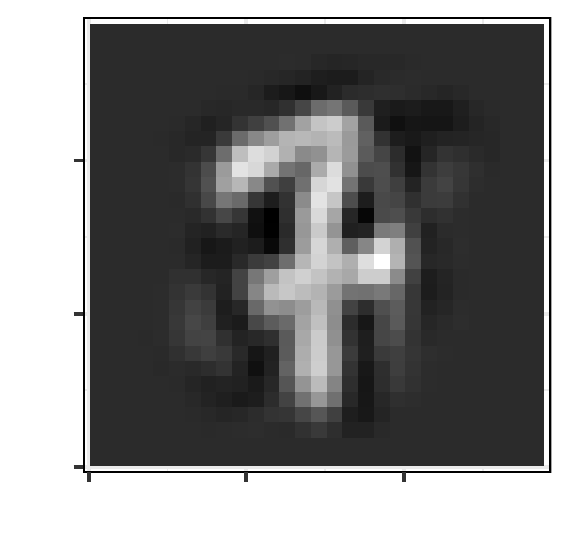}
\includegraphics[scale=0.4]{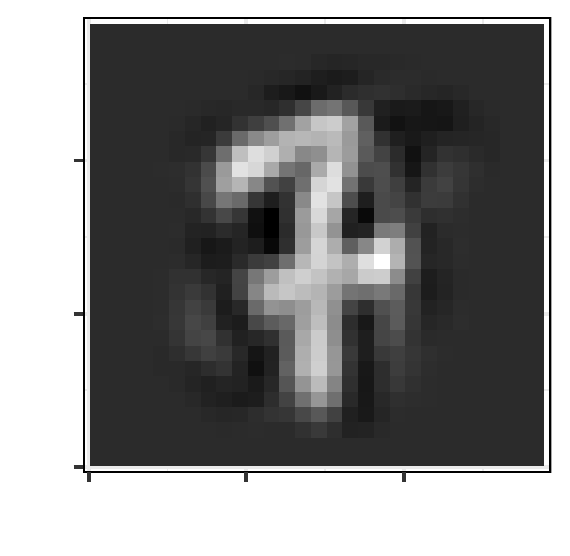}
\includegraphics[scale=0.4]{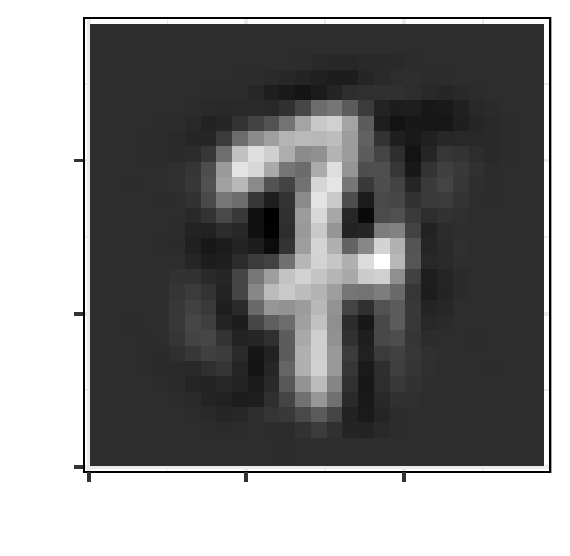}
\includegraphics[scale=0.4]{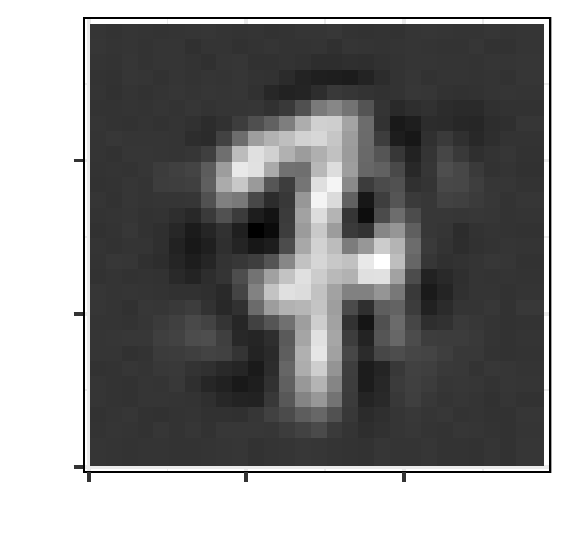}\\
\includegraphics[scale=0.4]{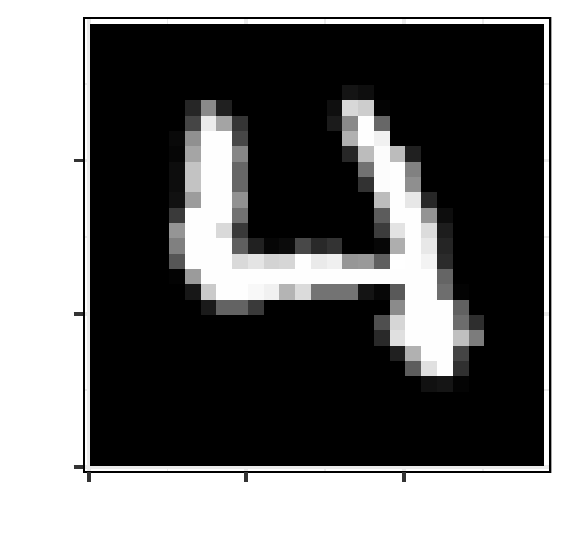}
\includegraphics[scale=0.4]{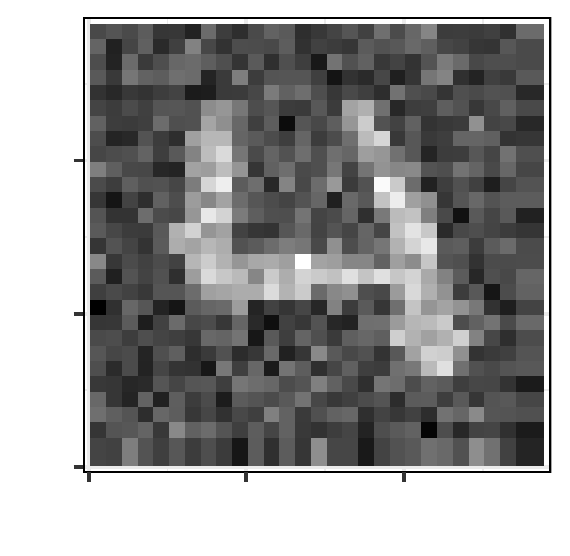}
\includegraphics[scale=0.4]{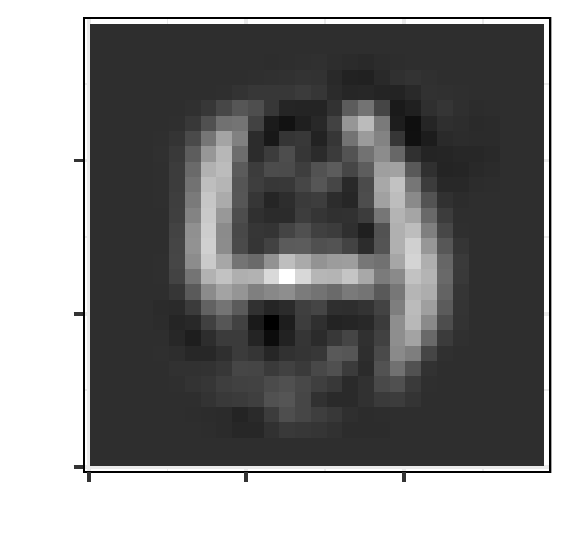}
\includegraphics[scale=0.4]{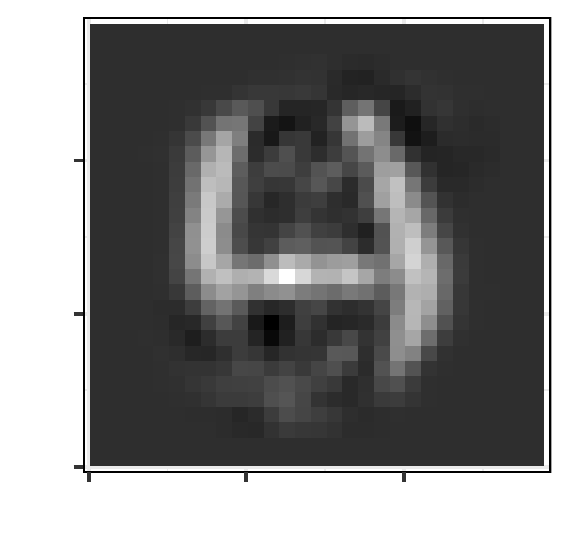}
\includegraphics[scale=0.4]{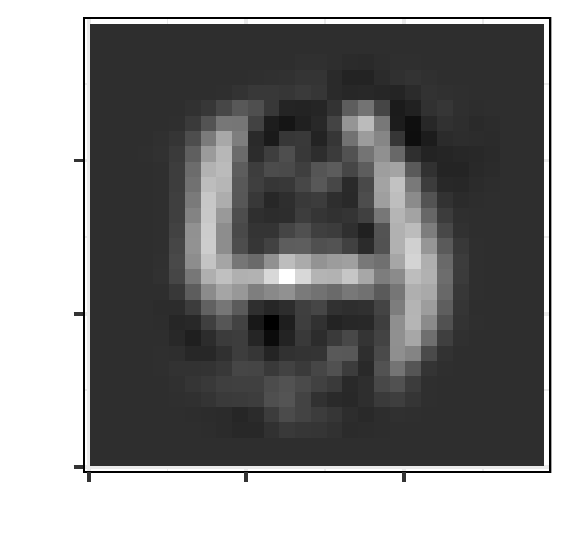}
\includegraphics[scale=0.4]{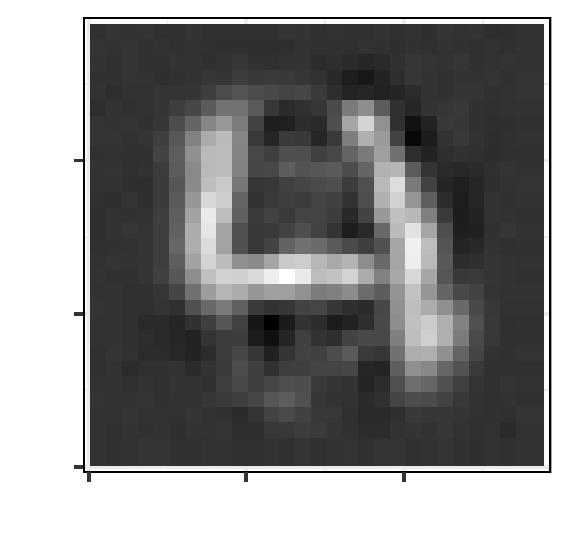}\\
\includegraphics[scale=0.4]{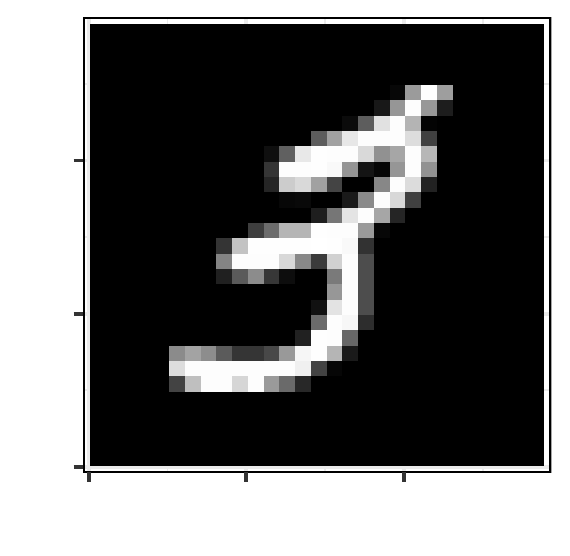}
\includegraphics[scale=0.4]{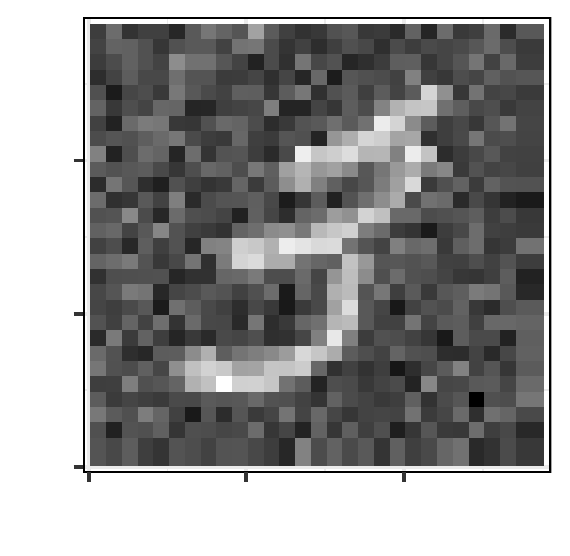}
\includegraphics[scale=0.4]{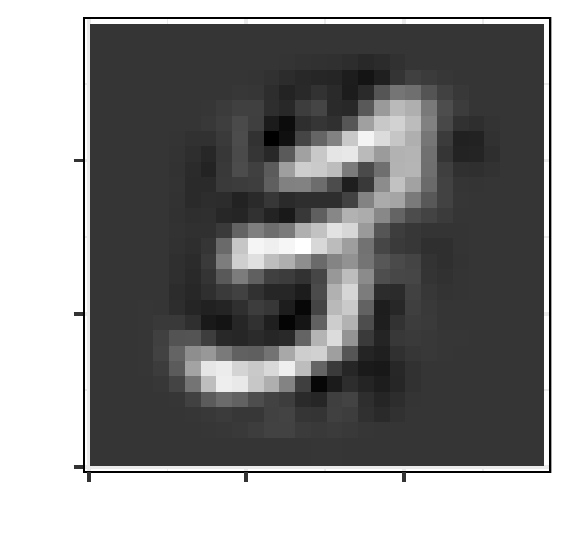}
\includegraphics[scale=0.4]{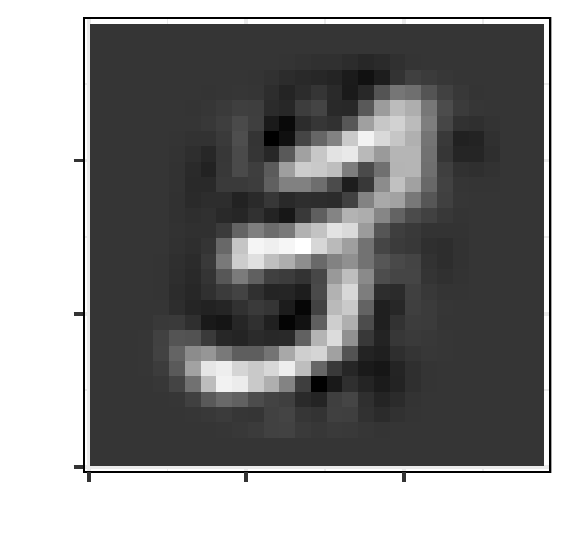}
\includegraphics[scale=0.4]{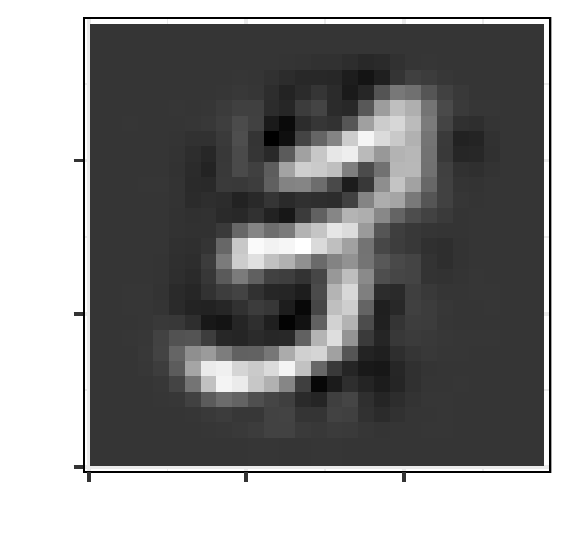}
\includegraphics[scale=0.4]{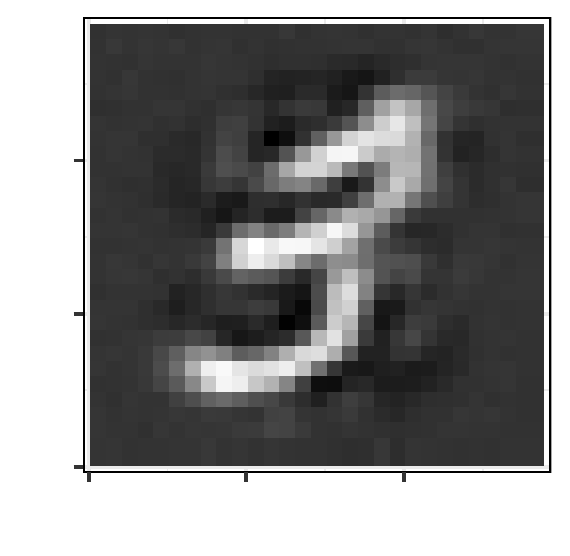}\\
\includegraphics[scale=0.4]{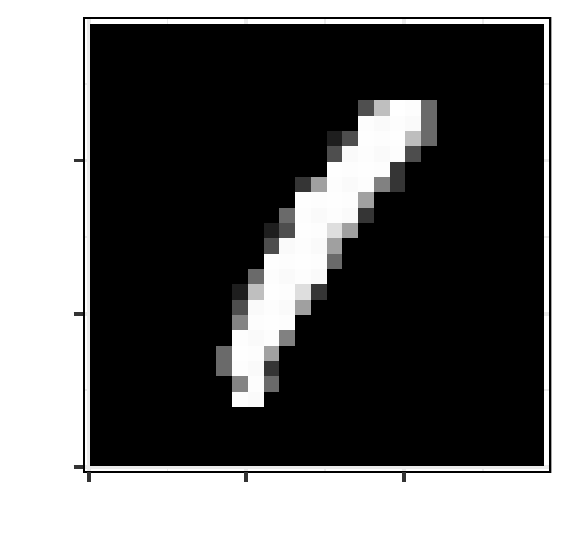}
\includegraphics[scale=0.4]{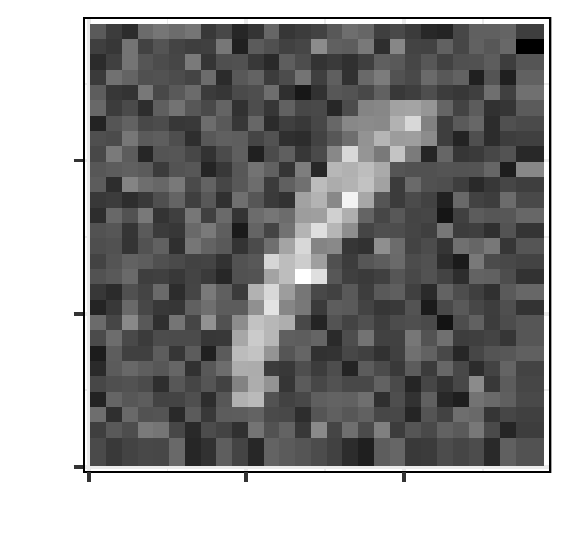}
\includegraphics[scale=0.4]{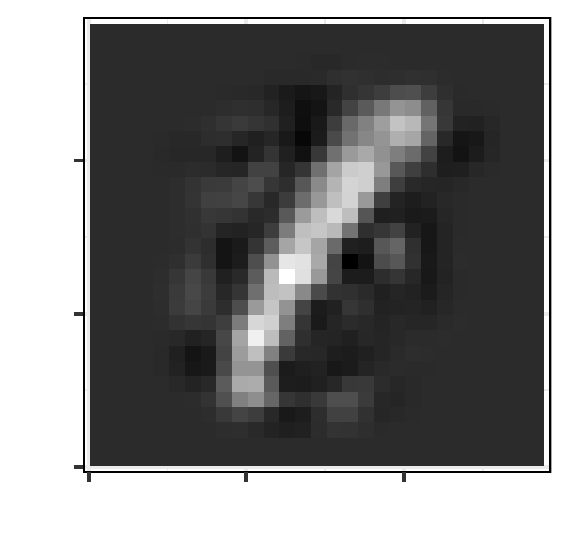}
\includegraphics[scale=0.4]{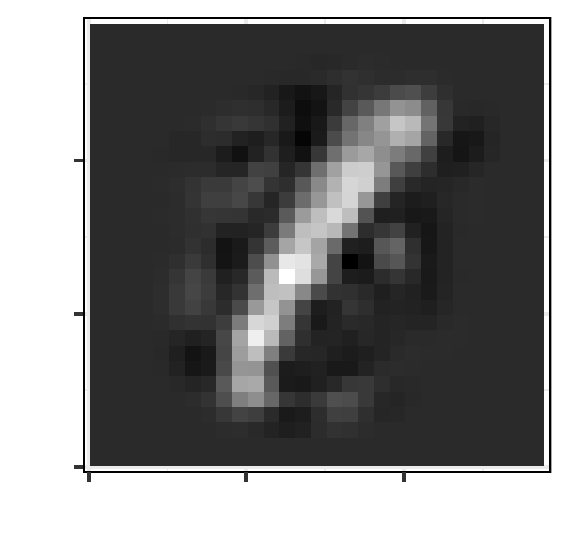}
\includegraphics[scale=0.4]{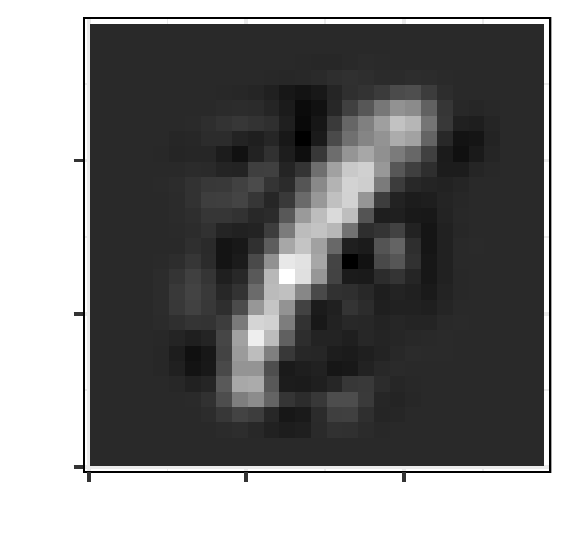}
\includegraphics[scale=0.4]{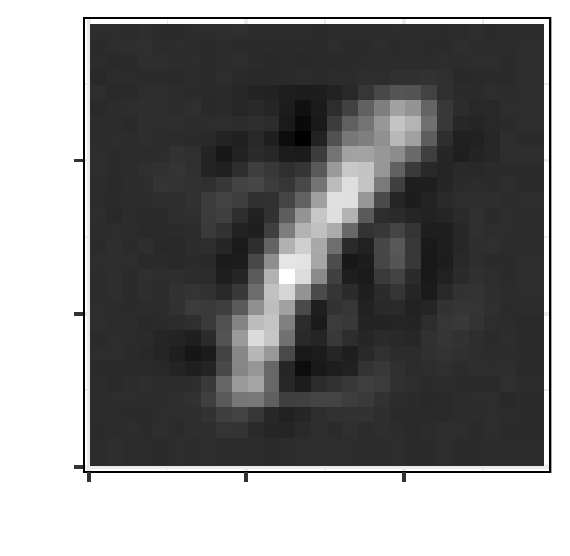}\\
\caption{The first and second columns show images before and after adding noise, respectively. The third to the sixth columns show denoised images obtained by $\texttt{FDB}_{\rm L_2}$, $\texttt{FDB}_{\rm pro}$, $\texttt{FDB}_{\rm pro1000}$, and Sample with $K=75$, respectively.}
\label{fig:Mnist}
\end{figure}

The detailed process is similar to the robust PCA in Section~\ref{sec:subsec:fb}. First, we apply $\texttt{FDB}_{\rm L_2}$, $\texttt{FDB}_{\rm pro}$, DetMCD, and Sample to training images and obtain the estimated location and scatter. $\texttt{FDB}_{\rm pro}$ with $k=1000$, denoted as $\texttt{FDB}_{\rm pro1000}$, is also used for better comparison. Here all methods have comparable computational time. Then, we calculate the eigenvectors of the robust estimated scatter. Next, we project the testing data to the subspace spanned by the first $K$ eigenvectors and then transform the projected data, i.e., scores, back to the original space. We denote the reconstructed data as $\hat{\boldsymbol{X}}_t$. Fig.~\ref{fig:Mnist} illustrates $\hat{\boldsymbol{X}}_t$ obtained by various methods with $K=75$. 
DetMCD is removed since it returns an error of high condition numbers and hence is not applicable to this example. We can see that the denoised images for the proposed methods are more clear than those for the Sample, which verifies the influence of adding noise on evaluating scatter and the efficiency of the proposed methods. As in \cite{Schreurs2021OutlierDI}, we also calculate the mean absolute error $\mathrm{MAE}=\sum_{i=1}^n\sum_{j=1}^p|x_{t(ij)}-\hat x_{t(ij)}|/(np)$ between the original and denoised images. Table~\ref{tab:1} shows that the MAE for Sample is obviously larger than those for proposed methods.

\begin{table}[!b]
    \centering
        \caption{MAEs for proposed robust methods and classical method with various $K$s.}
    \label{tab:1}
    \begin{tabular}{|l|c|c|c|c|c|}
    \hline
       $K$  & $15$ & $30$ & $45$ & $60$ & $75$\\
       \hline
       $\texttt{FDB}_{\rm L_2}$ & $23.04525$ & $19.27832$ & $17.52109$ & $16.63028$ & $16.26773$  \\\hline
       $\texttt{FDB}_{\rm pro}$ & $23.06536$ & $19.27085$ & $17.51270$ & $16.62653$ & $16.26147$\\\hline
       $\texttt{FDB}_{\rm pro1000}$ & $23.09654$ & $19.30584$ & $17.54031$ & $16.65916$ & $16.29452$\\\hline
       Sample & $24.58077$ & $20.60175$ & $18.92803$ & $18.16014$ & $17.95447$\\ 
       \hline
    \end{tabular}
\end{table}

\subsection{Outlier detection for Musk data}
\begin{figure}[h]
\centering
\subfigure[$\texttt{FDB}_{\rm L_2}$]{
\label{fig:subfig:Musk2} 
\includegraphics[scale=0.3]{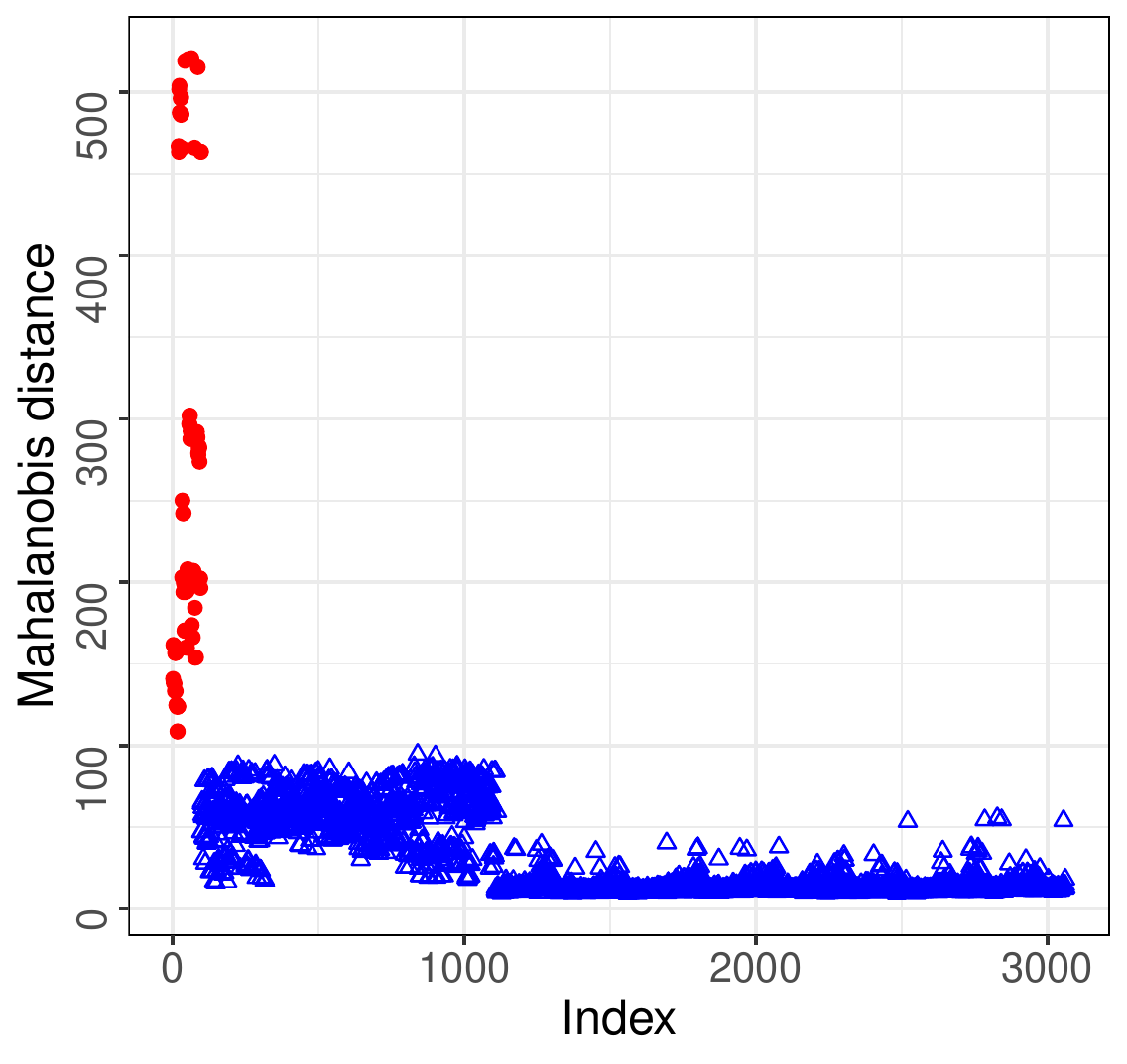}}
\subfigure[$\texttt{FDB}_{\rm pro}$]{
\label{fig:subfig:Muskp} 
\includegraphics[scale=0.3]{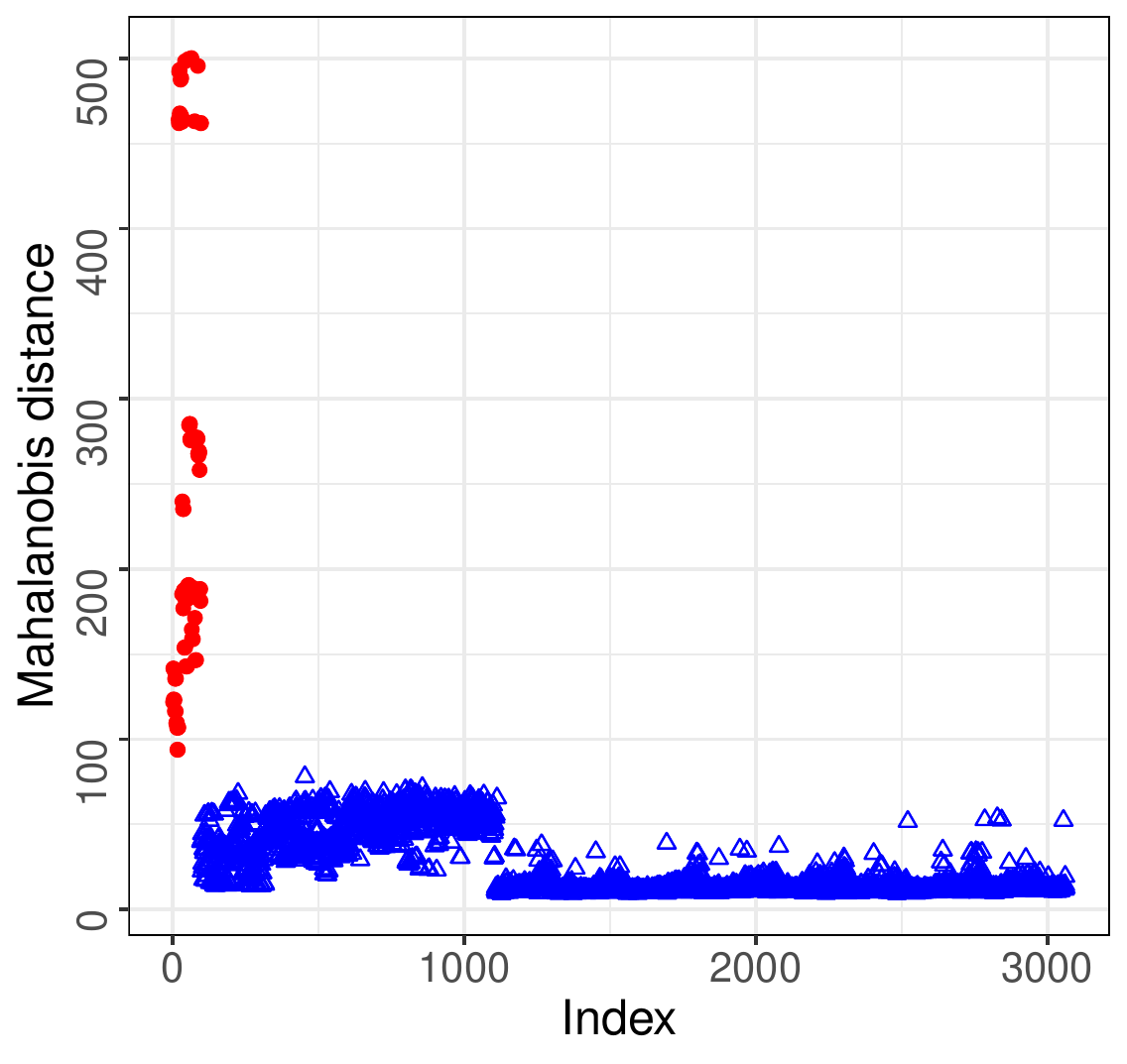}}
\subfigure[DetMCD]{
\label{fig:subfig:Muskd} 
\includegraphics[scale=0.3]{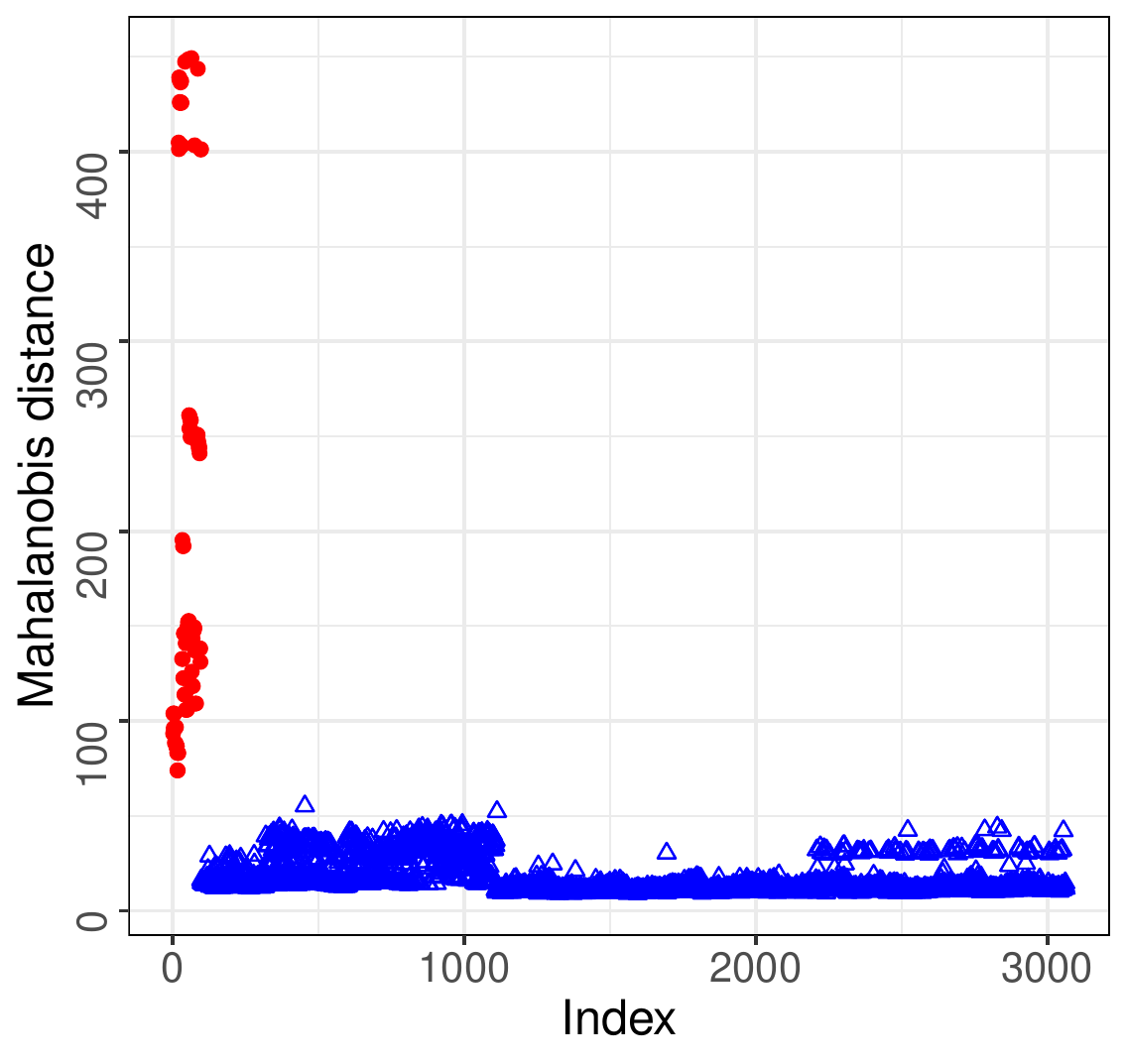}}
\subfigure[Sample]{
\label{fig:subfig:Muskn} 
\includegraphics[scale=0.3]{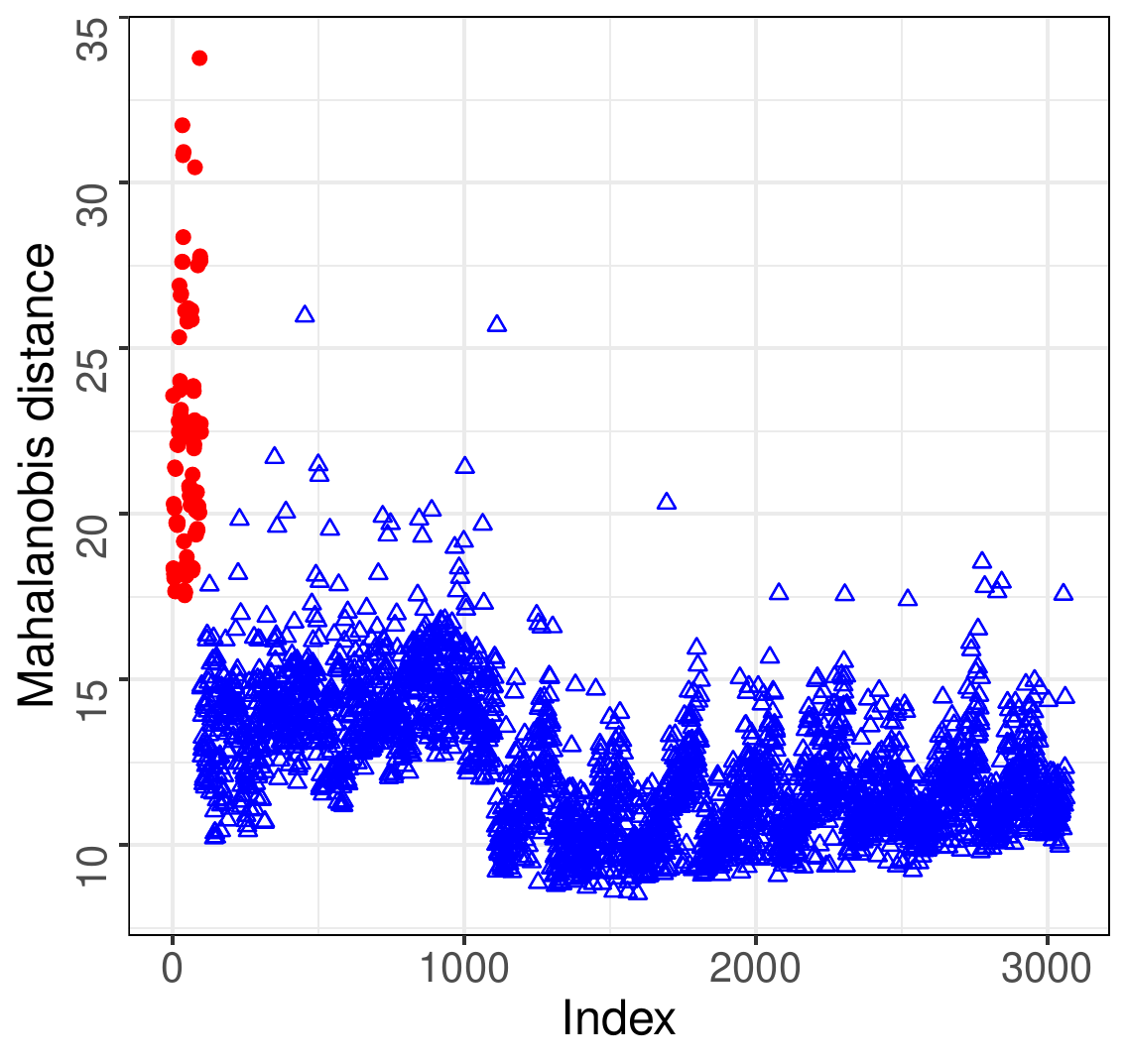}}
\caption{(a), (b), and (c) are robust distances obtained by $\texttt{FDB}_{\rm L_2}$,  $\texttt{FDB}_{\rm pro}$, and DetMCD, respectively. (d) Mahalanobis distance obtained by Sample.}
\label{fig:Musk}
\end{figure}

Musk data is commonly used in high-dimensional classification and outlier detection problems \citep{aggarwal2015theoretical,porwal2017outlier}. The original data, which can be found in \href{https://archive.ics.uci.edu/ml/datasets/Musk+(Version+2)}{UCI} includes $6598$ samples, divide into musk and non-musk classes. Each sample has $p=166$ features characterizing the molecule structure. Here we use a \href{http://odds.cs.stonybrook.edu/musk-dataset/}{preprocessed musk} dataset of $n=3062$, which consists of $2965$ non-musk samples as inliers and $97$ musk sample as outliers. Our task is to detect the outliers with the proposed methods, DetMCD and Sample. As shown in Fig.~\ref{fig:Musk}, our proposed methods and DetMCD can exactly pick out outliers, which are represented as red points, whereas Sample would take many regular samples as outliers. Furthermore, the computational time of $\texttt{FDB}_{\rm L_2}$, $\texttt{FDB}_{\rm pro}$, and DetMCD are $13.547$, $12.884$, and $53.391$ seconds, respectively. The proposed methods are computationally more efficient than DetMCD. 

\section{Discussion}
\label{sec:conc}
MCD-type methods suffer from high computational complexity due to the iteratively used C-step. To tackle the issue, we directly approximate the MCD subset with the trimmed subset induced by statistical depth. Two depth notions, the projection depth and the $L_2$ depth, are recommended due to high computational efficiency and robustness. 
In addition, we establish the equivalence between the desired MCD subset and the trimmed subset induced by the projection depth. Bypassing the iteration of the C-step, we manage to reduce the computational complexity from $O(\psi np^2+\psi p^3)$ to $O(knp)$ and $O(n^2p)$ with $\texttt{FDB}_{\rm pro}$ and $\texttt{FDB}_{\rm L_2}$, respectively. Moreover, the proposed estimators also reach the same level of robustness as the MCD estimator.
We conduct extensive simulation studies and show that our estimators are comparable with MCD-type estimators for low-dimensional data and significantly outperform MCD-type estimators for high-dimensional cases.

The real data examples provide strong evidence that \texttt{FDB} is a valuable complement to the toolset of robust multivariate analysis, including but not limited to PCA, LDA, image denoise and outlier detection. The \texttt{FDB} algorithm may benefit other applications that directly or indirectly rely on robust covariance matrix estimation, such as robust linear regression \citep{coakley1993bounded}, regression with continuous and categorical regressors \citep{hubert1997robust}, MCD-regression \citep{rousseeuw2004robust},  multivariate least trimmed squares estimation \citep{agullo2008multivariate}, and robust errors-in-variables regression \citep{fekri2004robust}. 

The present study primarily focuses on data from elliptical symmetric distributions with enough samples, which may be violated in practical scenarios \citep{Schreurs2021OutlierDI}. In Section S3.1 of the Supplementary Material, we evaluate the effectiveness of our estimators when dealing with skewed distributions. In addition, we did preliminary work to extend the proposed algorithm to the scenario of ``small $n$, large $p$'' and evaluate our idea with a real-world dataset in Section S3.2 of the Supplementary Material. 
Our preliminary exploration shows promising results for both cases. 
Further investigation is needed to address more general scenarios. One possible solution is to adapt depth notions applicable to more general distributions to obtain the $h$-subset and then apply the kernel trick to map the subset to a feature space, where outlier detection can be conducted. The computational time is significantly reduced for high-dimensional scenarios ($n>p$); however, the estimation accuracy still desires further improvement. Rather than a shrinkage estimator, it is also of interest to extend the MCD framework by considering a low-rank and sparse estimator to alleviate the curse of dimensionality.

\section*{Supplementary materials}

\begin{itemize}
    \item We provide an R-package named \texttt{FDB} and R codes of the \texttt{FDB} algorithm proposed in this paper.
    \item The file of supplement involves proofs of theoretical results, additional simulation results as well as preliminary explorations of potential extensions.
\end{itemize}

\section*{Acknowledgement}
We are very grateful to three anonymous referees, an associate editor, and the Editor for their valuable comments that have greatly improved the manuscript. The first two authors contribute equally to the paper.

\begin{spacing}{0.95}
\bibliographystyle{asa}
%\bibliographystyle{rss}
% \scriptsize
\setlength{\bibsep}{1pt}
\bibliography{fdb}
\end{spacing}
\end{document}